\RequirePackage[T1]{fontenc}
\documentclass[11pt,urlcolor=blue, linkcolor=blue]{article} 
\usepackage{cite}
\usepackage{float}
\usepackage[utf8]{inputenc}

\usepackage{tablefootnote}
\usepackage{amsmath, amsthm, amssymb,slashed}
\usepackage{ifpdf}
\ifpdf
  \usepackage[pdftex]{graphicx}
  \usepackage{epstopdf}
\else
  \usepackage[dvips]{graphicx}
\fi
\parskip=\baselineskip

\usepackage[usenames, dvipsnames]{color}
\usepackage[svgnames]{xcolor}
\definecolor{azure}{rgb}{0.0, 0.5, 1.0}
\definecolor{ballblue}{rgb}{0.13, 0.67, 0.8}
\definecolor{airforceblue}{rgb}{0.36, 0.54, 0.66}
\usepackage[colorlinks,citecolor=azure, urlcolor=azure, linkcolor=azure]{hyperref}

\allowdisplaybreaks[1]

\numberwithin{equation}{section}

\usepackage{booktabs}

\theoremstyle{definition}

\definecolor{mygray}{gray}{0.6}

\usepackage{upgreek}
\usepackage{bbm}

\newenvironment{myfont}[2][]{\csname#2\endcsname[#1]}{}

\usepackage{slashed}
\usepackage[makeroom]{cancel}
\usepackage[normalem]{ulem}
\usepackage{soul}
\newcommand{\stkout}[1]{\ifmmode\text{\sout{\ensuremath{#1}}}\else\sout{#1}\fi}

\usepackage{sseq}
\usepackage[all,cmtip]{xy}
\usepackage{tikz-cd}
\usepackage{tikz}
\usetikzlibrary{matrix}

\newcommand{\bea}{\begin{eqnarray}}
\newcommand{\eea}{\end{eqnarray}}
\def\be{\begin{equation}}
\def\ee{\end{equation}}

\newcommand{\ii}{\hspace{1pt}\mathrm{i}\hspace{1pt}}

\def\RP{{\mathbb{RP}}}

\newcommand{\nn}{\nonumber}

\usepackage{color}
\definecolor{red}{rgb}{1,0,0}
\definecolor{blue}{rgb}{0,0,1}
\definecolor{dblue}{rgb}{0,0,0.4}
\definecolor{green}{rgb}{0,1,0}
\definecolor{black}{rgb}{0,0,0}
\definecolor{white}{rgb}{1,1,1}

\definecolor{brn}{rgb}{.8,.4,.0}
\definecolor{redo}{rgb}{1,.5,.0}
\definecolor{ddgrn}{rgb}{0,0.4,0}
\definecolor{dgrn}{rgb}{0,0.55,0}
\definecolor{dbl}{rgb}{0,0,0.5}

\usepackage[bbgreekl]{mathbbol}
\usepackage{amscd}

\newcommand{\Z}{\mathbb{Z}}

\newcommand{\dd}{\hspace{1pt}\mathrm{d}}

\newcommand{\Eq}[1]{(\ref{#1})}

\newcommand{\Tr}{{\rm Tr}}

\newcommand{\bpm}{\begin{pmatrix}}
\newcommand{\epm}{\end{pmatrix}}
\newcommand{\bmm}{\begin{matrix}}
\newcommand{\emm}{\end{matrix}}

\newcommand{\cD}{ {\cal D} }

\newcommand{\cP}{ {\cal P} }

\def\CO{{\cal O}}
\def\CQ{{\cal Q}}

\def\CS{{\cal S}}
\def\CT{{\cal T}}

\def\Z{{\mathbb{Z}}}

\def\Tr{{\mathrm{Tr}}}

\newcommand{\CSg}{\mathrm{CS}_g}

\newcommand{\abk}{\mathrm{ABK}}

\def\Ext{\operatorname{Ext}}

\def \Hom{\operatorname{Hom}}

\def \H{\operatorname{H}}

\def \Z{\mathbb{Z}}
\def \Pin{\mathrm{Pin}}
\def \A{\mathcal{A}}
\def \RP{\mathbb{RP}}

\def\Ext{\operatorname{Ext}}

\usetikzlibrary{decorations.markings}
\usetikzlibrary{positioning}
\usetikzlibrary{shadings}

\newcommand{\SO}{{\rm SO}}
\newcommand{\Spin}{{\rm Spin}}
\newcommand{\U}{{\rm U}}
\newcommand{\SU}{{\rm SU}}
\newcommand{\PSU}{{\rm PSU}}
\newcommand{\Sp}{{\rm Sp}}
\renewcommand{\O}{{\rm O}}

\newcommand{\rE}{{\rm E}}
\newcommand{\rF}{{\rm F}}
\newcommand{\rG}{{\rm G}}

\newcommand{\rR}{\mathrm{R}}
\newcommand{\NS}{\mathrm{NS}}

\def\Sq{\mathrm{Sq}}

\def\B{\mathrm{B}}

\def\TP{\mathrm{TP}}

\newcommand{\arf}{\text{Arf}}

\usepackage{comment}

\newcommand{\tO}{{\rm O}}
\newcommand{\rO}{{\rm O}}

\usepackage{enumitem,cleveref}

\newcommand{\T}{\mathcal{T}}

\usepackage{datetime}
\usepackage{enumitem} 
\usepackage{moreenum}

\newcommand{\Wplusfootnote}[1]{%
\let\oldthefootnote=\thefootnote%
\stepcounter{mpfootnote}%
\addtocounter{footnote}{-1}%
\renewcommand{\thefootnote}{{W$^+$}}
\footnote{#1}%
\let\thefootnote=\oldthefootnote%
}

\newcommand{\Wminusfootnote}[1]{%
\let\oldthefootnote=\thefootnote%
\stepcounter{mpfootnote}%
\addtocounter{footnote}{-1}%
\renewcommand{\thefootnote}{{W$^-$}}
\footnote{#1}%
\let\thefootnote=\oldthefootnote%
}

\newcommand{\Zfootnote}[1]{%
	\let\oldthefootnote=\thefootnote%
	\stepcounter{mpfootnote}%
	\addtocounter{footnote}{-1}%
	\renewcommand{\thefootnote}{{Z$^0$}}
	\footnote{#1}%
	\let\thefootnote=\oldthefootnote%
}

\newcommand{\LG}{G_{\text{Lorentz}}}
\newcommand{\SPT}{\text{SPT}}
\newcommand{\dyn}{\text{dyn}}
\newcommand{\YM}{\text{YM}}
\newcommand{\SM}{{\rm SM}}

\newcommand{\DPin}{\mathrm{DPin}}
\newcommand{\EPin}{\mathrm{EPin}}

\begin{document}
\begin{titlepage}
\begin{center}

{\bf
 \LARGE{Higher Anomalies, Higher Symmetries, 
 and\\[3.75mm]  
 Cobordisms II:}\\[4.75mm]
\Large{Lorentz Symmetry Extension and\\[2mm] 
Enriched Bosonic/Fermionic Quantum 
Gauge Theory}\\[5.75mm]
}

\vskip0.5cm 
\Large{Zheyan Wan$^1$\Wminusfootnote{e-mail: {\tt wanzheyan@mail.tsinghua.edu.cn}}, 
 Juven Wang$^{2,3}$\Wplusfootnote{e-mail: {\tt jw@cmsa.fas.harvard.edu} 
}, 
and Yunqin Zheng$^{4,5,6}$\Zfootnote{e-mail: {\tt yqqzheng@gmail.com}}
\\[2.75mm]  
} 
\vskip.5cm
{\small{\textit{$^1${Yau Mathematical Sciences Center, Tsinghua University, Beijing 100084, China}\\}}
}
 \vskip.2cm
 {\small{\textit{$^2${Center of Mathematical Sciences and Applications, Harvard University,  Cambridge, MA 02138, USA}}\\}}
 \vskip.2cm
 {\small{\textit{$^3${School of Natural Sciences, Institute for Advanced Study, Princeton, NJ 08540, USA} \\}}
 	\vskip.2cm
}
{\small{\textit{$^4${Department of Physics, Princeton University, Princeton, NJ 08540, USA} \\}}
}
	\vskip.2cm
{\small{\textit{$^5${Kavli Institute for the Physics and Mathematics of the Universe, \\University of Tokyo,  Kashiwa, Chiba 277-8583, Japan} \\}}
}
{\small{\textit{$^6${Institute of Solid State Physics, University of Tokyo,  Kashiwa, Chiba 277-8583, Japan}}}
}


\end{center}

\vskip1.cm
\baselineskip 16pt
\begin{abstract}

We systematically study  Lorentz symmetry extensions in quantum field theories (QFTs)  and their 't Hooft anomalies via cobordism. 
The total symmetry $G'$ can be expressed in terms of the extension of Lorentz symmetry $\LG$ by an internal global symmetry $G$ as 
$1 \to G \to G' \to \LG \to 1$. 
By enumerating all possible $\LG$ and 
symmetry extensions, 
{other than the familiar SO/Spin/O/Pin$^{\pm}$ groups,}
we introduce a new EPin group (in contrast to DPin),
and provide natural physical interpretations to exotic groups E($d$),  EPin($d$), (SU(2)$\times$E(d))/$\Z_2$, (SU(2)$\times$EPin(d))/$\Z_2^{\pm}$, etc.
%
By Adams spectral sequence, we systematically classify all possible 
$d$d Symmetry Protected Topological states (SPTs as invertible TQFTs) 
and $(d-1)$d 't Hooft anomalies of QFTs
by co/bordism groups and invariants in $d\leq 5$.
%
%
We further gauge the internal   $G$, and study {Lorentz symmetry-enriched}
Yang-Mills theory 
with discrete theta terms given by gauged SPTs. 
{We not only
enlist familiar \emph{bosonic} Yang-Mills
but also discover new \emph{fermionic} Yang-Mills theories (when $\LG$ contains a graded fermion parity ${\Z_2^F}$),
applicable to bosonic (e.g., Quantum Spin Liquids) or fermionic (e.g., electrons) condensed matter systems.}
For a pure gauge theory, there is a one form symmetry {$I_{[1]}$} associated with the center of the gauge group $G$. We further study the anomalies of the emergent symmetry {$I_{[1]}\times \LG$} by higher cobordism invariants as well as QFT analysis. We focus on the simply connected $G=\SU(2)$ and briefly comment on non-simply connected $G=\SO(3)$, $\U(1)${, other simple Lie groups, and Standard Model gauge groups $(\SU(3)\times \SU(2)\times \U(1))/\Z_q$.} 
{We comment on SPTs protected by Lorentz symmetry, and the symmetry-extended trivialization for their boundary states.}

\end{abstract}

\end{titlepage}

  \pagenumbering{arabic}
    \setcounter{page}{2}
    

\tableofcontents


\section{Introduction and Summary}

Global symmetries and 't Hooft anomalies are key ingredients characterizing topological phases of quantum matters.
For a generic quantum field theory (QFT) denoted $\CQ_d$ in $d$ spacetime dimensions with global symmetry $G$, if one modifies $\CQ_d$ to $\CQ'_d$ by $G$-invariant deformations, it is commonly believed \cite{tHooft:1980xss} that the 't Hooft anomaly of $\CQ_d$ and $\CQ'_d$ are the same. In particular, the 't Hooft anomaly is invariant under renormalization group flow. Recently it has been conjectured \cite{Seiberg2019Strings} that given two quantum field theories $\CQ_d$ and $\CQ'_d$ in the same spacetime dimension, with the same global symmetries and 't Hooft anomalies, one can always add degrees of freedom at short distances to interpolate between $\CQ_d$ and $\CQ'_d$. Hence the global symmetries and the 't Hooft anomalies classifies the \emph{deformation classes} 
of quantum field theories.

It is widely believed that the 't Hooft anomaly of $\CQ_d$ can be cancelled by the anomaly inflow from an invertible topological quantum field theory 
(invertible TQFT or iTQFT) in $(d+1)$ dimensions.\footnote{We clarify several related notions. A QFT in $d$ dimensions with global symmetry $G$ can be anomalous. The anomaly can be captured by $(d+1)$ dimensional invertible TQFT. To emphasize that the $(d+1)$ dimensional invertible TQFT captures the anomaly of $d$ dimensional QFT, we also denote the $(d+1)$ dimensional invertible TQFT as the $(d+1)$ dimensional anomaly inflow action or partition function. The $(d+1)$ dimensional anomaly inflow action should be distinguished from the $(d+2)$ dimensional anomaly polynomial (when $G$ is continuous):  The $(d+2)$ anomaly polynomial is conventionally for the perturbative local anomaly classified by $\Z$ classes (known as free classes in mathematics).
However, 
we may abused the notion ``anomaly polynomial'' as the $(d+1)$d cobordism invariant for the non-perturbative global anomaly by $\Z_n$ classes (known as torsion classes in mathematics). } Physically, the invertible TQFT is characterized by a $G$-Symmetry Protected Topological state ($G$-SPTs)\cite{Chen:2011pg}. Mathematically, the invertible TQFT is characterized by a cobordism invariant \cite{Freed2016,  Kapustin:2014tfa, Kapustin:2014gma, Kapustin:2014dxa,
1711.11587GPW, Yonekura2018arXiv180310796Y, Wan2018bns1812.11967, Witten1909.08775}. 
Following a previous work \cite{Wan2018bns1812.11967},
the purpose of this work is to derive the cobordism invariants for various $G$ relevant for QFTs in various dimensions, hence potentially classifying the deformation classes of QFTs for given symmetries.

The examples we study in this work are the pure gauge theories whose gauge groups are small rank Lie groups $G$. In particular, we focus on $G=\SU(2)$.  Following \cite{1711.11587GPW, Wan2019oyr1904.00994}, there are  different $d$-dimensional gauge theories for the same gauge group $G$, which are obtained by gauging different $d$-dimensional $G$-SPTs. Thus in this work, we systematically construct the pure gauge theories in two steps: 
\begin{enumerate}
	\item We first classify the $d$-dimensional $G$-SPTs, by computing the cobordism invariants in the same dimension. Here $G$ is a unitary global symmetry, in particular we focus on $G=\SU(2)$. One can perform the similar exercise for other symmetry groups in the same spirit. 
	\item We further gauge the global symmetry $G$  to obtain a pure gauge theory with dynamical gauge group $G$. After gauging, there are emergent global symmetries. Because the gauge group that  we consider is simply connected, there is no magnetic emergent symmetry. However, since this paper focuses only on the pure gauge theories obtained by gauging SPTs, there is a one-form symmetry $I_{[1]}$ associated with the center of the dynamical gauge group $G$.  We further compute the 't Hooft anomalies involving the  emergent symmetry $I_{[1]}$, and match them with  the cobordism invariant of the emergent symmetries in $(d+1)$ dimensions. We find that if the nontrivial SPT involves $G$ gauge fields, the pure gauge theory obtained via gauging $G$ typically has a nontrivial anomaly involving the emergent one form symmetry. 
\end{enumerate}
In the rest of this introduction section, we explain the above two steps in detail.

\subsection{Lorentz Symmetry Enriched to $G'$-Symmetry, and  Their Extensions}
\label{secLorentz}

We first consider symmetry protected topological phases with the internal unitary global symmetry $G$. In this work, we focus on $G=\SU(2)$ which is a continuous small rank 0-form symmetry.  By internal, we demand that $G$ does not act on the coordinates. One can similarly consider  $G=\SU(3), \SO(3), \U(1)$ etc. These symmetries are related to the Standard Models of particle physics \cite{Aharony:2013hda, Tong:2017oea, Garcia-Etxebarria:2018ajm, 
Wang2018cai1809.11171,
Davighi:2019rcd,  2019arXiv191014668W, Ang:2019txy, JW2006.16996,JW2008.06499}. Beyond the internal symmetries, we assume the SPTs preserve  the Lorentz symmetry $G_{\text{Lorentz}}$. The Lorentz symmetry can be classified by the Stiefel-Whitney classes of the tangent bundle of the spacetime manifold $M$, which is denote as $w_i(TM)\equiv w_i$. Throughout this paper, we drop the $TM$ dependence and simply write it as $w_i$. Here we focus on the first and the second Stiefel-Whitney classes. By enumerating whether $w_1, w_1\cup w_1$ and $w_2$ are trivial or unconstrained, we find seven 
possibilities,\footnote{Here the $w_1\cup w_1$ has the cup product $\cup$.
 In general,  
 all the product notations between cohomology classes are cup product, such as $w_2 w_3  : =   w_2(TM)w_3(TM)=  w_2(TM)\cup w_3(TM)$.
All the product notations between a cohomology class $w_j$ and fermionic invariants such as Arf (or ABK, $\tilde{\eta}$  etc) , 
namely $w_j {\rm Arf}$,
means the value of Arf (or ABK, $\tilde{\eta}$  etc.) 
on the submanifold of $M$ which represents the Poincar\'e dual of $w_j$.
In other words, the  $w_j {\rm Arf} : = w_j \cup  {\rm Arf}={\rm Arf}  (\text{PD}(w_j))$.} 
listed in Table \ref{table:SWLG}. The Lorentz symmetries $\SO(d), \Spin(d), \O(d)$, and 
$\Pin^{\pm}(d)$ are extensively discussed in the literature \cite{1711.11587GPW, Cordova2017vab1711.10008}, but the two cases $\rE(d)$ and $\EPin(d)$ are less well-known. See \cite{Freed2016, Wan2018zql1812.11968, Wan2019oyr1904.00994} for an exploration of the $\rE(d)$ group and its application to physics and Yang-Mills gauge theory. As far as we know, the $\EPin(d)$ group is new in literature, though \cite{Kaidi:2019tyf} has an exploration of the analogous $\DPin(d)$ group and its string theory application. 
\footnote{The $\DPin(d)$ studied in \cite{Kaidi:2019tyf} is slightly different from the $\EPin(d)$ studied in this paper. In \cite{Kaidi:2019tyf}, the $\DPin(d)$ is defined by the extension $1\to \Z_2^+\times \Z_2^- \to \DPin(d) \to \O(d)\to 1$ which is very similar to \eqref{Z2Z2}, except that the extension is not central. The orientation reversal in $\O(d)$ acts on $\Z_2^+\times \Z_2^-$ by exchanging the two $\Z_2$ factors. In this paper, however,  the orientation reversal in $\O(d)$ does not act on $\Z_2^+\times \Z_2^-$ in the extension \eqref{Z2Z2}, i.e., the extension is central. In fact, all the extensions we will discuss in this paper are central.  }

\begin{table}
	\centering
	\begin{tabular}{|c | c  c  c |}
		\hline
		Group & $w_1$ & $w_1\cup w_1$ & $w_2$ \\ 
		\hline
		$\SO(d)$ & 0 & 0 & unrestricted \\
		$\Spin(d)$ & 0 & 0 & 0 \\
		$\O(d)$ & unrestricted & unrestricted & unrestricted \\
		$\rE(d)$ & unrestricted & 0 & unrestricted \\
		$\Pin^+(d)$ & unrestricted & unrestricted & 0 \\
		$\Pin^-(d)$ & unrestricted & \multicolumn{2}{c|}{$w_1\cup w_1=w_2$} \\
		$\EPin(d)$ & unrestricted & 0 & 0 \\
		\hline
	\end{tabular}
	\caption{{The bundle constraints for} Stiefel-Whitney classes of the Lorentz symmetries. The $w_i\equiv w_i(TM)$
	are Stiefel-Whitney (SW) classes of the tangent bundle $TM$ of the spacetime manifold $M$. 
	{The SW classes are mod 2 cohomology classes, thus all enlisted constraints are also mod 2.
		In $d=2$, the {$w_2 + w_1\cup w_1 = 0$ mod 2} is always true, therefore all 2d smooth manifolds have $\Pin^-$ structures. Notice that for both $\Pin^+(2)$ and $\EPin(2)$, the bundle constraints are $w_2=w_1^2=0$. But this does not imply $\Pin^+(2)$ and $\EPin(2)$ are the same. See discussions around \eqref{1.4} for more details. 
	}
	}
	\label{table:SWLG}
\end{table}

Whether $w_i$ is trivial or unconstrained has significant physical consequences. If $w_2$ is unconstrained, the manifold does not admit a spin/Pin$^{\pm}$ structure, and a quantum field theory that can be defined on such a manifold can not contain a fermion in the operator spectrum. We denote such a theory to be bosonic.   On the other hand, if $w_2=0$, the theory allows a fermion in the operator spectrum and we denote it as fermionic. When there are additional internal global symmetries apart from the Lorentz symmetry, a modification will occur, which will be discussed in section \ref{secSU2}. 

If $w_1=0$, the spacetime manifold is oriented. A quantum field theory that can only be defined on such an oriented manifold does not have time reversal symmetry. On the other hand, if $w_i=0$, the spacetime manifold is unorientable. A quantum field theory that can be defined on such an unorientable manifold is time reversal symmetric. Depending on whether $w_1\cup w_1$ is restricted or not, the Kramers parity of certain local operator in the operator spectrum can vary. We enumerate all different possibilities of $\LG$ as 
follows:
\begin{eqnarray}\label{Glorentz}
\LG=
\left\{\begin{array}{llll}
\SO(d), & ~~~ \text{bosonic}, &\\
\Spin(d), &~~~\text{fermionic}, & &  \Z_2^{F}.\\
\O(d), &~~~ \text{bosonic}, & \CT^2=1, & \Z_2^{T}.\\
\mathrm{E}(d), &~~~ \text{bosonic}, & \CT^2=-1,  & \Z_4^{TB}.\\
\Pin^+(d), &~~~ \text{fermionic}, & \CT^2=(-1)^F, & \Z_4^{TF}.\\
\Pin^-(d), &~~~ \text{fermionic}, & \CT^2=1, & \Z_2^{T} \times  \Z_2^{F}.\\
\EPin(d), &~~~ \text{fermionic}, & \CT^2(\psi_+, \psi_-)=(\psi_+, -\psi_-), & \Z_4^{TF} \text{ and }  \Z_2^{T} \times  \Z_2^{F}.
\end{array}\right.
\end{eqnarray}
{In the last column of \Eq{Glorentz}. we also summarize the discrete part of the extended Lorentz groups in terms of the notations friendly to condensed matter people.
\\
$\bullet$ The $\Z_2^{F}$ has a fermion parity generator $(-1)^F$ whose square is $+1$.\footnote{{One way 
to see whether we have bosonic or fermions theory in  \Eq{Glorentz} is based on whether the $\LG$ contains a 
graded fermion parity ${\Z_2^F}$ symmetry as a normal subgroup.
The $\Spin(d)$ and $\Pin^{\pm}(d)$ contain a normal ${\Z_2^F}$, 
while $\EPin(d)$ contains two fermion parity symmetries ${\Z_2^{F_+}} \times {\Z_2^{F_-}}$.   
}}
 \\
$\bullet$ The $\Z_2^{T}$ has a time reversal symmetry generator $\CT$ whose square is $\CT^2=+1$. \\
$\bullet$ The $\Z_4^{TF}$ has a $\CT$  generator  whose square is $\CT^2=(-1)^F$ of the fermion number $F$. \\
$\bullet$ The $\Z_4^{TB}$ has a $\CT$  generator whose square is $\CT^2=(-1)^B$ of the boson number $B$. \\
The normal subgroup 
of  $\Z_4^{TF}$ is  $\Z_2^{F}$, 
so a short exact sequence says $1 \to  \Z_2^{F} \to \Z_4^{TF} \to \Z_2^{T}  \to 1$. 
The normal subgroup 
of  $\Z_4^{TB}$ is  $\Z_2^{B}$, 
so $1 \to  \Z_2^{B} \to \Z_4^{TB} \to \Z_2^{T}  \to 1$.} 
In the last case, the $\EPin(d)$ symmetry can be regarded as being both the $\Pin^+(d)$ and $\Pin^-(d)$ symmetry simultaneously, because $w_1^2=w_2=0$ implies $w_1^2+w_2=0$ ($\Pin^-$ condition) and $w_2=0$ ($\Pin^+$ condition). Because each fermion can only transform with a definite Kramers parity (either Kramers singlet or Kramers doublet), being simultaneously $\Pin^+(d)$ and $\Pin^-(d)$ means there are two species of fermions presented in the $\EPin$ structure, $\psi_+$ and $\psi_-$, being Kramers singlet $\CT^2=+1$ and doublet $ \CT^2=-1$ respectively.  The discrete parts of the $\EPin$ symmetry correspond to $\Z_2^{T} \times  \Z_2^{F}$ for $\psi_+$ and $\Z_4^{TF}$ for $\psi_-$.  \footnote{We thank R.Thorngren for clarifying the physical interpretation of the EPin structure. } Mathematically, the $\EPin(d)$ group appears in the extension of $\O(d)$ by  $\Z_2^+\times \Z_2^-$. 
\begin{eqnarray}\label{Z2Z2}
{
1\to \Z_2^+\times \Z_2^- \to \EPin(d) \to \O(d)\to 1.
}
\end{eqnarray}
Hence $\Z_2^+\times \Z_2^-$ belongs to the center of $\EPin(d)$. Taking the quotient of $\Z_2^+$ or $\Z_2^-$ yields $\Pin^{\pm}(d)$ respectively, 
\begin{eqnarray}
\Pin^+(d)  = \frac{\EPin(d)}{\Z_2^+}, ~~~\Pin^-(d)  = \frac{\EPin(d)}{\Z_2^-}.
\end{eqnarray}

When $d=2$, there is a relation $w_2=w_1^2$ for every 2d manifold. This implies that every 2d manifold is a $\Pin^-$ manifold, and furthermore when $w_2$ is trivial, the manifold is a $\Pin^+$ manifold. Hence for 2d $\Pin^+$ manifold, both $w_1^2$ and $w_2$ are trivial. This relation is the same as that of the 2d $\EPin$ manifold. However, this does not imply $\Pin^+(2)$ and $\EPin(2)$ are the same. As we argue below, $w_1^2$ and $w_2$ are trivialized in two different ways for $\Pin^+$ and $\EPin$ manifolds respectively. For 2d $\Pin^+$ manifold, we have the extension 
\begin{eqnarray}\label{1.4}
1\to \Z_2 \to \Pin^+(2) \to \O(2) \to 1
\end{eqnarray}
Suppose the 1-cochain in $\Z_2$ (i.e. the generator of $\Z_2$) is $a$, then the $w_2=w_1^2$ in $\O(2)$ in trivialized in $\Pin^+(2)$ via
\begin{eqnarray}
\Pin^+(2): ~~~w_2=w_1^2= \delta a
\end{eqnarray}
where $a$ parameterizes the $\Pin^+$ structure. 
Moreover, for 2d $\EPin$ manifold, we have the extension \eqref{Z2Z2}. Suppose the 1-cochains in $\Z_2\times \Z_2$ are $a$ and $b$, then the $w_2, w_1^2$ in $\O(2)$ are trivialized in $\EPin(2)$ via
\begin{eqnarray}
\EPin(2): ~~~w_2=\delta a, ~~~ w_1^2+w_2=\delta b
\end{eqnarray}
where the pair $(a,b)$ is the $\EPin$ structure. In components, $a$ is the $\Pin^+$ structure, and $b$ is the $\Pin^-$ structure. Thus we have seen explicitly that $w_1^2$ and $w_2$ are trivialized in different ways in $\Pin^+(2)$ and $\EPin(2)$.

Given the internal global symmetry group $G$, and the Lorentz symmetry group $\LG$, the total group is given by the extension 
\begin{eqnarray}\label{groupext}
1\to G\to G'\to \LG \to 1.
\end{eqnarray}
In particular, the total group $G'$ does not have to be the {simple} direct product of the internal and Lorentz groups $G\times \LG$. For example, let us consider a bosonic theory without time reversal, hence $\LG=\SO(d)$, and let us take the internal symmetry to be $\SU(2)$. There are two choices of the total group, $G'= \SU(2)\times \SO(d)$, and $G'= (\SU(2)\times\Spin(d))/\Z_2$. 
{Physically,} the latter means that the field carrying charge 1 under $\Z_2\subset \SU(2)$ also transforms under the fermion parity $\Z_2^F\subset \Spin(d)$.\footnote{{Another way
to rephrase is that the spacetime spinor (as fermions) must be in the 
even integer representation of SU(2) (say {\bf 2}, {\bf 4}, {\bf 6}, etc., namely isospin 1/2, 3/2, 5/2 , etc.) instead of the 
odd integer representation of SU(2) (say {\bf 1}, {\bf 3}, {\bf 5}, etc., which is also an integer spin
representation of SO(3) (isospin 1, 2, 3, etc.).}  
} 
{Such a relation is named the {Spin$^h$ relation} or {Spin-SU(2) relation}\cite{Cordova2018acb1806.09592, Wang2018cai1809.11171, Wang:2018qoyWWW}.}
Notice that although there is a $\Spin(d)$ in the total group, it does not mean that the theory is fermionic. If the internal symmetry $\SU(2)$ is 
{entirely}
broken, the total group is still $\Spin(d)/\Z_2= \SO(d)$ hence  it is still bosonic. 

{Follow the terminology of \emph{Lorentz symmetry fractionalization} \cite{Hsin1909.07383} and \emph{Lorentz symmetry enrichment} \cite{Wan2018zql1812.11968, Wan2019oyr1904.00994} to $G'$ from the $\LG$ by an internal symmetry $G$, we will call these QFTs with the internal-spacetime symmetry structure $G'$ in \Eq{groupext}
as the \emph{Lorentz symmetry enriched QFTs}.}

\subsection{Bordism Group, Cobordism Group, and SPTs}

Given the total group $G'$, the SPTs protected by the total symmetry $G'$ is {classified} by the cobordism group \cite{Freed2016}
\begin{eqnarray}\label{cobor}
\TP_d(G')
\end{eqnarray}
{and their invariants are given by the cobordism group generators: cobordism invariants.}
Here $\TP$ is a shorthand for topological phase. 
Practically, it is useful to first compute the bordism group $\Omega_d^{G'}$. $\TP_d(G')$ is 
	 related to the bordism group $\Omega_d^{G'}$ by the short exact sequence
	\bea\label{eq:TPexact}
	0\to\Ext^1(\Omega_d^{G'},\Z)\to\TP_d(G')\to\Hom(\Omega_{d+1}^{G'},\Z)\to0.
	\eea
{See \cite{Freed2016,  1711.11587GPW, Wan2018bns1812.11967} for the notations:
\begin{itemize}
\item the $\Ext^1(\Omega_d^{G'},\Z)$ includes the finite abelian group classification $\Z_n$ part (the torsion part), classifying
nonperturbative global anomalies (not captured by Feynman diagram).
Examples include Witten SU(2) anomaly \cite{Witten:1982fp} and the new SU(2) anomaly \cite{Wang:2018qoyWWW}.
\item the $\Hom(\Omega_{d+1}^{G'},\Z)$ includes
the infinite abelian group classification $\Z$ part (the free part), classifying
perturbative local anomalies (captured by Feynman diagram).
Examples include Adler-Bell-Jackiw anomalies \cite{Adler1969gkABJ, Bell1969tsABJ} captured by one-loop Feynman diagrams.
\end{itemize}
}
The majority part of this work is to use the Adams spectral sequence to compute $ \Omega_d^{G'}$ from which $\TP_d(G')$ can be inferred. Physically, the $G'$-SPT given by $\TP_d(G')$ is the precursor of the  gauge theory with dynamical gauge group $G$. We can also view such $G$-SPT as the anomaly inflow action  for $(d-1)$ dimensional quantum field theory with the 0-form global symmetry $G$. 

The following fact is useful to infer $\TP_d(G') $ from  $\Omega_d^{G'}$.  For all the examples in this work, $\Omega_d^{G'}$ is a tensor product of finite order cyclic groups $\Z_p$ and infinite order cyclic group $\Z$.  If $\Z_p$ is a subgroup of $\Omega_d^{G'}$, then $\Z_p$ is also a subgroup of $\TP_d(G')$. If $\Z$ is a subgroup of $\Omega_d^{G'}$, then {this $\Z$ becomes} a subgroup of $\TP_{d-1}(G')$ instead. In summary, 
\begin{eqnarray}
\begin{split}
\Z_p \subset \Omega_d^{G'} ~~&\Rightarrow~~ \Z_p \subset \TP_d(G'), \\
\Z \subset \Omega_d^{G'} ~~&\Rightarrow~~ \Z \subset \TP_{d-1}(G').
\end{split}
\end{eqnarray}

\subsection{A Mathematical Primer For $\Omega_d^{G'}$}

We give a mathematical primer for computing the bordism group $\Omega_d^{G'}$, see \cite{Freed2016, Wan2018bns1812.11967, 2019arXiv191014668W}  for details. 
We will use the generalized Pontryagin-Thom isomorphism,
\begin{eqnarray}
\Omega_d^{G'}=\pi_d(MTG')
\end{eqnarray}
to identify the cobordism group $\Omega_d^{G'}$ as the homotopy group of the Madsen-Tillmann spectrum $MTG'$. Hence computing $\Omega_d^{G'}$ is equivalent to computing $\pi_d(MTG')$.

To compute $\pi_d(MTG')$, we use the Adams spectral sequence
\begin{eqnarray}\label{ADSS}
E_{2}^{s,t}\equiv \Ext_{\A_p}^{s,t}(\H^*(Y,\Z_p),\Z_p)~~\Rightarrow~~\pi_{t-s}(Y)_p^{\wedge}.
\end{eqnarray}
We explain the notations below. 
Here $\A_p$ is the mod $p$ Steenrod algebra. When $p=2$, $\A_2$ is generated by Steenrod squares $Sq^i$.  $Y$ is any spectrum, which will be identified with the Madsen-Tillmann spectrum $Y=MTG'$. For any finitely generated abelian group $G$, we define  $G_p^{\wedge}=\lim_{n\to\infty}G/p^nG$ to be the $p$-completion of $G$. $\Ext_{\A_p}^{s,t}(\H^*(Y,\Z_p),\Z_p)$ is the second page, i.e., the $E_2$ page,  of the Adam spectral sequence. Using the differential $d_r^{s,t}: E_r^{s,t}\to E_r^{s+r, t+r-1}$, one can {determine} the $r+1$-th page $E_{r+1}$ from the data in the $r$-th page $E_r$, via $E_{r+1}^{s,t}= \frac{\ker d_r^{s,t}}{\mathrm{im} d_r^{s-r, t-r+1}}$. Thus from the $E_2$ page, one can determine the $E_3$ page, $E_4$ page, etc subsequently. The sequence $\{E_r, r=2,3,4...\}$  will stabilize until certain $r$, and one denotes the stabilized page as $E_{\infty}^{s,t}$.  The double arrow in \Eq{ADSS} means that the sequence will finally converge to the stabilized page $E_{\infty}^{s,t}$, and one can use the data in $E_{\infty}^{s,t}$ to extract $\pi_{t-s}(Y)_p^{\wedge}$. 

In all the examples considered in this paper, the $p=2$ suffices for computing the while cobordism group, i.e., $\pi_{t-s}(Y)_2^{\wedge} = \pi_{t-s}(Y)$. Hence we claim that for $p=2$, the Adams spectral sequence \eqref{ADSS} with $Y=MTG'$, i.e., $ \Ext_{\A_2}^{s,t}(\H^*(MTG',\Z_2),\Z_2)~~\Rightarrow~~\pi_{t-s}(MTG')_2^{\wedge}.$ completely determines $\Omega_d^{G'}$.\footnote{{See some counter examples in \cite{Wan2018bns1812.11967, Guo2018vij1812.11959} that $p=2$ is \emph{not} enough to determines $\Omega_d^{G'}$.}} 


Let us focus on an example. If $MTG'=M\Spin\wedge X$ where $X$ is an  arbitrary topological space,
by Corollary 5.1.2 of \cite{2018arXiv180107530B}, we have
\bea
\Ext_{\A_2}^{s,t}(\H^*(M\Spin\wedge X,\Z_2),\Z_2)=\Ext_{\A_2(1)}^{s,t}(\H^*(X,\Z_2),\Z_2)
\eea
for $t-s<8$.
Here $\A_2(1)$ is the subalgebra of $\A_2$ generated by $\Sq^1$ and $\Sq^2$.
Hence for the dimension $d=t-s<8$, the Adams spectral sequence \eqref{ADSS} reduces to 
\bea\label{eq:ExtA_2(1)}
\Ext_{\A_2(1)}^{s,t}(\H^*(X,\Z_2),\Z_2)\Rightarrow(\Omega_{t-s}^{G'})_2^{\wedge}.
\eea
The $\H^*(X,\Z_2)$ is an $\A_2(1)$-module whose internal degree $t$ is given by the $*$.
Our computation of $E_2$ pages of $\A_2(1)$-modules is based on 
Lemma 11 of \cite{Wan2018bns1812.11967}.
More precisely, we find a short exact sequence of $\A_2(1)$-modules $0\to L_1\to L_2\to L_3\to 0$, then apply Lemma 11 of \cite{Wan2018bns1812.11967} to compute $\Ext_{\A_2(1)}^{s,t}(L_2,\Z_2)$ by the data of $\Ext_{\A_2(1)}^{s,t}(L_1,\Z_2)$ and $\Ext_{\A_2(1)}^{s,t}(L_3,\Z_2)$. 
Our strategy is choosing $L_1$ to be the direct sum of suspensions of $\Z_2$ on which $\Sq^1$ and $\Sq^2$ act trivially, then we take $L_3$ to be the quotient of $L_2$ by $L_1$.
{If $\Ext_{\A_2(1)}^{s,t}(L_3,\Z_2)$ is undetermined, then we take $L_3$ to be the new $L_2$ and repeat this procedure.
	We can use this procedure again and again until $\Ext_{\A_2(1)}^{s,t}(L_2,\Z_2)$ is determined.}

\subsection{Gauging $G$, Emergent Symmetries, and Anomalies}

We further gauge the global symmetry $G$ in the $G$-SPT computed from \eqref{cobor}. Suppose the $G$-background field is $A$, 
{where $A$ is a connection of the $G$-bundle}.  Denote the partition function of the SPT $Z_{\SPT}[A]$. Because $G$-SPT is also an iTQFT, the $Z_{\SPT}[A]$ is a {complex} 
$\U(1)$ phase, {whose inverse
$Z_{\SPT}[A]^{-1}$ represents the inverted iTQFT that cancels the original $Z_{\SPT}[A]$ iTQFT}. Gauging $G$ amounts to summing over $A$ in the path integral. The partition function after gauging $G$ is\cite{Seiberg:2010qd, Aharony:2013hda, Freed:2017rlk, Cordova2017vab1711.10008, 1711.11587GPW, Ang:2019txy, Kaidi:2019tyf, Kaidi:2019pzj}
\begin{eqnarray}\label{pathint}
Z = \int [\cD A] ~Z_{\SPT}[A] ~Z_{\dyn}[A],
\end{eqnarray}
where $Z_{\dyn}[A]$ is the partition function of a non-topological theory, which has nontrivial coupling constants and can flow under the renormalization group. Since in this work we focus on pure gauge theories, we will exclusively take $Z_{\dyn}[A]$ to be the Yang-Mills action $Z_{\YM}[A]$, where
\begin{eqnarray}
Z_{\YM}[A]= \exp\left(-\frac{  \ii}{4 g^2} \int \Tr \left(F \wedge \star F\right) \right).
\end{eqnarray}

After gauging the 0-form symmetry $G$, the dynamical gauge theory exhibits emergent global symmetries. If $G$ has nontrivial center, there is an emergent 1-form global symmetry $G_{e,[1]}$.
For instance, when $G=\SU(2)$, $G_{e,[1]}=\Z_{2, [1]}$; when $G=\SO(3)$, $G_{e, [1]} = 0$. 
Notice that this is no longer true if one includes matter in the fundamental representation of $G$.

Although this paper mainly focus on  $G=\SU(2)$, it is beneficial to comment on a few more examples of emergent symmetries which is not featured in the $\SU(2)$ example. 
(See Table \ref{table:1-sym} and \ref{table:standardmodel} for instances.)
In particular, 
if either $u_2(V_G)\in \H^2(\B G, \U(1))$ or $u_2(V_{G})\in \H^2(\B G, \Z_n)$ for some integer $n$ is nontrivial, then there is an emergent $(d-3)$-form global symmetry, i.e. $G_{m,[d-3]}$. \cite{Bhardwaj:2017xup, Tachikawa:2017gyf, Delacretaz:2019brr}  This is to be contrasted with the case when $G$ is discrete. If $G$ is a discrete 0-form symmetry, gauging $G$ in $d$ dimensions will induce a $(d-2)$ form global symmetry. In our case, $G$ is a continuous 0-form symmetry, and gauging it will induce a $(d-3)$-form global symmetry. For example, when $G=\U(1)$, $G_{m,[d-3]}=\U(1)_{[d-3]}$; when $G=\SO(3)$, $G_{m,[d-3]}= \Z_{2,[d-3]}$.

In all the examples considered in this paper, after gauging $G$, different choices of extensions \eqref{groupext} give the same trivial product algebra between $G_{e,[1]}$ and $\LG$, i.e., the emergent total symmetry is $G_{e,[1]}\times \LG$.\footnote{If $G_{m,[d-3]}$ is also nontrivial, the total gauge group can be nontrivial extension of $\LG$ by $G_{m,[d-3]}$. } However, the distinct extensions \eqref{groupext} yield distinct anomalies of $G_{e,[1]}\times \LG$, which 
 can be systematically enumerated  from the cobordism group
\begin{eqnarray}\label{highcobor}
\TP_d(G_{e,[1]}\times \LG). 
\end{eqnarray}
This should be distinguished from \eqref{cobor} because in \eqref{highcobor} the $G'=G_{e,[1]}\times \LG$ is a higher group. We emphasize that different choices of $Z_{\SPT}[A]$ in the path integral \eqref{pathint} can dramatically affect the dynamics of gauge theories after gauging $G$, in particular, the 't Hooft anomalies are in general different for different $Z_{\SPT}[A]$.  For example, for $d=4$, we consider $G=\SU(2)$ and $\LG=\O(4)$. The $G$-SPT includes a 4d theta term with $\theta=0$ and $\pi$. It has been extensively discussed \cite{Gaiotto2017yupZoharTTT1703.00501} and \cite{Wan2019oyr1904.00994, Wang:2019obe} that after $\SU(2)$ is gauged, there is a mixed anomaly between the emergent 1-form symmetry $\Z_{2,[1]}$ and $\O(4)$ for $\theta=\pi$ while no anomaly for $\theta=0$. Hence as a consequence, the dynamics for $\theta=0$ and $\pi$ are dramatically different. See \cite{Gaiotto2017yupZoharTTT1703.00501, 2019arXiv191014668W, Wang:2019obe, AnberPoppitz2018tcj1805.12290, 
Cordova2018acb1806.09592, Wan2018djlW21812.11955, Anber_2020, Anber_2019,Tanizaki_2020,Aitken_2018, Wan2019oyr1904.00994} for constraints of gauge theory dynamics from symmetries and anomalies.

\subsection{Summary in Tables: Higher Symmetries in Gauge Theory and  (Co)bordism Groups}

For the convenience of readers, for various gauge group $G$,
we summarize various results on one-form symmetries
in Tables \ref{table:1-sym} and \ref{table:standardmodel}, 
bordism groups in Table \ref{table:tableOmega}, 
and cobordism groups in Table \ref{table:tableTP}.

 In Table \ref{table:1-sym},
  for various gauge group $G$
we summarize the associated one-form electric symmetry $G_{[1],e}$  
(relevant for the center of the gauge group $Z(G)$, in Table \ref{table:1-sym}'s second column)
and one-form magnetic symmetry $G_{[1],m}$ 
(relevant for the first homotopy group $\pi_1(G)$, 
in Table \ref{table:1-sym}'s third column) for pure gauge theories without matter fields in spacetime dimensions $d=4$.
We also list down one-form symmetry $G_{[1]}$ when gauged matter fields in representation (Rep ${\bf R}$) present, in Table \ref{table:1-sym}'s last column.
See also a related discussion on one-form symmetries in a recent work \cite{Hsin2007.05915}. 

In Table \ref{table:standardmodel}, as an example, 
for gauge theories with a Standard Model gauge group
\bea \label{eq:GSM}
G_{\SM_q}\equiv \frac{\SU(3)\times \SU(2)\times \U(1))}{\Z_q},
\eea
where $q=1,2,3,6$ (in the first column),
we show the one-form electric and magnetic symmetries 
for pure gauge theories (in the second and the third columns).
We also show the one-form symmetries for SM with matter fields in the SM Rep (in the last column). 
See related recent work 
 \cite{Tong:2017oea, Garcia-Etxebarria:2018ajm, 
Wang2018cai1809.11171, Davighi:2019rcd,  2019arXiv191014668W, JW2006.16996,JW2008.06499}
for an overview:
\bea \label{eq:rep-3generations}
&&\text{(SU(3) representation, SU(2) representation, hypercharge $Y$)}\Rightarrow\\
&&\Bigg( ({\bf 3},{\bf 2}, 1/6)_L,({\bf 3},{\bf 1}, 2/3)_R,({\bf 3},{\bf 1},-1/3)_R,({\bf 1},{\bf 2},-1/2)_L,({\bf 1},{\bf 1},-1)_R  \Bigg)\times \text{3 generations}. \nn
\eea
The subscript $L$ and $R$ here are left-handed and right-handed Lorentz spacetime spinors.
In fact, each of the triplet given above \Eq{eq:rep-3generations} are in the smallest $\frac{\SU(3)\times \SU(2)\times \U(1))}{\Z_6}$ representations, although 
there are still four choices of gauge groups  $G_{\SM_q}$ \Eq{eq:GSM} with $q=1,2,3,6$.

\begin{table}[!h]
	\centering
	{
	\hspace{-16.mm}
	\;\;\;
	\fontsize{11}{10pt} \selectfont
	\begin{tabular}{| c c c c |}
		\hline
		\multicolumn{4}{|c|}{One-form symmetries of $G$-gauge theories at $d=4$, with or without matter fields}\\
		\hline
		 gauge group $G$ & 
		1-form $e$ sym $G_{[1],e}$  &
		1-form $m$ sym $G_{[1],m}$
		& 
		$
         	\begin{array}{l}
		\text{$G_{[1]}$ with matter}\\
		\text{in Rep {\bf R}}: G_{[1]}
		\end{array}
		$
		 \\
		\hline
		\SU(2)& ${\Z_2}_{[1],e}$& no & 
		$
		\begin{array}{rl}
		\text{fund $\bf 2$:}& \text{no}. \\ 
		\text{adjoint $\bf 3$:}& {\Z_2}_{[1],e}.  
		\end{array}
		$ 
		 \\
		 \hline
		\SO(3)& no & ${\Z_2}_{[1],m}$ &  \text{vector $\bf 3$}: ${\Z_2}_{[1],m}$. \\
		\hline
		$\SU(N), N\geq 2$ &${\Z_N}_{[1],e}$ & no  &
		$
		\begin{array}{rl}
		\text{fund $\bf N$}:& \text{no.} \\ 
		\text{adjoint $\bf N^2-1$}:&  {\Z_N}_{[1],e}.
		\end{array}
		$ \\
		\hline
               $\PSU(N), N\geq 2$ & no & ${\Z_N}_{[1],m}$ & {\text{adjoint} $\bf N^2-1$}: $ {\Z_N}_{[1],m}$.\\
               \hline
               $\U(1)$ & ${\U(1)}_{[1],e}$ & ${\U(1)}_{[1],m}$  &
               $
               \begin{array}{l}
		{\text{charge $q$: ${\Z_q}_{[1],e}$, ${\U(1)}_{[1],m}$.}} 
		\end{array}
		$
		\\
               \hline
               $\U(N), N> 1$ & ${\U(1)}_{[1],e}$  & ${\U(1)}_{[1],m}$  &
               {$
               \begin{array}{l}
		\text{Given a U(1) charge $q$}\\
		\text{$q=1$, fund $\bf N$: ${\U(1)}_{[1],m}$.} \\ 
		\text{$q=0$, adjoint $\bf N^2-1$:}\\ 
		\quad\quad \text{${\U(1)}_{[1],e}, {\U(1)}_{[1],m}$.}
		\end{array}
		$}
		\\
               \hline
		$\SO(N), N>2$ & 
		$
		\begin{array}{rl}
			{\Z_2}_{[1],e}, & N=0\mod 2\\
			\text{no}, & N=1\mod 2
		\end{array}
		$ & ${\Z_2}_{[1],m}$ &
		$ 
		\begin{array}{l}
                \text{vector $\bf N$: ${\Z_2}_{[1],m}$.}
		\end{array}$
		 \\
		 \hline
		$\Spin(N), N>2$  & 
		$
		\begin{array}{rl}
		{\Z_4}_{[1],e},& N=2 \mod 4\\ 
		({\Z_2}_{[1],e})^2,& N=0 \mod 4\\ 
		{\Z_2}_{[1],e},& N=1 \mod 2 
		\end{array}
		$
		 &  no & 
		  $
		\begin{array}{l}
		\text{spinor irrep $\bf 2^{[\frac{N-1}{2}]}$}:\text{no}.
		\end{array}
		$
		 \\
		 \hline
		$\Sp(N), N \geq 1$   & $\Z_{2[1],e}$ &  no &  fund: no.\\
		 \hline
		$\rG_2$   & 0 & 0  & \\
				 \hline
		$\rF_4$   &  0& 0  & \\
		\hline
		$\rE_6$   & 
		$\begin{array}{r l}
		\Z_{3[1],e}, &\text{for $\rE_6^{\text{simp.cn}}$}
		\\
		\text{no}, &\text{for $\rE_6^{\text{adjoint}}$}
		\end{array}
		$ &
		$\begin{array}{r l}
		\text{no} , &\text{for $\rE_6^{\text{simp.cn}}$}
		\\
		\Z_{3[1],m}, &\text{for $\rE_6^{\text{adjoint}}$}
		\end{array}
		$
		 &
		 \\
		\hline
		$\rE_7$   & 
		$\begin{array}{r l}
		\Z_{2[1],e}, &\text{for $\rE_7^{\text{simp.cn}}$}
		\\
		\text{no}, &\text{for $\rE_7^{\text{adjoint}}$}
		\end{array}
		$ 
		 &
		 $\begin{array}{r l}
		\text{no} , &\text{for $\rE_7^{\text{simp.cn}}$}
		\\
		\Z_{2[1],m}, &\text{for $\rE_7^{\text{adjoint}}$}
		\end{array}
		$
		 & \\
			\hline
		$\rE_8$   & 0 & 0& \\
		\hline
	\end{tabular}\;\;
	}
	\caption{{The one-form symmetry $G_{[1]}$ for various pure gauge theories, or gauge theories with the matter charged in a certain representation (Rep, denoted 
		{\bf R})  of the gauge group $G$. Here ``fundamental'' is shorthand as ``fund.''
		The first column lists the gauge group $G$.
		The second column lists the electric 1-form symmetry $G_{[1],e}$ for a pure $G$ gauge theory.
		The third column lists the magnetic 1-form symmetry $G_{[1],m}$ for a pure $G$ gauge theory.
		The fourth column lists the $G_{[1],e}$ and $G_{[1],m}$ for a $G$ gauge theory coupled to matter in various Rep {\bf R}.
		Here we consider the standard choices of Lie groups $\rG_2, \rF_4, \rE_6, \rE_7, \rE_8$ for the exceptional Lie algebras
		$\mathfrak{g}_{2}$, ${\mathfrak {f}}_{4}$,${\mathfrak {e}}_{6}$,
${\mathfrak {e}}_{7}$, ${\mathfrak {e}}_{8}$; of course, there may be other possible Lie groups for give Lie algebras.
The $\rE_6$ and $\rE_7$ have the adjoint type or simply connected type of Lie groups \cite{Bourbaki2008LieGroups}, denoted as $G^{\text{adjoint}}$ and $G^{\text{simp.cn}}$,
their centers and homotopy groups are related by $Z(G^{\text{adjoint}})=\pi_1(G^{\text{simp.cn}})$. 
(The simplest familiar example is the Lie algebra $\mathfrak{su}_2 = \mathfrak{so}_3$ whose Lie groups are 
$G^{\text{simp.cn}}=\SU(2)$ and $G^{\text{simp.cn}}=\SO(3)$.)
} 
	}
	\label{table:1-sym}
\end{table}

 \begin{table}[!h]
 	\hspace{-6.8mm}
	\centering
	\begin{tabular}{|c c c c|}
	\hline
		\multicolumn{4}{|c|}{One-form symmetries of 4d $\frac{\SU(3)\times \SU(2)\times \U(1))}{\Z_q}$-gauge theories at $d=4$, with or without matter fields}\\
		\hline
		$q$ & 
		1-form $e$ sym $G_{[1],e}$  &
		1-form $m$ sym $G_{[1],m}$ &
		{$G_{[1]}$ with Standard Model matter}
		{in Rep {\bf R}}: $G_{[1]}$
		\\
		\hline
		1 & ${({\Z_3}\times {\Z_2}\times \U(1))}_{[1],e}$ & ${\U(1)}_{[1],m}$ & $G_{[1],e}:{(\Z_3'\times \Z_2')}_{[1],e}$. $G_{[1],m}:{\U(1)}_{[1],m}$. \\
		\hline 
		2 & ${(\Z_3\times \U(1))}_{[1],e}$ & ${\U(1)}_{[1],m}$ & $G_{[1],e}:{(\Z_3')}_{[1],e} $. $G_{[1],m}:{\U(1)}_{[1],m}$.\\
		\hline 
		3 & ${(\Z_2\times \U(1))}_{[1],e}$ & ${\U(1)}_{[1],m}$  & $G_{[1],e}: {(\Z_2')}_{[1],e}$. $G_{[1],m}:{\U(1)}_{[1],m}$.\\
		\hline 
		6 & ${\U(1)}_{[1],e}$ &  ${\U(1)}_{[1],m}$ & $G_{[1],e}:$ no. $G_{[1],m}:{\U(1)}_{[1],m}$.\\
		\hline
	\end{tabular}
    \caption{One-form symmetries for 4d $(\SU(3)\times \SU(2)\times \U(1))/\Z_q$ pure gauge theory (the second and third columns)
    or $(\SU(3)\times \SU(2)\times \U(1))/\Z_q$ gauge theory with Standard Model (SM) matter field representation {\bf R} as \Eq{eq:rep-3generations}
    (the last column). 
    {Here the ${(\Z_2')}_{[1],e}$ (similarly ${(\Z_3')}_{[1],e}$) means that the unbroken one-form symmetries from the diagonal combination of
    the original ${(\Z_2\times \U(1))}_{[1],e}$ (same for ${(\Z_3\times \U(1))}_{[1],e}$).}
    }
	\label{table:standardmodel}
\end{table}


Let us make a few more remarks in particular focusing on 4d gauge theories:
\begin{enumerate}[leftmargin=2.mm, label=\textcolor{blue}{(\arabic*)}., ref={(\arabic*)}]
\item 	
\emph{Electric 1-form symmetry} $G_{[1],e}$: For pure gauge theory, the electric 1-form symmetry $G_{[1],e}$ is identified with the center of the gauge group $G$, called $Z(G)$. 
After coupled to matter fields in Rep $\bf R$ (the electric Rep for the gauge group $G$), the electric 1-form symmetry may be reduced. 
Here are some ways to determine $G_{[1],e}$:

$\bullet$ The $G_{[1],e}$ is the \emph{maximal subgroup} sub$(Z(G))$ (or more generally, the \emph{stabilizer}) of the center $Z(G)$ which does not transform Rep $\bf R$. 

$\bullet$ The $G_{[1],e}$ is the \emph{maximal subgroup} sub$(Z(G))$ such that the associated Wilson line (or Polyakov line if the line is along the time direction) cannot be opened up by matter fields in Rep $\bf R$ on two ends of the line operator.

\item 	
\emph{Magnetic 1-form symmetry} $G_{[1],m}$:
The $G_{[1],m}$ exists if there is a degree-2 characteristic class of the gauge bundle serving as the conserved current. 
Here are some ways to determine $G_{[1],m}$:

$\bullet$ The $G_{[1],m}$ can be determined by the first homotopy group $\pi_1(G)$ of the gauge group $G$.

$\bullet$ The $G_{[1],m}$ can be associated with the $\pi_2(G'/G_{\rm{sub}})$ if we embed the $G$-gauge theory into
a higher energy Yang-Mills-Higgs theory with an ultraviolet (UV) gauge group $G'$, broken down to $G_{\rm{sub}}$ by Higgsing,
where there can be 't Hooft-Polyakov monopole as dynamical objects.
%
We provide one example from the 't Hooft-Polyakov monopole viewpoint:

\begin{enumerate}
	\item In Standard Model (SM) physics, given $G_{\SM_q}$ \Eq{eq:GSM} and matter Rep \Eq{eq:rep-3generations}, we can determine the
	$G_{[1],e}$ and $G_{[1],m}$
	(the last column of Table \ref{table:standardmodel}). 
	In particular, the $G_{[1],m} = {\U(1)}_{[1],m}$ because the (electrically) gauged charged matter does not break ${\U(1)}_{[1],m}$ from 
	the $\Z$ class
	of 't Hooft lines
	\bea
	\pi_1(G_{\SM_q})= \pi_1( \frac{\SU(3)\times \SU(2)\times \U(1)}{\Z_q})=\Z.
	\eea 
	We can also embed the SM to Georgi-Glashow SU(5) grand unification or grand unified theory (GUT) \cite{Georgi1974syUnityofAllElementaryParticleForces},
	such that $G'=\SU(5)$ and $G_{\rm{sub}}=G_{\SM_6}$ broken down by GUT Higgs field, we can determine the 
	't Hooft-Polyakov monopole from this SU(5) GUT via\footnote{Here we use the
	long exact sequence of a fibration 
	$F \rightarrow E \to B=E/F$ where $F$ is the fiber or normal subgroup and $E$ is the total space or the total group,
	then we have the long exact sequence
	\bea
	 \dots \to \pi_n F\to \pi_nE\to \pi_n E/F\to \pi_{n-1}F \to \pi_{n-1}E \to \dots .
	\eea
	So for $E=\SU(5)$ and $F=G_{\SM_6}= \frac{\SU(3)\times \SU(2)\times \U(1)}{\Z_6}$, we have
	\bea
	 \dots \to \pi_2 (G_{\SM_6}) \to \pi_2 (\SU(5)) \to \pi_2 (\SU(5)/G_{\SM_6})\to \pi_{1}(G_{\SM_6}) \to \pi_{1}(\SU(5))  \to \dots ,
	\eea
	where $ \pi_2 (\SU(5))=0$ and $ \pi_1 (\SU(5))=0$ for the simply connected simple Lie group SU(5), 
	and $ \pi_{1}(G_{\SM_6})=\Z$.
	So $0  \to \pi_2 (\SU(5)/G_{\SM_6})\to \pi_{1}(G_{\SM_6}) \to 0 $ implies $\pi_2 (\SU(5)/G_{\SM_6})= \pi_{1}(G_{\SM_6})$.
	}
	 \bea
	 \pi_2(G'/G_{\rm{sub}})=\pi_2(\frac{\SU(5)}{G_{\SM_6}}) =\pi_2(\frac{\SU(5)}{ ({\SU(3)\times \SU(2)\times \U(1)})/{\Z_6}}) =\Z.
          \eea
          Thus the classification of 't Hooft-Polyakov monopole from SU(5) GUT agrees with 't Hooft lines of the low energy SM.
\end{enumerate}

$\bullet$ When $G$ is  connected (namely, $\pi_0(G)=0$), the $G_{[1],m}$ can also be determined by the second cohomology group $\H^2(\B G,R)$ for $R=\U(1)$ or $\Z_n$ with some $n$. To see this, we first use the Hurewicz theorem\cite{Hatcher:478079}. The theorem shows that if $G$ is connected, i.e., $\pi_1(\B G)= \pi_0(G)=0$, then $\H_1(\B G, \Z)=0$ and $\pi_1(G)= \pi_2(\B G)= \H_2(\B G, \Z)$. Then we use the universal coefficient theorem, which shows that the  exact sequence $0\to \Ext(\H_1 (\B G, \Z), R) \to \H^2(\B G, R)\to \Hom(\H_2(\B G, \Z), R)\to 0$ holds. Since the Hurewicz theorem guarantees that $\H_1(\B G, \Z)=0$, the universal coefficient theorem tells us that
\begin{eqnarray}
\H^2(\B G, R)= \Hom(\H_2(\B G, \Z), R) = \Hom(\pi_1(G), R).
\end{eqnarray}
This relates $\pi_1(G)$, which classifies the magnetic 1-form symmetry and $H^2(\B G, R)$, for connected $G$. We need to find the coefficient $R$'s such that $\H^2(\B G, R)= \Hom(\pi_1(G), R)$ is nontrivial. We provide two examples. 
\begin{enumerate}
	\item For $G=\U(1)$,  $\pi_1(G)= \Z$, then 
	\begin{eqnarray}
	\H^2(\B \U(1), R)= \Hom(\pi_1(\U(1)), R) = \Hom(\Z, R) = R.
	\end{eqnarray}
	Thus one can choose $R=\U(1)$ which gives $\H^2(\B \U(1), \U(1))=\U(1)$, whose generator is simply the first Chern class $c_1(V_{\U(1)})$. The first Chern class serves as a conserved current of the magnetic 1-form symmetry, which couples to the background field $B_{m,[2]}$ via $\exp\left(\ii  \int_{M_4} \frac{c_1(V_{\U(1)})}{2\pi} B_{2,[m]}\right)$ added in the partition function. 
	One can also choose $R= \Z_n$ for arbitrary $n$, but this does not give additional characteristic classes and the magnetic 1-form symmetry. 
	\item For $G=\SO(3)$, thus $\pi_1(G)=\Z_2$, then 
	\begin{eqnarray}
	\H^2(\B\SO(3), R) = \Hom (\pi_1(\SO(3)), R)=\Hom (\Z_2, R) 
	= 
	\begin{cases}
	\Z_2, & R= \U(1) ~\text{or} ~ \Z_{2n}.\\
	0, & R= \Z_{2n+1}.
	\end{cases}
	\end{eqnarray}
	Thus the minimal coefficient is $R=\Z_2$. Other non-minimal coefficients $R=\Z_{2n} (n>1)$ or $\U(1)$ do not give additional characteristic classes and the magnetic 1-form  symmetry. For $R=\Z_2$, the generator is  $w_2(V_{\SO(3)})\in \H^2(\B\SO(3), \Z_2)$, which serves as a conserved current of the magnetic 1-form symmetry, which couples to the background field $B_{2,[m]}$ via $\exp(\ii \pi \int_{M_4} w_2(V_{\SO(3)}) B_{m,[2]})$ added in the partition function. 
\end{enumerate}
To summarize, we find that the magnetic 1-form symmetry, which is defined to be classified by $\pi_1(G)$, can also be captured by $\H^2(\B G, R)$ with suitable choices of the coefficient ring $R=\U(1)$ or $\Z_n$.


%
%
%
%
%
%
%
%

%

$\bullet$ After the theory is coupled to matter fields in Rep $\bf R$, the magnetic 1-form symmetry remains the same.
\color{black}



Let us illuminate Table \ref{table:standardmodel} in the gauge bundle viewpoint.
We specify the conserved current for the $\U(1)$  magnetic 1-form symmetries in the four cases of the \emph{pure} gauge theories with the  standard model gauge group $(\SU(3)\times \SU(2)\times \U(1))/\Z_q$. Denote the $\U(1)$ gauge field in the {numerator} of the gauge group as $a$.  For $q=1$, the current is $\dd a$. For $q=2$, the current is $2 \dd a$. It correlates with the $\PSU(2)=\SO(3)$ 
bundle via $\frac{2\dd a}{2\pi}= w_2(V_{\PSU(2)})\mod 2$.  For $q=3$, the current is $3 \dd a$.  It correlates with the $\PSU(3)$ bundle via $\frac{3\dd a}{2\pi}= w_2(V_{\PSU(3)})\mod 3$. For $q=6$, the current is $6  \dd a$.  It correlates with the $\PSU(2)$ and $\PSU(3)$ bundles via $\frac{6\dd a}{2\pi}= 3 w_2(V_{\PSU(2)}) + 2 w_2(V_{\PSU(3)})\mod 6$.  

\end{enumerate}

In Table \ref{table:tableOmega}, for various dimensions $d$, 
we summarize partial results of bordism groups $\Omega_d$ discussed in later sections.

 In Table \ref{table:tableTP}, for various dimensions $d$, 
we summarize partial results of cobordism groups $\TP_d$ discussed in later sections.

\subsection{Outline}

The rest of this work computes the bordism groups, cobordism group and cobordism invariants for the internal symmetries $G=\SU(2)$ and various Lorentz symmetries $\LG$, as well as their extensions. We further gauge $G$ to obtain $G$ dynamical gauge theories and compute the cobordism invariants for the emergent symmetry $G_{e,[1]}\times \LG$. In section \ref{secLG}, we review the cobordism invariants for the Lorentz symmetry, without the internal symmetry. In section \ref{secSU2}, study the extension of various choices of $\LG$ by the global symmetry $\SU(2)$,  compute the cobordism invariants and discuss their physical interpretations. In section \ref{GaugingSU2}, we promote $\SU(2)$ global symmetry to be a dynamical gauge group, and study the emergent symmetry, and bundle constraints and the anomalies for the resulting $\SU(2)$ gauge theories. In the appendix \ref{app} and \ref{app2}, we compute the bordism and cobordism groups and invariants for symmetries involving $\LG=\rE(d)$ and $\LG= \EPin(d)$ respectively. 

\pagebreak

\begin{table}[!h]
{
\[ \arraycolsep=5pt\def\arraystretch{1.5} 
\hspace{-22mm}
\begin{array}{|c|ccccccccccc c c c c|}
\hline
d & 0 & 1 & 2 & 3 & 4 & 5 & 6 & 7 & 8 & 9 & 10 & 11 & 12 & 13 & 14 \\
\hline
\Omega_d^{\SO} & \Z& 0 & 0 & 0 & \Z &\Z_2    & 0 & 0& \Z^2 &  & & &  &  &\\
\Omega_d^\Spin & \Z & \Z_2 & \Z_2 & 0 & \Z & 0 & 0 & 0 & \Z^2 & \Z_2^2 & \Z_2^3 & &  &  & \\
\Omega_d^{\O} & \Z_2 &  0& \Z_2 & 0 & \Z_2^2 & \Z_2 & \Z_2^3 &\Z_2 &\Z_2^5  & \Z_2^3 & \Z_2^8&\Z_2^5 &\Z_2^{11}  &\Z_2^9  &\Z_2^{17}\\
      \Omega_d^{\rE} &  \Z_2 & 0  & \Z_2  & 0  & \Z_2^2  & \Z_2  &  & & & &  & &  &  & \\
\Omega_d^{\Pin^+} & \Z_2 & 0 & \Z_2 & \Z_2 & \Z_{16} & 0 & 0 & 0 & \Z_2 \times \Z_{32} & 0 & \Z_2^3 & &  &  &\\
\Omega_d^{\Pin^-} & \Z_2 & \Z_2  & \Z_8 & 0 & 0 & 0 & \Z_{16} & 0 & \Z_2^2 &  \Z_2^2& \Z_2\times\Z_8\times\Z_{128}& &  &  &\\
{\Omega_d^{\DPin}} & \Z_2 & \Z_2  & {\Z_2^2}  & {\Z_8} & \Z_2^2 & 0 &  \Z_2^2  &  &  &  & & &  &  &\\
{\Omega_d^{\EPin}} & \Z_2 & \Z_2  & {\Z_2\times\Z_4}  & {\Z_2} & \Z_2^2 & 0 &  &  &  &  & & &  &  &\\
\Omega_d^{\Spin^c} & \Z & 0 & \Z & 0 & \Z^2 & 0 & \Z^2 & 0 & \Z^4 & 0 & \Z^4\times\Z_2 & &  &  & \\
\Omega_d^{\Spin^h} & \Z & 0 & 0 & 0 & \Z^2 & \Z_2^2  & \Z_2^2 & 0 & \Z^4 &0  &\Z_2 & &  &  &\\
\Omega_d^{\rm{String}} & \Z & \Z_2 & \Z_2 & \Z_{24} & 0 & 0 & \Z^2 & 0 & \Z_2 \times \Z & \Z_2^2 & \Z_6 & &  &  & \\
 \Omega_d^{\SO\times\SU(2)} & \Z & 0 & 0 & 0 & \Z^2  & \Z_2 &  & & & &  & &  &  & \\
  \Omega_d^{\Spin\times_{\Z_2}\SU(2)} &  \Z& 0& 0 &0 & \Z^2  &\Z_2^2  &  & & & &  & &  &  & \\
  \Omega_d^{\Spin\times\SU(2)} &  \Z& \Z_2 &\Z_2  &0 &\Z^2  &\Z_2  &  & & & &  & &  &  & \\
  \Omega_d^{\O\times\SU(2)} &\Z_2  & 0 &\Z_2  &0 &\Z_2^3  & \Z_2  &  & & & &  & &  &  & \\
  \Omega_d^{\rE\times_{\Z_2}\SU(2)} & \Z_2 & 0 &\Z_2  &0 &\Z_2^3  & \Z_2   &  & & & &  & &  &  & \\
      \Omega_d^{\rE\times\SU(2)} & \Z_2 & 0 &\Z_2  &0 &\Z_2^3  & \Z_2  &  & & & &  & &  &  & \\
   \Omega_d^{\Pin^+\times_{\Z_2}\SU(2)} & \Z_2 & 0 & \Z_2 & 0& \Z_2\times \Z_4 & \Z_2   &  & & & &  & &  &  & \\
    \Omega_d^{\Pin^-\times_{\Z_2}\SU(2)} &\Z_2  & 0& \Z_2 & 0& \Z_2^3 &\Z_2^2  &  & & & &  & &  &  & \\
    \Omega_d^{\SO\times\Z_{2,[1]}} & \Z &0 &\Z_2  &0 &\Z\times \Z_4  &\Z_2^2  &  & & & &  & &  &  & \\
     \Omega_d^{\Spin\times\Z_{2,[1]}} & \Z & \Z_2 &\Z_2^2  &0 &\Z\times \Z_2  &0  &  & & & &  & &  &  & \\
     \Omega_d^{\O\times\Z_{2,[1]}} & \Z_2 & 0 &\Z_2^2  & \Z_2 & \Z_2^4  & \Z_2^4  &  & & & &  & &  &  & \\
      \Omega_d^{\rE\times\Z_{2,[1]}} &\Z_2  & 0&\Z_2^2  &\Z_2 & \Z_2^4 &\Z_2^4  &  & & & &  & &  &  & \\
     \Omega_d^{\Pin^+\times\Z_{2,[1]}} &  \Z_2& 0& \Z_2^2 &\Z_2^2 &\Z_2\times\Z_{16}   &\Z_2^2  &  & & & &  & &  &  & \\
     \Omega_d^{\Pin^-\times\Z_{2,[1]}}  &\Z_2 &\Z_2 &\Z_2\times \Z_8  &\Z_2 &\Z_2  &\Z_2&  & & & &  & &  &  & \\
     \Omega_d^{\EPin\times\Z_{2,[1]}}  &\Z_2 &\Z_2 &\Z_2^2\times \Z_4  &\Z_2^2 &\Z_2^4  &\Z_2^2&  & & & &  & &  &  & \\
     \Omega_d^{\DPin\times\Z_{2,[1]}}  &\Z_2 &\Z_2 &\Z_2^3 &\Z_2^2\times \Z_8 &\Z_2^5  &\Z_2^4&  & & & &  & &  &  & \\
 \hline
  \end{array} 
 \]
  }
\caption{We summarize partial results of bordism groups $\Omega_d^G$ obtained in later sections. 
${\Spin^h=\Spin\times_{\Z_2}\SU(2)}$
and $\Spin^c={\Spin\times_{\Z_2}\U(1)}$.
}
\label{table:tableOmega}
 \end{table}

 \pagebreak

 \begin{table}[!h]
{
\[ \arraycolsep=5pt\def\arraystretch{1.5} 
\hspace{-22mm}
\begin{array}{|c|ccccccccccc c c c c|}
\hline
d & 0 & 1 & 2 & 3 & 4 & 5 & 6 & 7 & 8 & 9 & 10 & 11 & 12 & 13 & 14 \\
\hline
\TP_d^{\SO} &0 & 0 & 0 & \Z & 0 &\Z_2    & 0 &  \Z^2&0 &  & & &  &  &\\
\TP_d^\Spin &  0& \Z_2 & \Z_2 & \Z & 0 & 0 & 0 & \Z^2  & 0& \Z_2^2 & \Z_2^3 & &  &  & \\
\TP_d^{\O} & \Z_2 &  0& \Z_2 & 0 & \Z_2^2 & \Z_2 & \Z_2^3 &\Z_2 &\Z_2^5  & \Z_2^3 & \Z_2^8&\Z_2^5 &\Z_2^{11}  &\Z_2^9  &\Z_2^{17}\\
      \TP_d^{\rE} &  \Z_2 & 0  & \Z_2  & 0  & \Z_2^2  & \Z_2  &  & & & &  & &  &  & \\
\TP_d^{\Pin^+} & \Z_2 & 0 & \Z_2 & \Z_2 & \Z_{16} & 0 & 0 & 0 & \Z_2 \times \Z_{32} & 0 & \Z_2^3 & &  &  &\\
\TP_d^{\Pin^-} & \Z_2 & \Z_2  & \Z_8 & 0 & 0 & 0 & \Z_{16} & 0 & \Z_2^2 &  \Z_2^2& \Z_2\times\Z_8\times\Z_{128}& &  &  &\\
{\TP_d^{\DPin}} & \Z_2 & \Z_2  & {\Z_2^2}  & {\Z_8} & \Z_2^2 & 0 &  \Z_2^2  &  &  &  & & &  &  &\\
{\TP_d^{\EPin}} &\Z_2  & \Z_2  & {\Z_2\times\Z_4}  & {\Z_2} & \Z_2^2 & 0 &  &  &  &  & & &  &  &\\
\TP_d^{\Spin^c} &0  & \Z & 0  & \Z^2 & 0  & \Z^2 & 0  & \Z^4 & 0  &  \Z^4 &\Z_2 & &  &  & \\
\TP_d^{\Spin^h} &  0& 0  & 0  & \Z^2 & 0  & \Z_2^2  & \Z_2^2 & \Z^4 & 0  &0  &\Z_2 & &  &  &\\
\TP_d^{\rm{String}} &0  & \Z_2 & \Z_2 & \Z_{24}  & 0 & \Z^2 & 0 & \Z   & \Z_2   & \Z_2^2 & \Z_6 & &  &  & \\
 \TP_d^{\SO\times\SU(2)} &0  & 0 & 0 & \Z^2 & 0   & \Z_2  &  & & & &  & &  &  & \\
   \TP_d^{\Spin\times_{\Z_2}\SU(2)} &0  & 0& 0  & \Z^2 & 0 &\Z_2^2  &  & & & &  & &  &  & \\
   \TP_d^{\Spin\times\SU(2)} & 0 & \Z_2 &\Z_2  &\Z^2 & 0  &\Z_2  &  & & & &  & &  &  & \\
   \TP_d^{\O\times\SU(2)} &\Z_2  & 0 &\Z_2  &0 &\Z_2^3  & \Z_2  &  & & & &  & &  &  & \\
  \TP_d^{\rE\times_{\Z_2}\SU(2)} & \Z_2 & 0 &\Z_2  &0 &\Z_2^3  & \Z_2   &  & & & &  & &  &  & \\
       \TP_d^{\rE\times\SU(2)} & \Z_2 & 0 &\Z_2  &0 &\Z_2^3  & \Z_2   &  & & & &  & &  &  & \\
    \TP_d^{\Pin^+\times_{\Z_2}\SU(2)} & \Z_2 & 0 & \Z_2 & 0& \Z_2\times \Z_4 & \Z_2  &  & & & &  & &  &  & \\
     \TP_d^{\Pin^-\times_{\Z_2}\SU(2)} &  \Z_2 & 0& \Z_2 & 0& \Z_2^3 &\Z_2^2 &  & & & &  & &  &  & \\
   \TP_d^{\SO\times\Z_{2,[1]}} &  0&0 & \Z_2 &\Z &\Z_4  &\Z_2^2  &  & & & &  & &  &  & \\
      \TP_d^{\Spin\times\Z_{2,[1]}} &0  &\Z_2 &\Z_2^2  &\Z &\Z_2  &0  &  & & & &  & &  &  & \\
      \TP_d^{\O\times\Z_{2,[1]}} & \Z_2 & 0 &\Z_2^2  & \Z_2 & \Z_2^4  & \Z_2^4  &  & & & &  & &  &  & \\
       \TP_d^{\rE\times\Z_{2,[1]}} & \Z_2 & 0&\Z_2^2  &\Z_2 & \Z_2^4 &\Z_2^4 &  & & & &  & &  &  & \\
      \TP_d^{\Pin^+\times\Z_{2,[1]}} &  \Z_2& 0& \Z_2^2 &\Z_2^2 &\Z_2\times\Z_{16}   &\Z_2^2  &  & & & &  & &  &  & \\
     \TP_d^{\Pin^-\times\Z_{2,[1]}} &\Z_2 &\Z_2 &\Z_2\times \Z_8  &\Z_2 &\Z_2  &\Z_2  &  & & & &  & &  &  & \\
     \TP_d^{\EPin\times\Z_{2,[1]}}  &\Z_2 &\Z_2 &\Z_2^2\times \Z_4  &\Z_2^2 &\Z_2^4  &\Z_2^2&  & & & &  & &  &  & \\
     \TP_d^{\DPin\times\Z_{2,[1]}}  &\Z_2 &\Z_2 &\Z_2^3 &\Z_2^2\times \Z_8 &\Z_2^5  &\Z_2^4&  & & & &  & &  &  & \\
 \hline
  \end{array} 
 \]
  }
\caption{We summarize partial results of cobordism groups $\TP_d^G$  obtained in later sections. 
See also the caption in Table \ref{table:tableOmega}.
}
\label{table:tableTP}
 \end{table}


\section{Anomalies from Lorentz Symmetry $\LG$}
\label{secLG}

In this section, we demand the global symmetry $G$ to be trivial, and the total symmetry is just $\LG$. We will systematically discuss the anomalies for each choice of $\LG$ in \eqref{Glorentz}. The cobordism invariants have been discussed in \cite{Wan2018bns1812.11967}. These cobordism invariants will appear repeatedly when the internal unitary symmetry $G$ is non-trivial. Hence for simplicity, we once for all discuss  bordism group $\Omega_d^{\LG}$  and the associated cobordism invariants here, and will not repeat the discussion in the following sections. 

The group $\LG= \rE(d)$ and $\EPin(d)$ are relatively unfamilar, we leave  the (co)bordism groups and the cobordism invariants of these groups in a separate work. 

\subsection{$\LG= \SO(d)$}
\label{subsecSO}

When the spacetime symmetry is $\LG=\SO(d)$, one can formulate the quantum field theory on arbitrary non-spin oriented manifold. In particular, the quantum field theory does not have time reversal symmetry, and does not depend on the spin structure.  In the condensed matter language, such a quantum field theory can be understood as emerging from a UV lattice model with only bosonic local degrees of freedom, i.e., the theory is bosonic. 

We enumerate the bordism and cobordism groups and list the cobordism invariants, in Table \ref{table:SOBordism}. Below, we comment on the nontrivial cobordism invariants, and interpret them as the anomaly inflow actions  of QFTs living in one dimension lower. 
\begin{enumerate}
	\item When $d=3$, the anomaly $16\CSg$ can be saturated by the 1+1d chiral boson CFT with the K-matrix being the rank-8 Cartan matrix of $\rE_8$.\footnote{{This 1+1d CFT is also known as the boundary CFT of 2+1d $\rE_8$ quantum Hall state.}} 
	Such a theory has chiral central charge $c_-= 8$, thus has gravitational anomaly $16 \CSg$ \cite{PhysRevB.91.054406}. Due to the nontrivial chiral central charge, such a theory must be gapless. 

	\item When $d=5$, the anomaly $w_2 w_3$ can be saturated by either gapless or gapped theories in $3+1$d. The gapless theory is the all fermion electrodynamics, where both the $\U(1)$ charge and $\U(1)$ monopole are fermions \cite{Kravec_2015, Wang:2018qoyWWW}.
	The gapped theory is the $\Z_2$ gauge theory with $\Z_2$ fermionic charge and $\Z_2$ fermionic strings \cite{Thorngren_2015, PhysRevB.87.235122}.
	
\end{enumerate}

\begin{table}[t]
	\centering
	\begin{tabular}{ c c c c }
		\hline
		\multicolumn{4}{c}{Bordism and Cobordism group}\\
		\hline
		$d$ & 
		$\Omega^{\SO }_d$ &
		$\TP^{\SO }_d$ &
		Cobordism Invariant \\
		\hline
		1& 0 & 0&\\
		2& $0$ & 0&\\
		3 & $0$ & $\Z$ & $16 \CSg$\\
		4 & $\Z$& 0 & \\
		5 & $\Z_2$ & $\Z_2$&  $w_2w_3$\\
		\hline
	\end{tabular}
	\caption{
		The bordism and cobordism groups of the symmetry  $\SO(d)$. The cobordism invariants, representing the $d$ dimensional anomaly inflow actions  of the $d-1$ dimensional quantum field theories,  are also enumerated. $\CSg$ is the gravitational Chern Simons term. It can be defined by lifting to the 4d integral as $\CSg[X]= \pi \int_Y \widehat{A}=2\pi \sigma[Y]$ where $\partial Y= X$. $w_i$ is the $i$-th Stiefel-Whitney (SW) class of the tangent bundle of the spacetime manifold. In this paper, we will denote the $w_i$ as the $i$-th SW class for the spacetime manifold $M$, short for $w_i(TM)$. When referring to the SW class for the vector bundle $V_G$ associated with certain group $G$ (where $G$ can be gauge group or global symmetry group), we will denote it as $w_i(V_G)$. 
	{The $16 \CSg$ is also the invertible TQFT at the low energy of  2+1d $\rE_8$ quantum Hall state whose boundary edge modes have 1+1d CFT
	of chiral central charge $c_-= 8$.}
	}
	\label{table:SOBordism}
\end{table}

\subsection{$\LG= \Spin(d)$}
\label{subsecSpin}

When the spacetime symmetry is $\LG=\Spin(d)$, we need to consider the manifold that allows a spin structure, i.e., the spacetime manifold is a spin manifold. In particular $w_2=0$. In the condensed matter language, a quantum field theory that can only be defined on a spin manifold should be understood as flowing from a UV lattice system with local fermionic degrees of freedom (i.e., the theory is fermionic, in terms of a condensed matter language). 

\begin{table}[t]
	\centering
	\begin{tabular}{ c c c c }
		\hline
		\multicolumn{4}{c}{Bordism and Cobordism group}\\
		\hline
		$d$ & 
		$\Omega^{\Spin }_d$ &
		$\TP^{\Spin }_d$ &
		Cobordism Invariant \\
		\hline
		1& $\Z_2$ & $\Z_2$& $\widetilde{\eta}$\\
		2& $\Z_2$ & $\Z_2$& $\text{Arf}$\\
		3 & $0$ & $\Z$ & $\CSg$\\
		4 & $\Z$& 0 & \\
		5 & $0$ & $0$&  \\
		\hline
	\end{tabular}
	\caption{
		The bordism and cobordism groups of the symmetry  $\Spin(d)$. The cobordism invariants, representing the $d$ dimensional anomaly inflow actions  of the $d-1$ dimensional quantum field theories,  are also enumerated. $\widetilde{\eta}$ is the mod 2 reduction of the index of the 1d Dirac operator. $\arf$ is the Arf invariant.  $\CSg$ is the gravitational Chern Simons term defined in Table \ref{table:SOBordism},
		{The $\CSg$ is also the invertible TQFT at the low energy of  2+1d chiral $p$-wave ($p_x +\ii p_y$) superconductor whose boundary edge modes have 1+1d CFT
	of chiral central charge $c_-= 1/2$.} 
	}
	\label{table:SpinBordism}
\end{table}

We enumerate the bordism and cobordism groups and list the cobordism invariants, in Table \ref{table:SpinBordism}. Below, we comment on the nontrivial cobordism invariants, and interpret them as the anomaly inflow actions  of QFTs living in one dimension lower. 
\begin{enumerate}
	\item When $d=1$, $\widetilde{\eta}$ is the mod 2 reduction of the index of the 1d Dirac operator, i.e., it counts the number of zero modes modulo 2. Notice that the most general 1d connected spacetime manifold is a circle. The index on the circle depends on the spin structure $\rho$, i.e., whether the boundary condition of the fermion is Ramond (R) or Neveu-Schwarz (NS). \footnote{In the literature,  Ramond and Neveu-Schwarz boundary conditions are also termed Periodic (P) and Anti-Periodic (AP) boundary conditions. }
	\begin{eqnarray}\label{tildeeta}
	\tilde{\eta}(\rho)= 
	\begin{cases}
      1,& ~~~ \rho=\rR.\\
	0, & ~~~ \rho=\NS.
	\end{cases}
	\end{eqnarray}
	
	\item When $d=2$, the $\arf$ invariant can be constructed as follows. A 2d oriented manifold without boundary, equipped with a metric, can be described as the Riemann surface. The Riemann surface can be classified by its genus $g$. Since any 2d Riemann surface is a spin manifold, it can be equipped with a spin structure $\rho$, encoded by the boundary condition of the spinors along each non-contractible cycle. For example, when the Riemann surface is a torus $g=1$, there are two non-contracible cycles. A spinor can have Ramond (R) or Neveu-Schwarz (NS) boundary condition along each cycle.  $\arf$ is the Arf invariant of the spin structure $\rho$, 
	\begin{eqnarray}
	\arf(\rho)=
	\begin{cases}
		1, &~~~ \rho=(\rR, \rR).\\
	0, &~~~ \rho= (\rR,\NS), (\NS, \rR), (\NS, \NS).
	\end{cases}
	\end{eqnarray}
	As we can see, $\arf$ is a natural 2d generalization of the 1d mod 2 Dirac index $\tilde{\eta}$. We denote the spin structure $(R, R)$ as odd, and other three spin structures as even. For the Riemann surface with other genus $g$, there are $2^{g-1}(2^g-1)$ odd spin structures, and $2^{g-1}(2^g+1)$ even ones. Physically partition function involving the Arf invariant, $e^{i \pi \arf}$, can be understood as the partition function of the {1+1d Kitaev chain \cite{Kitaev2001chain0010440}}. Hence the boundary anomalous quantum field theory saturating the $\arf$ anomaly is a single {0+1d} Majorana fermion. Indeed, as discussed in \cite{Dijkgraaf:2018vnm}, a Majorana fermion with odd spin structure (Romand boundary condition) is anomalous under fermi parity. A physical imprint of the fermi parity anomaly is the existence of fermion zero mode, hence the fermion expectation value is nontrivial $\langle \chi \rangle\neq 0$. 
	
	\item  When $d=3$, the gravitational Chern Simons term $\CSg$ has been defined in Table \ref{table:SOBordism}. Notice that in the case $\LG=\SO(3)$, only $16\CSg$ is well defined. This is because to define Chern Simons term on 3d manifold, one needs to define it as an integral on a 4d manifold with a boundary, $\CSg= \pi \int_Y \widehat{A}$ and demand that the 4d integral does not depend on the choice of the 4d manifold, but only depends on the boundary $\partial Y$. For $\LG= \SO(3)$, the manifold is nonspin, and only $16\CSg= 16\pi \int_Y \widehat{A}$ does not depend on the choice of the bulk manifold. However, for $\LG=\Spin(d)$, the manifold is a spin manifold, and $\CSg= \pi \int_Y \widehat{A}$ already does not depend on the choice of the bulk manifold. The anomaly $\CSg$ implies that the theory living on the $1+1$d boundary should have chiral central charge $c_-=1/2$. One example of such a theory is a left moving  Majorana fermion. It is a fermionic CFT with the left and right central charge $c_L=1/2$ and $c_R=0$. Hence the chiral central charge is $c_-=c_L-c_R=1/2$.

\end{enumerate}


\subsection{$\LG=\O(d)$}
\label{subsecO}

When the spacetime symmetry is $\LG=\O(d)$, the spacetime manifold should be unorientable. A quantum field theory that can be formulated on an unorientable manifold should have time reversal symmetry. Moreover, since a generic unorientable manifold may not have a $\Pin^{\pm}$ structure, the theory should be regarded as a bosonic, in particular $\T^2=1$ on physical local operators. (As we will see, in certain situations where there are onsite unitary global symmetry, one can formulate a fermionic theory on a manifold with $\LG=\O(d)$.)

\begin{table}[t]
	\centering
	\begin{tabular}{ c c c c }
		\hline
		\multicolumn{4}{c}{Bordism and Cobordism group}\\
		\hline
		$d$ & 
		$\Omega^{\O }_d$ &
		$\TP^{\O }_d$ &
		Cobordism Invariant \\
		\hline
		1& $0$ & 0 & \\
		2& $\Z_2$ & $\Z_2$& $w_1^2$\\
		3 & $0$ & 0 & \\
		4 & $\Z_2\times \Z_2$& $\Z_2\times \Z_2$ & $w_1^4, w_2^2$\\
		5 & $\Z_2$ & $\Z_2$& $w_2 w_3$ \\
		\hline
	\end{tabular}
	\caption{
		The bordism and cobordism groups of the symmetry  $\O(d)$. The cobordism invariants, representing the $d$ dimensional anomaly inflow actions of the $d-1$ dimensional quantum field theories, are also enumerated.  $w_i$ is the $i$-th Stiefel-Whitney (SW) class of the tangent bundle of the spacetime manifold, as defined in Table.\ref{table:SOBordism}. 
	}
	\label{table:OBordism}
\end{table}

We enumerate the bordism and cobordism groups and list the cobordism invariants, in Table \ref{table:OBordism}. Below, we comment on the nontrivial cobordism invariants, and interpret them as the anomaly  inflow actions of QFTs living in one dimension lower. 
\begin{enumerate}
	\item When $d=2$, the $w_1^2$ is a cobordism invariant describing the invertible TQFT in $1+1$d. In condensed matter language, such an invertible TQFT describes a bosonic symmetry protected topological phase with time reversal symmetry, i.e., the Haldane chain. The boundary of it supports nontrivial bosonic degrees of freedom that transforms under time reversal as a Kramers doublet. 
	
	\item When $d=4$, there are two generators of cobordism invariants.  The first invariant $w_1^4$ describes the invertible TQFT in $3+1$d, which also describes time reversal anomaly for a $2+1$d quantum field theory. A well known example that saturates such an anomaly is a $\Z_2$ gauge theory where both the $\Z_2$ charge and $\Z_2$ flux are Kramers doublets\cite{PhysRevB.87.235122}. 
	
	\item When $d=4$,  the second cobordism  invariant $w_2^2$ describes the invertible TQFT in $3+1$d, which also describes  the anomaly of $\SO(3)$ Lorentz symmetry for a $2+1$d quantum field theory.  A well known example that saturates such an anomaly is a $\Z_2$ gauge theory where both the $\Z_2$ charge and $\Z_2$ flux are fermions\cite{PhysRevB.87.235122}.
	
	\item When $d=5$, the anomaly $w_2 w_3$ has already been discussed in the case $\LG= \SO(d)$, which will not be repeated here. 
\end{enumerate}

%
%

%
%
%

\subsection{$\LG= \Pin^+(d)$}

When the spacetime symmetry is $\LG=\Pin^+(d)$, the spacetime manifold should be unorientable, and allows a $\Pin^+$ structure. In particular the manifold has nontrivial $w_1$, but a trivial $w_2$.  A quantum field theory that can be formulated on a $\Pin^+$ manifold should have time reversal symmetry, and there exists a fermion in the operator spectrum. Time reversal acts on the fermion as $\T^2=(-1)^F$ where $F$ measures the fermion number, $F=0$ for boson and $F=1$ for fermion. 


\begin{table}[t]
	\centering
	\begin{tabular}{ c c c c }
		\hline
		\multicolumn{4}{c}{Bordism and Cobordism group}\\
		\hline
		$d$ & 
		$\Omega^{\Pin^+ }_d$ &
		$\TP^{\Pin^+ }_d$ &
		Cobordism Invariant \\
		\hline
		1& $0$ & 0 & \\
		2& $\Z_2$ & $\Z_2$& $w_1 \cup \tilde{\eta}$\\
		3 & $\Z_2$ & $\Z_2$ &  $w_1\cup \arf$ \\
		4 & $\Z_{16}$& $\Z_{16}$ & $\eta$\\
		5 & $0$ & $0$&  \\
		\hline
	\end{tabular}
	\caption{
		The bordism and cobordism groups of the symmetry  $\Pin^+(d)$. The cobordism invariants, representing the $d$ dimensional anomaly inflow actions of the $d-1$ dimensional quantum field theories, are also enumerated. 
	}
	\label{table:Pin+Bordism}
\end{table}

We enumerate the bordism and cobordism groups and list the cobordism invariants, in Table \ref{table:Pin+Bordism}. Below, we comment on the nontrivial cobordism invariants, and interpret them as the anomaly inflow actions of QFTs living in one dimension lower. 
\begin{enumerate}
	\item When $d=2$, $w_1\cup \tilde{\eta}$ is a cobordism invariant describing a fermionic invertible TQFT with time reversal symmetry on an unorientable manifold. In fact, this invertible TQFT is the effective field theory of a Kitaev Chain (with two Majorana fermions per unit cell) protected by time reversal symmetry with $\T^2=(-1)^F$, already existing in the free fermion classification in the ten-fold way. One quantum field theory description of the Kitaev chain is a 2-component Dirac fermion with large and negative mass term. To relate this QFT description to the cobordism invariant $w_1\cup \tilde{\eta}$, it is convenient to consider the $0+1$d boundary theory and discuss its anomaly. The $0+1$d theory is a massless 2-component Majorana fermion $\chi$ with the Lagrangian $-\ii \chi^T \frac{\dd}{\dd t} \chi$. Under time reversal, $\T: \chi(t) \to \gamma^0 \chi(-t)$ where $\gamma^0= \ii \sigma^2$, and it is obvious that the classical Lagrangian is $\T$ invariant. Denoting the path integral of this Majorana fermion by $Z[\rho]$, where $\rho=\rR, \NS$ is the spin structure on the circle. Then under time reversal, the path integral transforms as 
	\begin{eqnarray}
	\T: Z[\rho] \to 
	\begin{cases}
	Z[\rho], & ~~~ \rho=\NS\\
	-Z[\rho], &~~~ \rho=\rR
	\end{cases}
	\end{eqnarray}
	which can be compactly organized as $\T: Z[\rho]\to Z[\rho] \exp(\ii \pi \tilde{\eta})$. This indicates precisely the anomaly $w_1\cup \tilde{\eta}$. \footnote{This anomaly can not be changed by modifying the partition function with a counter term $(-1)^{\tilde{\eta}}$. Same comment also applies to the anomaly $w_1 \cup \arf$ in $d=3$. }
	
	{Let us also explain why $2 w_1\cup \tilde{\eta}$ is trivial to get the $\Z_2$ class. Note that
	$2 w_1\cup \tilde{\eta}= w_1^2$, moreover $w_1^2=w_2$ is true for all 2d smooth manifolds (always $\Pin^-$ in 2d), and
	$w_2=0$ for $\Pin^+$ thus
	 $w_1^2=w_2=0$ mod 2.}
	
	\item When $d=3$, $w_1\cup \arf$ is a cobordism invariant describing a fermionic invertible TQFT with time reversal symmetry on an unorientable manifold. In fact, such an invertible TQFT is described by the time reversal invariant topological superconductor, and appears in the free fermion classification in the ten-fold way. Similar to $d=2$ case, one can still view $w_1\cup \arf$ as the mixed anomaly between time reversal and fermion parity of a $1+1$d Majorana fermion. This anomaly has been systematically discussed in \cite{Karch_2019}, 
	\begin{eqnarray}
	\T: Z[\rho] \to 
	\begin{cases}
	Z[\rho], & ~~~ \rho=(\rR,\NS), (\NS, \rR), (\NS, \NS),\\
	-Z[\rho], &~~~ \rho=(\rR,\rR),
	\end{cases}
	\end{eqnarray}
	which can be compactly organized as $\T: Z[\rho]\to Z[\rho] \exp(\ii \pi \arf)$. This indicates precisely the anomaly $w_1\cup \arf$. This anomaly is $\mod$ 2 anomaly because both $w_1$ and $\arf$ are mod 2 quantities, hence $2w_1 \cup \arf$ is trivial. 
	
	
	\item When $d=4$, $\eta$ is a cobordism invariant describing a fermionic invertible TQFT with time reversal symmetry. Physically, the $\eta$ invariant describes the $\nu=1$ (the fundamental) time reversal invariant  topological superconductor (TSC) with $\T^2=(-1)^F$. The quantum field theory for this TSc is a 4d complex Dirac fermion with time reversal acting on the fermion as $\T: \psi(t,\vec x)\to \gamma^0 \psi^*(-t,\vec x)$, and turn on a large negative mass.  \footnote{This was usually denoted as $\mathcal{CT}$ in the condensed matter literature. However, for the consistency of notations presented in this paper, we will call this $\T$. }   At the free fermion level, such a SPT is $\Z$ classified, which is labeled by the integer $\nu$ by taking $\nu$ copies of $\nu=1$ TSc's together. However, in the presence of interaction, $\nu=16$ TSc can be driven to a trivial TSc by $\T$ preserving interactions without closing the gap\cite{Kapustin:2014dxa, Witten:2015aba, Metlitski:2015yqa, 1711.11587GPW, Witten2016cio1605.02391, Bi:2018xvr}.  Hence interaction reduces the classification to $\Z_{16}$. 
\end{enumerate}

\subsection{$\LG= \Pin^-(d)$}

When the spacetime symmetry is $\LG=\Pin^-(d)$, the spacetime manifold should be unorientable, and allows a $\Pin^-$ structure. In particular $w_2+w_1^2$ is trivial.  A quantum field theory that can be formulated on a $\Pin^-$ manifold should have time reversal symmetry, and there exists a fermion in the operator spectrum. Time reversal acts on the fermion as $\T^2=1$.   


\begin{table}[t]
	\centering
	\begin{tabular}{ c c c c }
		\hline
		\multicolumn{4}{c}{Bordism and Cobordism group}\\
		\hline
		$d$ & 
		$\Omega^{\Pin^- }_d$ &
		$\TP^{\Pin^- }_d$ &
		Cobordism Invariant \\
		\hline
		1& $\Z_2$ & $\Z_2$ & $\tilde{\eta}$\\
		2& $\Z_8$ & $\Z_8$& $\abk$\\
		3 & $0$ & $0$ &   \\
		4 & $0$& $0$ & \\
		5 & $0$ & $0$&  \\
		\hline
	\end{tabular}
	\caption{
		The bordism and cobordism groups of the symmetry  $\Pin^-(d)$. The cobordism invariants, representing the $d$ dimensional anomaly inflow actions of the $d-1$ dimensional quantum field theories, are also enumerated. 
	}
	\label{table:Pin-Bordism}
\end{table}

We enumerate the bordism and cobordism groups and list the cobordism invariants, in Table \ref{table:Pin-Bordism}. Below, we comment on the nontrivial cobordism invariants, and interpret them as the anomaly inflow actions of QFTs living in one dimension lower. 
\begin{enumerate}
	\item When $d=1$, the only topology of a connected manifold is a circle, hence $w_1=0$. Thus the $\Pin^-$ condition $w_1^2+w_2=0$ is trivially satisfied. Thus the cobordism invariant for the $\Pin^-$ manifold reduces to the cobordism invariant for the Spin manifold, and is given by $\tilde{\eta}$, as we have discussed in section \ref{subsecSpin}. 
	
	\item When $d=2$,  $\abk$  is a cobordism invariant describing a fermionic invertible TQFT with time reversal symmetry satisfying $\T^2=1$. $\abk$ depends on the choice of the $\Z_4$ valued $\Pin^-$ structure, which is abstract to describe (as opposed to the $\Spin$ structure where one can specify by the boundary condition of fermions across each cycle on $T^2$). See \cite{Kapustin:2014dxa} for the partition function of ABK invariant.  However, the physical meaning of this invariant is clear. A quantum field theory realizing this invertible TQFT is a Dirac fermion $\psi$ in $1+1$d with large negative mass. The fermion $\psi$ transforms under time reversal as $\T: \psi(t,x)\to \gamma^0 \psi^*(-t,x)$, which is usually denoted as $\mathcal{CT}$ in the literature. Without interaction, it corresponds to the symmetry class BDI, and the SPT classification is $\Z$. In the presence of interaction, the classification collapses to $\Z_8$, which is precisely described by the $\abk$ invariant. Another way to understand the $\abk$ invertible TQFT is via the Smith isomorphism, $\Omega_3^{\Spin}(\B\Z_2)\simeq \Omega_2^{\Pin^-}$\cite{Kapustin:2014dxa, Hason_2020, Cordova:2019wpi}. This isomorphism relates a 3d invertible  spin TQFT with a $\Z_2$ unitary global symmetry to a 2d invertible $\Pin^-$ TQFT without any symmetry. Indeed, the former also has $\Z_8$ classification, given by $\SO(n)_1$ Chern Simons\cite{Cordova2017vab1711.10008, Seiberg:2016rsg, Gu_2014, Kapustin:2014dxa, Wang_2017, Kapustin_2017}. The physical meaning of this isomorphism is that the domain wall of spontaneously broken $\Z_2$ symmetry supports a 2d $\abk$ TQFT. 
\end{enumerate}

\section{Global Symmetry: $G=\SU(2)$}
\label{secSU2}

In this section, we demand the internal global symmetry $G$ to be $\SU(2)$, and allow different choices of the Lorentz symmetry $\LG$. For a given $G$ and $\LG$, there can be multiple choices of $G'$ determined by the exact sequence \eqref{groupext}.  Throughout this section, we will assume $G=\SU(2)$ to be the global symmetry. In section \ref{GaugingSU2}, we will promote the global symmetry $G$ to be  dynamical, and the physical interpretation for the same exact sequence \eqref{groupext} will be different. 

\subsection{$\LG=\SO(d)$}

We first specify $\LG= \SO(d)$. It means that the spacetime manifold is oriented, and generically does not allow a spin structure. We will consider a continuous quantum field theory that can be defined on the such a non-spin oriented manifold. As it will be discussed below, such a quantum field theory can allow a fermion in the operator spectrum, as long as the fermion carries $j= \Z+1/2$ $\SU(2)$ isospin. This should be contrasted to the case where $\SU(2)$ is absent in section \ref{subsecSO}. 

\subsubsection{Two Classes of Lorentz Symmetry Extensions}

Demanding $\LG=\SO(d)$ in the exact sequence \eqref{groupext}, we obtain
\begin{eqnarray}
1\to \SU(2)\to G' \to \SO(d)\to 1.
\end{eqnarray}
Because $\SO(d)$ tangent bundle has an unconstrained $w_2$, there are two choices of extensions $G'$, by either identifying the second Stiefel-Whitney class of the $\SU(2)/\Z_2=\SO(3)$ bundle  $w_2(V_{\SO(3)})$ to be either zero or $w_2$, 
\begin{eqnarray}\label{gbcSO}
G'= 
\begin{cases}
\SU(2)\times \SO(d), &~~~ w_2(V_{\SO(3)})=0,\\
({\SU(2)\times \Spin(d)})/{\Z_2}, &~~~ w_2(V_{\SO(3)})=w_2.
\end{cases}
\end{eqnarray}
The two choices of symmetry extensions have different physical interpretations.

For $G'= \SU(2)\times \SO(d)$, any local operator $\CO$ transforming in the $j$ isospin representation of $\SU(2)$, for any $j$, is a boson. There is no fermionic operator in the operator spectrum. In the condensed matter language, a continuous quantum field theory with Lorentz symmetry $\SU(2)\times \SO(d)$ can be understood as emerging from a lattice model whose fundamental degrees of freedom are bosonic. As an example, a quantum field theory with $\SU(2)\times \SO(d)$ global symmetry is the $2$ complex scalars with a degenerate mass.

For $G'= (\SU(2)\times \Spin(d))/\Z_2$, the operator spectrum is allowed to contain fermionic operators. However, the quotient $\Z_2$ implies that the statistics of the operator is correlated with its $\SU(2)$ isospin. Specifically, a fermionic operator should transform under $j\in \Z+ \frac{1}{2}$ representation of $\SU(2)$, while a bosonic operator should transform under $j\in  \Z$ representation of $\SU(2)$. This Spin-$\SU(2)$ relation  is analogous to the Spin-Charge relation when the global symmetry is  $\U(1)$\cite{Seiberg:2016rsg}. In the condensed matter language, a theory with the global symmetry $G'=(\SU(2)\times \Spin(d))/\Z_2$ emerges from a lattice model whose fundamental degrees of freedom are fermionic and transforms in $j= \Z+ \frac{1}{2}$ representation of $\SU(2)$, and is regarded as a fermionic theory. Formally, the correlation between the $\SU(2)$ and Lorentz quantum numbers also implies nontrivial constraints between the $\SU(2)/\Z_2= \SO(3)$ bundle and the $\Spin(d)/\Z_2=\SO(d)$ tangent bundle of the spacetime, 
$w_2(V_{\SO(3)}) = w_2$ as in \eqref{gbcSO}. 
We emphasize that although the theory is fermionic, one can still place the theory on a nonspin manifold, by demanding the $\SU(2)$ background field to be a $\Spin$-$\SU(2)$ connection. As an example, a fermionic quantum field theory with $(\SU(2)\times \Spin(d))/\Z_2$ global symmetry is the two free Dirac fermions with a degenerate mass.

\subsubsection{(Co)bordism Groups and Invariants of $\SO(d)\times \SU(2)$}

As introduced in the introduction, to compute $\Omega_d^{\SO\times \SU(2)}$, we need to compute the Adams spectral sequence \eqref{ADSS} with $MTG'= MT(\SO\times\SU(2))=M\SO\wedge(\B\SU(2))_+$. Here $X_+$ is the disjoint union of the topological space $X$ and a point.  

\begin{figure}[H]
	\begin{center}
		\begin{tikzpicture}
		\node at (0,-1) {0};
		\node at (1,-1) {1};
		\node at (2,-1) {2};
		\node at (3,-1) {3};
		\node at (4,-1) {4};
		\node at (5,-1) {5};
		\node at (6,-1) {$t-s$};
		\node at (-1,0) {0};
		\node at (-1,1) {1};
		\node at (-1,2) {2};
		\node at (-1,3) {3};
		\node at (-1,4) {4};
		\node at (-1,5) {5};
		\node at (-1,6) {$s$};
		
		\draw[->] (-0.5,-0.5) -- (-0.5,6);
		\draw[->] (-0.5,-0.5) -- (6,-0.5);

		\draw (0,0) -- (0,5);
		\draw (4,0) -- (4,5);

		\draw (4.1,0) -- (4.1,5);

		\draw[fill] (5,0) circle(0.05);
		
		\end{tikzpicture}
	\end{center}
	\caption{$E_2$ page of the Adams spectral sequence with symmetry $\SO(d)\times \SU(2)$. The Bordism group $\Omega_*^{\SO \times \SU(2)}$ and the invariants can be read off from the this chart. }
	\label{fig:E_2SOSU2}
\end{figure}
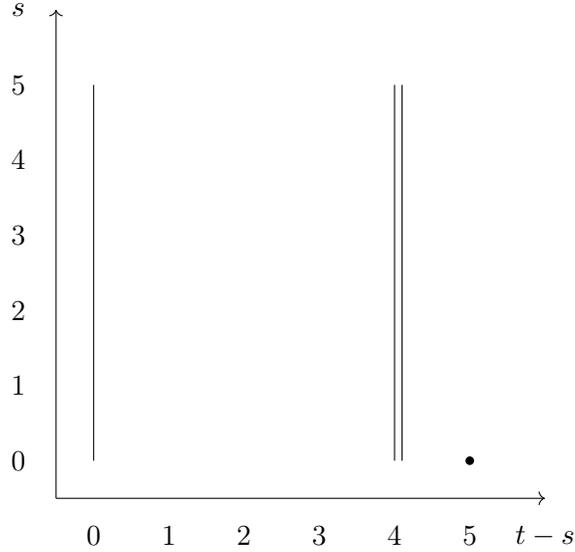

By K\"unneth formula, 
\bea
\H^*(M\SO\wedge(\B\SU(2))_+,\Z_2)=\H^*(M\SO,\Z_2)\otimes\H^*(\B\SU(2),\Z_2).
\eea
We have used the reduced version, note that the reduced cohomology of $X_+$ is exactly the ordinary cohomology of $X$. Since there is no odd torsion, the Adams spectral sequence is
\bea\label{seq}
\Ext_{\A_2}^{s,t}(\H^*(M\SO,\Z_2)\otimes\H^*(\B\SU(2),\Z_2),\Z_2)\Rightarrow\Omega_{t-s}^{\SO\times\SU(2)}.
\eea
We need to evaluate $\H^*(M\SO,\Z_2)$ and $\H^*(\B\SU(2),\Z_2)$ respectively. 
\begin{enumerate}
	\item To evaluate $\H^*(M\SO,\Z_2)$, we use the fact that the localization of $M\SO$ at the prime $2$ is $M\SO_{(2)}=H\Z_{(2)}\vee \Sigma^4H\Z_{(2)}\vee\Sigma^5H\Z_2\vee\cdots$.  Here $HG$ is the Eilenberg-MacLane spectrum of the group $G$, $\Sigma$ is the suspension, and $\vee$ is the wedge sum. So the mod 2 cohomology of $M\SO$ is 
	\bea
	\H^*(M\SO,\Z_2)=\A_2/\A_2\Sq^1\oplus\Sigma^4\A_2/\A_2\Sq^1\oplus\Sigma^5\A_2\oplus\cdots.
	\eea
	In the above, the projective $\A_2$-resolution of $\A_2/\A_2\Sq^1$ (denoted by $P_{\bullet}$) is
	\bea
	\cdots\to\Sigma^3\A_2\to\Sigma^2\A_2\to\Sigma\A_2\to\A_2\to\A_2/\A_2\Sq^1
	\eea
	where the differentials $d_1$ are induced from $\Sq^1$.
	\item To evaluate $\H^*(\B\SU(2),\Z_2)$, we find
	\bea\label{HBSU2}
	\H^*(\B\SU(2),\Z_2)=\Z_2[c_2].
	\eea
	Here $c_2$ is the second Chern class of the $\SU(2)$ bundle.
\end{enumerate}
Next, we combine the results. Since $P_{\bullet}$ is actually a free $\A_2$-resolution of $\A_2/\A_2\Sq^1$,  $P_{\bullet}\otimes\H^*(\B\SU(2),\Z_2)$ is also a free $\A_2$-resolution of $\A_2/\A_2\Sq^1\otimes\H^*(\B\SU(2),\Z_2)$. The $E_2$ page of the Adams spectral sequence is shown in Figure \ref{fig:E_2SOSU2}, from which we can read off the bordism groups and invariants, as shown in Table \ref{table:SOSU2Bordism}.

\subsubsection{(Co)bordism Groups and Invariants of $(\Spin(d)\times \SU(2))/\Z_2$}

Let $G'=(\Spin(d)\times \SU(2))/\Z_2$, then by \cite{Freed2016}, we have $MTG'=M\Spin\wedge\Sigma^{-3}M\SO(3)$. For $t-s<8$, since there is no odd torsion, we have the Adams spectral sequence
\bea
\Ext_{\A_2(1)}^{s,t}(\H^{*+3}(M\SO(3),\Z_2),\Z_2)\Rightarrow\Omega_{t-s}^{\frac{\Spin \times \SU(2)}{\Z_2^F}}
\eea
which can be used to determine the bordism group and invariants. Here, the $\A_2(1)$-module structure of $\H^{*+3}(M\SO(3),\Z_2)$ below degree 5 is shown in Figure \ref{fig:A_2(1)MSO3}. From Figure \ref{fig:A_2(1)MSO3}, one can obtain the $E_2$ page shown in Figure \ref{fig:E_2SpinSU2Z2}. From the $E_2$ page, one can read off the (co)bordism group and invariants.

\begin{figure}[H]
	\begin{center}
		\begin{tikzpicture}[scale=0.5]
		
		\node[below] at (0,0) {$U$};
		\node[right] at (0,2) {$w_2U$};
		\node[right] at (0,3) {$w_3U$};
		\node[right] at (0,4) {$w_2^2U$};
		\node[left] at (0,5) {$w_2w_3U$};
		\node[left] at (0,6) {$w_3^2U$};
		\node[right] at (1,6) {$w_3^2U+w_2^3U$};
		\node[right] at (1,7) {$w_2^2w_3U$};
		\node[left] at (1,8) {$w_3^2w_2U$};
		\node[above] at (1,9) {$w_3^3U$};

		\draw[fill] (0,0) circle(.1);
		\draw[fill] (0,2) circle(.1);
		\draw (0,0) to [out=150,in=150] (0,2);
		\draw[fill] (0,3) circle(.1);
		\draw (0,2) -- (0,3);
		\draw[fill] (0,4) circle(.1);
		\draw[fill] (0,5) circle(.1);
		\draw[fill] (0,6) circle(.1);
		\draw (0,5) -- (0,6);
		\draw[fill] (1,6) circle(.1);
		\draw (0,4) to [out=30,in=150] (1,6);
		\draw[fill] (1,7) circle(.1);
		\draw (1,6) -- (1,7);
		\draw (0,5) to [out=30,in=150] (1,7);
		\draw[fill] (1,8) circle(.1);
		\draw (0,6) to [out=30,in=150] (1,8);
		\draw[fill] (1,9) circle(.1);
		\draw (1,8) -- (1,9);
		\draw (1,7) to [out=30,in=30] (1,9);
		\end{tikzpicture}
	\end{center}
	\caption{The $\A_2(1)$-module structure of $\H^{*+3}(M\SO(3),\Z_2)$ below degree 5.}
	\label{fig:A_2(1)MSO3}
\end{figure}

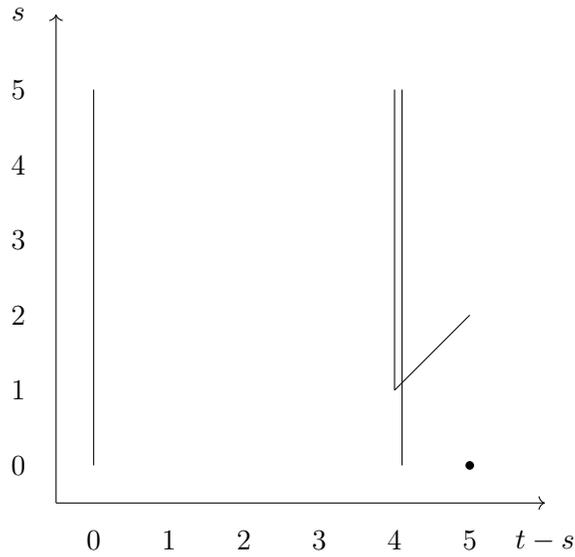
\begin{figure}[H]
	\begin{center}
		\begin{tikzpicture}
		\node at (0,-1) {0};
		\node at (1,-1) {1};
		\node at (2,-1) {2};
		\node at (3,-1) {3};
		\node at (4,-1) {4};
		\node at (5,-1) {5};
		\node at (6,-1) {$t-s$};
		\node at (-1,0) {0};
		\node at (-1,1) {1};
		\node at (-1,2) {2};
		\node at (-1,3) {3};
		\node at (-1,4) {4};
		\node at (-1,5) {5};
		\node at (-1,6) {$s$};
		
		\draw[->] (-0.5,-0.5) -- (-0.5,6);
		\draw[->] (-0.5,-0.5) -- (6,-0.5);

		\draw (0,0) -- (0,5);
		\draw (4,1) -- (4,5);
		
		\draw (4,1) -- (5,2);
		
		\draw (4.1,0) -- (4.1,5);
		
		\draw[fill] (5,0) circle(0.05);
		
		\end{tikzpicture}
	\end{center}
	\caption{$E_2$ page of the Adams spectral sequence with symmetry $(\Spin(d) \times \SU(2))/\Z_2$. The Bordism group $\Omega_*^{({\Spin \times \SU(2)})/{\Z_2}}$ and the invariants can be read off from the this chart. }
	\label{fig:E_2SpinSU2Z2}
\end{figure}

\subsubsection{Physical Interpretations of the Cobordism Invariants and Classification of Anomalies}
\label{anomSOSU2}

The bordism and cobordism groups and the cobordism invariants are shown in Table \ref{table:SOSU2Bordism} and \ref{table:SpinSU2Bordism}.  Below, we comment on the nontrivial cobordism invariants, and interpret them as the anomaly inflow actions of QFTs living in one dimension lower.



\begin{table}[t]
	\centering
	\begin{tabular}{ c c c c }
		\hline
		\multicolumn{4}{c}{Bordism and Cobordism group}\\
		\hline
		$d$ & 
		$\Omega^{\SO \times \SU(2)}_d$ &
		$\TP^{\SO \times \SU(2)}_d$ &
		Cobordism Invariant \\
		\hline
		1& 0 & 0&\\
		2& 0 & 0&\\
		3 & $0$ & $\Z^2$ & $\mathrm{CS}^{\SU(2)}_3, 16 \CSg$\\
		4 & $\Z^2$& 0 & \\
		5 & $\Z_2$ & $\Z_2$&  $w_2w_3$\\
		\hline
	\end{tabular}
	\caption{
		The bordism and cobordism groups of the symmetry  $\SU(2)\times \SO(d)$. 
	}
	\label{table:SOSU2Bordism}
\end{table}

\subsubsection*{Nontrivial Anomalies from $\TP^{\SO \times \SU(2)}_d$}

We first discuss the anomalies associated with the global symmetry $\SO(d)\times \SU(2)$ in Table \ref{table:SOSU2Bordism}. The anomaly $16 \CSg$ and $w_2 w_3$ persists even when the $\SU(2)$ symmetry is explicitly broken, and appeared already in  $\TP^{\SO}_d$. Hence we will not discuss these cobordism invariants here.

The remaining nontrivial cobordism invariant is $\mathrm{CS}_3^{\SU(2)}$.  The anomaly $\mathrm{CS}_3^{\SU(2)}$ can be saturated by bosonic gapless Wess-Zumino-Witten (WZW) CFT with the target space $\SU(2)$ in 2d. There is a WZW term, $k \mathrm{WZW}_{\SU(2)}$ that can be defined only by introducing an auxiliary 3d manifold that is bounded by the 2d spacetime. Demanding that the WZW term $k \mathrm{WZW}_{\SU(2)}$ to be independent of the choice of 3d manifold enforces  $k\in \Z$.  When coupled to the $\SU(2)$ background field, the WZW term depends on the choice of 3d manifold, hence it contributes a nontrivial anomaly $k \mathrm{CS}_3^{\SU(2)}$. Different choice of $k$ corresponds to different anomaly, hence $k$ parameterizes the integer classification $\Z$.




We remark that one can not write down a 5d $\SU(2)$ Chern Simons term simply because the rank of the symmetry group is too low. Explicitly, one can check for instance 
$\Tr(A\dd A\dd A+\cdots) \sim \sum_{a,b,c=1}^3 \sum_{\mu, \nu, \rho, \sigma, \lambda=0}^4 \epsilon_{abc}\epsilon^{\mu\nu\rho\sigma\lambda}A^a_{\mu}\partial_{\nu}A^b_{\rho}\partial_{\sigma}A^c_\lambda + \cdots$ which vanishes because the epsilon tensors enforce a minus sign if we exchange $b\leftrightarrow c, \nu \leftrightarrow \sigma, \rho\leftrightarrow \lambda$. This explains that there is no perturbative anomaly for $\SU(2)^3$ triangle diagram. However, for $\SU(N)$ with  $N\geq 3$, a nontrivial 5d Chern Simons term $\mathrm{CS}_5^{\SU(N)} = \frac{1}{24\pi^2}\int \Tr\left(A\dd A\dd A-\frac{3\ii}{2}A^3 \dd A -\frac{3}{5}A^5\right)$ exists.

\begin{table}[t]
	\centering
	\begin{tabular}{ c c c c }
		\hline
		\multicolumn{4}{c}{Bordism and Cobordism group}\\
		\hline
		$d$ & 
		$\Omega^{(\Spin\times \SU(2))/\Z_2}_d$ &
		$\TP^{(\Spin\times \SU(2))/\Z_2}_d$ &
		Cobordism Invariant \\
		\hline
		1& 0 & 0&\\
		2& 0 & 0&\\
		3 & $0$ & $\Z^2$ & $\widehat{\mathrm{CS}}_3^{\SU(2)}, 16 \CSg$\\
		4 & $\Z^2$& 0 & \\
		5 & $\Z_2\times \Z_2$ & $\Z_2\times \Z_2$&  $w_2w_3, \widehat{\mathcal{I}}_{1/2}$\\
		\hline
	\end{tabular}
	\caption{
		The bordism and cobordism groups of the symmetry  $(\Spin(d)\times \SU(2))/\Z_2$. 
	}
	\label{table:SpinSU2Bordism}
\end{table}

\subsubsection*{Nontrivial Anomalies from $\TP^{(\Spin\times \SU(2))/\Z_2}_d$}

We further discuss the anomalies associated with the global symmetry $(\Spin\times \SU(2))/\Z_2$ in Table \ref{table:SpinSU2Bordism}. Similar to the case $\SO(d)\times \SU(2)$, the anomaly $16 \CSg$ and $w_2 w_3$ persist even when the $\SU(2)$ symmetry is explicitly broken, and appear already in $\Omega_d^{\SO}$ and $\TP^{\SO}_d$. Hence we will not discuss these cobordism invariants here.

The remaining  nontrivial cobordism invariants are $\widehat{\mathrm{CS}}_3^{\SU(2)}$ and $\mathcal{I}_{1/2}$, which we discuss below. 
The cobordism invariant $\widehat{\mathrm{CS}}_3^{\SU(2)}$ is obtained from $\mathrm{CS}_3^{\SU(2)}$ via replacing the $\SU(2)$ gauge field by the Spin-SU(2) connection, with suitable gravitational Chern Simons term $4\CSg$. We will derive such gravitational term in section \ref{secsu2gauge}.  This is reminiscent to the $\widehat{\mathrm{CS}}_3^{\U(1)}$ Chern Simons theory where the gauge field is $\Spin_c$ connection. Such a theory is well-defined on a non-spin manifold only when $2\CSg$ is appended.

The remaining nontrivial cobordism invariant is $\widehat{\mathcal{I}}_{1/2}$, which is the mod 2 reduction of the Dirac operator of a fermion in the $j=1/2$ $\SU(2)$ isospin representation,  coupled to Spin-$\SU(2)$ connection and formulated on a non-spin manifold in 5d\cite{Wang:2018qoyWWW}.  If formulating $\widehat{\mathcal{I}}_{1/2}$ on the spin manifold and couple to an ordinary $\SU(2)$ gauge field, the anomaly $\widehat{\mathcal{I}}_{1/2}$ reduces to the standard Witten $\SU(2)$ anomaly ${\mathcal{I}}_{1/2}$\cite{Wang:2018qoyWWW}.  In \cite{Wan2019oyr1904.00994}, it has been suggested that $\widehat{\mathcal{I}}_{1/2}$ can be expressed in terms of a twisted version of Stiefel-Whitney class $w_3'$ and the $\arf$ invariant. A more precise relation will be discussed in \cite{JuvenWIP}.


\subsection{$\LG=\Spin(d)$}
\label{subsecSpinSU2}

When $\LG=\Spin(d)$, the spacetime manifold is an oriented manifold that allows a spin structure. Thus the quantum field theory on such a manifold contains a fermion in the local operator spectrum.

\subsubsection{Total Symmetry and Classification of  Anomalies}

Demanding $\LG=\Spin(d)$ in the exact sequence \eqref{groupext}, we obtain
\begin{eqnarray}
1\to \SU(2) \to G' \to \Spin(d) \to 1.
\end{eqnarray}
Since both $w_1$ and $w_2$ vanish for the $\Spin(d)$ bundle, $w_2(V_{\SO(3)})$ can not be correlated with anything. So $w_2(V_{\SO(3)})=0$, which implies that the $\SO(3)$ bundle is lifted to the $\SU(2)$ bundle. There is only one choice of extension
\begin{eqnarray}\label{Spintotalsym}
G'= \SU(2)\times \Spin(d), ~~~ w_2(V_{\SO(3)})=0.
\end{eqnarray}
The total symmetry \eqref{Spintotalsym} implies that the fermion carries $\SU(2)$ isospin $j$ with either $j$ being integer or half-integer. This should be contrasted with the total symmetry $G'=(\SU(2)\times \Spin(d))/\Z_2$ where fermion can only transform under $j=\Z+ 1/2$ $\SU(2)$ isospin. The product structure of the total symmetry implies that there is no nontrivial bundle constraints between $\SU(2)$ and the tangent bundle of the spacetime manifold.

\subsubsection{(Co)bordism Groups and Invariants of $\Spin(d)\times \SU(2)$}

We need to compute the Adams spectral sequence \eqref{ADSS} with  $MTG'=MT(\Spin\times\SU(2))=M\Spin\wedge(\B\SU(2))_+$. For $t-s<8$, since there is no odd torsion, the Adams spectral sequence is
\bea
\Ext_{\A_2(1)}^{s,t}(\H^*(\B\SU(2),\Z_2),\Z_2)\Rightarrow\Omega_{t-s}^{\Spin}(\B\SU(2)).
\eea
The $\A_2(1)$-module structure of $\H^*(\B\SU(2),\Z_2)$ below degree 5 is shown in Figure \ref{fig:A_2(1)BSU2}, from which one can find the $E_2$ page as shown in Figure  \ref{fig:E_2SpinSU2}. One can read off the (co)bordism invariants from the $E_2$ page as in Table \ref{table:SpinxSU2Bordism}.

\begin{figure}
	\begin{center}
		\begin{tikzpicture}[scale=0.5]
		
		\node[below] at (0,0) {$1$};
		
		\draw[fill] (0,0) circle(.1);
		
		\node[below] at (0,4) {$c_2$};
		
		\draw[fill] (0,4) circle(.1);
		
		\end{tikzpicture}
	\end{center}
	\caption{The $\A_2(1)$-module structure of $\H^*(\B\SU(2),\Z_2)$ below degree 5.}
	\label{fig:A_2(1)BSU2}
\end{figure}

\begin{figure}
	\begin{center}
		\begin{tikzpicture}
		\node at (0,-1) {0};
		\node at (1,-1) {1};
		\node at (2,-1) {2};
		\node at (3,-1) {3};
		\node at (4,-1) {4};
		\node at (5,-1) {5};
		\node at (6,-1) {$t-s$};
		\node at (-1,0) {0};
		\node at (-1,1) {1};
		\node at (-1,2) {2};
		\node at (-1,3) {3};
		\node at (-1,4) {4};
		\node at (-1,5) {5};
		\node at (-1,6) {$s$};
		
		\draw[->] (-0.5,-0.5) -- (-0.5,6);
		\draw[->] (-0.5,-0.5) -- (6,-0.5);

		\draw (0,0) -- (0,5);
		\draw (0,0) -- (2,2);
		\draw (4,3) -- (4,5);

		\draw (4.1,0) -- (4.1,5);
		
		\draw (4.1,0) -- (5.1,1);
		
		\end{tikzpicture}
	\end{center}
	\caption{$E_2$ page of the Adams spectral sequence with symmetry $\Spin(d) \times \SU(2)$. The Bordism group $\Omega_*^{{\Spin \times \SU(2)}}$ and the invariants can be read off from the this chart. }
	\label{fig:E_2SpinSU2}
\end{figure}

\begin{table}[t]
	\centering
	\begin{tabular}{ c c c c }
		\hline
		\multicolumn{4}{c}{Bordism and Cobordism group}\\
		\hline
		$d$ & 
		$\Omega^{\Spin\times \SU(2)}_d$ &
		$\TP^{\Spin\times \SU(2)}_d$ &
		Cobordism Invariant \\
		\hline
		1& $\Z_2$& $\Z_2$& $\tilde{\eta}$\\
		2& $\Z_2$ & $\Z_2$&  $\arf$\\
		3 & $0$ & $\Z\times \Z$ & $\mathrm{CS}_3^{\SU(2)}, \CSg$\\
		4 & $\Z\times \Z$& 0 & \\
		5 & $\Z_2$ & $\Z_2$&  $\mathcal{I}_{1/2}$\\
		\hline
	\end{tabular}
	\caption{
		The bordism and cobordism groups of the symmetry  $\Spin(d)\times \SU(2)$. 
	}
	\label{table:SpinxSU2Bordism}
\end{table}

\subsubsection{Physical Interpretations of the Cobordism Invariants and Classification of Anomalies}

The bordism and cobordism groups and  the cobordism invariants are enumerated in Table \ref{table:SpinxSU2Bordism}. The anomaly $\tilde{\eta}, \arf, \CSg$ already appear in $\TP^{\Spin}_d$, which have been discussed already in section \ref{subsecSpin}. $\mathrm{CS}^{\SU(2)}_3$ is the 3d $\SU(2)$ Chern Simons term at level 1, discussed already in section \ref{anomSOSU2}. $\mathcal{I}_{1/2}$ is the mod 2 reduction of the Dirac operator of fermion in the $j=1/2$ $\SU(2)$ isospin representation,coupled to $\SU(2)$ background field and formulated on the spin manifold. This is obtained from $\widehat{\mathcal{I}}_{1/2}$ in section \ref{anomSOSU2} by taking $w_2=0$.

Another way of representing the cobordism invariant $\mathcal{I}_{1/2}$ is $\tilde{\eta} c_2^{\SU(2)} \mod 2$, where $\tilde{\eta}$ is defined in \eqref{tildeeta} and $c_2^{\SU(2)}$ is the second Chern class of the $\SU(2)$ vector bundle. The physical meaning of the SU(2) Witten anomaly in this representation is more transparent: If we couple the system to the $\SU(2)$ background field $A$ with nontrivial instanton number $\int_M c_2^{\SU(2)}$, the anomaly $\tilde{\eta} c_2^{\SU(2)} \mod 2$ means under fermi-parity transformation, the partition function acquires a minus sign, 
\begin{eqnarray}
Z[A] \to (-1)^{\int_M c_2^{\SU(2)}} Z[A] 
\end{eqnarray}
which further means that the core of the $\SU(2)$ instanton traps a nontrivial fermion zero mode. This is precisely the signature of the $\SU(2)$ Witten anomaly.

\subsection{$\LG= \O(d)$}

When $\LG= \O(d)$, the spacetime manifold is an unorientable manifold, in particular $w_1$ and $w_2$ are both un-constrained. A quantum field theory that can be formulated on the most generic unorientable manifold must be time reversal symmetric. However, $\T^2$ does not have to be identity. The eigenvalue of $\T^2$ on a  local operator depends on the choice of symmetry extension, as we will discuss below. 

\subsubsection{Four Lorentz Symmetry Extensions}

Demanding $\LG= \O(d)$ in the exact sequence \eqref{groupext}, we obtain
\begin{eqnarray}
1\to \SU(2) \to G' \to \O(d) \to 1.
\end{eqnarray}
For $\O(d)$,  $w_1$, $w_1^2$ and $w_2$ are all unconstrained. Hence one can identify $w_2(V_{\SO(3)})$ of the $\SO(3)$ vector bundle  with  $K_1 w_1^2 + K_2 w_2$ for all four choices of $(K_1, K_2)\in \Z_2\times \Z_2$. This means that there   are four choices of extensions, 
\begin{eqnarray}\label{Osym}
\begin{split}
G'= 
\begin{cases}
\SU(2)\times \O(d), & ~~~ w_2(V_{\SO(3)})=0,\\
({\SU(2)\times \rE(d)})/{\Z_2}, &~~~ w_2(V_{\SO(3)})=w_1^2,\\
({\SU(2)\times \Pin^+(d)})/{\Z_2}, &~~~ w_2(V_{\SO(3)})=w_2,\\
({\SU(2)\times \Pin^-(d)})/{\Z_2}, &~~~ w_2(V_{\SO(3)})=w_1^2+ w_2.\\
\end{cases}
\end{split}
\end{eqnarray}
The four choices of symmetry extensions have different physical interpretations. 

For $G'= \SU(2)\times \O(d)$, which corresponds to trivial extension, any local operator in the operator spectrum should be bosonic, and allows to have arbitrary SU(2) isospin $j$. Time reversal transformation satisfies $\T^2= 1$. Formally, there are no gauge bundle constraints between the $\SU(2)$ bundle and the tangent bundle of the spacetime manifold. 

For $G'= ({\SU(2)\times \rE(d)})/{\Z_2}$, any local operator in the operator spectrum should still be bosonic. However, the $\SU(2)$ isospin $j$ should be correlated with its Kramers parity: any operator  with $j=\Z$ transforms as a Kramers singlet, while any operator with $j=Z+ 1/2$ transforms as a Kramers doublet. In summary, $\T^2=(-1)^{2j}$. Formally, the $\SU(2)/\Z_2= \SO(3)$ bundle and the tangent bundle of the spacetime manifold are correlated: 
$w_2(V_{\SO(3)}) = w_1^2$.

For $G'= ({\SU(2)\times \Pin^+(d)})/{\Z_2}$, the statistics and Kramers parity of a local operator are correlated with the $\SU(2)$ isospin $j$.  If the operator is a fermion, it transforms as a Kramers doublet under time reversal, and carries $j=\Z+1/2$ $\SU(2)$ isospin.  If the operator is a boson, it transforms as a Kramers singlet under time reversal, and carries $j=\Z$ $\SU(2)$ isospin. In summary, $\T^2= (-1)^{(2j+1)F}$. Formally, the  $\SU(2)/\Z_2= \SO(3)$ bundle and the tangent bundle of the spacetime manifold are correlated: 
$w_2(V_{\SO(3)}) = w_2$.

For  $G'= ({\SU(2)\times \Pin^-(d)})/{\Z_2}$, the statistics of a local operator are correlated with the $\SU(2)$ isospin $j$, while the Kramers parity does not.  If the operator is a fermion, it  carries $j=\Z+1/2$ $\SU(2)$ isospin.  If the operator is a boson, it  carries $j=\Z$ $\SU(2)$ isospin. In both cases, the operator should transform as a Kramers singlet. In summary, $\T^2=1$. Formally, the  $\SU(2)/\Z_2= \SO(3)$ bundle and the tangent bundle of the spacetime manifold are correlated: 
$w_2(V_{\SO(3)}) = w_1^2+ w_2$.

\subsubsection{(Co)bordism Groups and Invariants of $\O(d)\times \SU(2)$}

We need to compute the Adams spectral sequence \eqref{ADSS} with   $MTG'=MT(\rO\times\SU(2))=M\rO\wedge(\B\SU(2))_+$. By K\"unneth formula, 
\bea
\H^*(M\rO\wedge(\B\SU(2))_+,\Z_2)=\H^*(M\rO,\Z_2)\otimes\H^*(\B\SU(2),\Z_2).
\eea
Since there is no odd torsion,  the Adams spectral sequence \eqref{ADSS} can be written as
\bea
\Ext_{\A_2}^{s,t}(\H^*(M\rO,\Z_2)\otimes\H^*(\B\SU(2),\Z_2),\Z_2)\Rightarrow\Omega_{t-s}^{\rO\times\SU(2)}.
\eea
We need to compute $\H^*(M\rO,\Z_2)$ and $\H^*(\B\SU(2),\Z_2)$ separately. 
\begin{enumerate}
	\item $\H^*(M\rO,\Z_2)$  can be computed as follows. The Thom spectrum $M\rO$ is the wedge sum of suspensions of the mod 2 Eilenberg-MacLane spectrum $H\Z_2$, hence $\H^*(M\tO,\Z_2)$ is the direct sum of suspensions of the mod 2 Steenrod algebra $\A_2$, actually Thom proved
	that
	\bea
	\pi_*(M\rO)=\Omega_*^{\rO}=\Z_2[x_2,x_4,x_5,x_6,x_8,\dots]
	\eea
	where the generators are in each degree other than $2^n-1$. So $M\tO=H\Z_2\vee\Sigma^2H\Z_2\vee2\Sigma^4H\Z_2\vee\Sigma^5H\Z_2\vee\cdots$ and
	\begin{eqnarray}
	\H^*(M\tO,\Z_2)=\A_2\oplus\Sigma^2\A_2\oplus2\Sigma^4\A_2\oplus\Sigma^5\A_2\oplus\cdots
	\end{eqnarray}
	\item $\H^*(\B\SU(2),\Z_2)$ is already computed in \eqref{HBSU2}
	\begin{eqnarray}
	\H^*(\B\SU(2),\Z_2)= \Z_2[c_2].
	\end{eqnarray}
\end{enumerate}
Combining the above results, we find
\bea\label{Ho}
&&\H^*(M\rO,\Z_2)\otimes\H^*(\B \SU(2),\Z_2)\nn\\
&=&(\A_2\oplus\Sigma^2\A_2\oplus2\Sigma^4\A_2\oplus\Sigma^5\A_2\oplus\cdots)\otimes\Z_2[c_2] \nonumber\\
&=&\A_2\oplus\Sigma^2\A_2\oplus3\Sigma^4\A_2\oplus\Sigma^5\A_2\oplus\cdots.
\eea
Using
\bea
\Ext_{\A_2}^{s,t}(\Sigma^r\A_2,\Z_2)=\left\{\begin{array}{ll}\Hom_{\A_2}^t(\Sigma^r\A_2,\Z_2)=\Z_2&\text{ if }t=r, s=0\\ 0 &\text{ else}\end{array}\right.,
\eea 
and apply it to every component in \eqref{Ho}, we obtain the bordism group and invariants as shown in Table \ref{table:OSU2Bordism}. The physical interpretation of cobordism invariants will be discussed in section \ref{secPhyInt}.

\begin{table}[t]
	\centering
	\begin{tabular}{ c c c c }
		\hline
		\multicolumn{4}{c}{Bordism and Cobordism group}\\
		\hline
		$d$ & 
		$\Omega^{\O\times \SU(2)}_d$ &
		$\TP^{\O\times \SU(2)}_d$ &
		Cobordism Invariant \\
		\hline
		1& $0$& $0$& \\
		2& $\Z_2$ & $\Z_2$&  $w_1^2$\\
		3 & $0$ & $0$ & \\
		4 & $\Z_2\times \Z_2\times \Z_2$& $\Z_2\times \Z_2\times \Z_2$ & $w_1^4, w_2^2, c_2^{\SU(2)}\mod 2$\\
		5 & $\Z_2$ & $\Z_2$&  $w_2w_3$\\
		\hline
	\end{tabular}
	\caption{
		The bordism and cobordism groups of the symmetry  $\O(d)\times \SU(2)$. 
	}
	\label{table:OSU2Bordism}
\end{table}

\subsubsection{(Co)bordism Groups and Invariants of $(\rE(d)\times \SU(2))/\Z_2$}

We first derive a useful equivalent expression for $MTG'= MT((\rE(d)\times \SU(2))/\Z_2)$.

First, notice that
$\rE$ is defined to be the subgroup of $\rO\times\Z_4$ consisting of the pairs $(A,j)$ such that $\det A=j^2$, there is a fibration $\B\rE\to\B\rO\xrightarrow{w_1^2}\B^2\Z_2$.  We can also think of the space $\B\rE$ as the fiber of $w_1 + x: \B\rO \times \B\Z_4 \to \B\Z_2$, where $x$ is the generator of $\H^1(\B\Z_4, \Z_2)$.  Note that $\SU(2)\times_{\Z_2}\Z_4=\Pin^+(3)$, we can think of the space $\B(\rE\times_{\Z_2}\SU(2))$ as the fiber of $w_1+w_1':\B\rO\times\B\Pin^+(3)\to\B\Z_2$, where $w_1'$ is the generator of $\H^1(\B\Pin^+(3),\Z_2)$. Take $W$ to be the rank 3 vector bundle on $\B\Pin^+(3)$ determined by $\B\Pin^+(3)\to\B\rO(3)$. 
Define a map $f: \B\rO \times \B\Pin^+(3) \to \B\rO \times \B\Pin^+(3)$ by $(V, V') \to (V + W - 3, V')$, with inverse $(V, V') \to (V - W + 3, V')$.  
Observe $f^*(w_1) = w_1 + w_1'$, so that $\B\rE$ is homotopy equivalent to $\B\SO \times \B\Pin^+(3)$.  
The canonical bundle $\B\rE \to \B\rO$ corresponds to $V - W + 3$ on $\B\SO \times \B\Pin^+(3)$.  
So $MT(\rE\times_{\Z_2}\SU(2)) = MT\SO \wedge \text{Thom}(\B\Pin^+(3), 3 - W) = M\SO\wedge\Sigma^{-3}M\Pin^+(3)$.  
$M\Pin^+(3)$ can be further simplified, by using $M\Pin^+(3)=MT\Pin^-(3)=MT(\Spin(3)\times\Z_2)=MT\Spin(3)\wedge MT\Z_2$.
\begin{eqnarray}
MT((\rE\times\SU(2))/\Z_2) = M\SO\wedge\Sigma^{-4}M\SU(2)\wedge\Sigma^1MT\rO(1)\simeq M\rO\wedge\Sigma^{-4}M\SU(2).
\end{eqnarray}

By K\"unneth formula, 
\bea
\H^*(M\rO\wedge\Sigma^{-4}M\SU(2),\Z_2)=\H^*(M\rO,\Z_2)\otimes\H^{*+4}(M\SU(2),\Z_2).
\eea
Since there is no odd torsion, we have the Adams spectral sequence
\bea
\Ext_{\A_2}^{s,t}(\H^*(M\rO,\Z_2)\otimes\H^{*+4}(M\SU(2),\Z_2),\Z_2)\Rightarrow\Omega_{t-s}^{\rE\times_{\Z_2}\SU(2)}.
\eea
Using
\bea
\H^{*+4}(M\SU(2),\Z_2)=\Z_2[c_2]U
\eea
where $c_2$ is the Chern class of the $\SU(2)$ bundle and $U$ is the Thom class, we find 
\bea\label{Ho1}
&&\H^*(M\rO,\Z_2)\otimes\H^{*+4}(M\SU(2),\Z_2)\nn\\
&=&(\A_2\oplus\Sigma^2\A_2\oplus2\Sigma^4\A_2\oplus\Sigma^5\A_2\oplus\cdots)\otimes\Z_2[c_2]U\nn\\
&=&\A_2\oplus\Sigma^2\A_2\oplus3\Sigma^4\A_2\oplus\Sigma^5\A_2\oplus\cdots.
\eea
From \eqref{Ho1}, one can read off the bordism group and invariant as shown in Table \ref{table:ESU2Bordism}.  The physical interpretation of cobordism invariants will be discussed in section \ref{secPhyInt}.

\begin{table}[t]
	\centering
	\begin{tabular}{ c c c c }
		\hline
		\multicolumn{4}{c}{Bordism and Cobordism group}\\
		\hline
		$d$ & 
		$\Omega^{(\rE\times \SU(2))/\Z_2}_d$ &
		$\TP^{(\rE\times \SU(2))/\Z_2}_d$ &
		Cobordism Invariant \\
		\hline
		1& $0$& $0$& \\
		2& $\Z_2$ & $\Z_2$&  $w_1^2$\\
		3 & $0$ & $0$ & \\
		4 & $\Z_2\times \Z_2\times \Z_2$& $\Z_2\times \Z_2\times \Z_2$ & $w_1^4, w_2^2, c_2^{\SU(2)}\mod 2$\\
		5 & $\Z_2$ & $\Z_2$&  $w_2w_3$\\
		\hline
	\end{tabular}
	\caption{
		The bordism and cobordism groups of the symmetry  $(\rE(d)\times \SU(2))/\Z_2$. 
	}
	\label{table:ESU2Bordism}
\end{table}

\subsubsection{(Co)bordism Groups and Invariants of $(\Pin^+(d)\times \SU(2))/\Z_2$}

Let $G'=(\Pin^+(d)\times \SU(2))/\Z_2$, then by \cite{Freed2016,1711.11587GPW}, we have $MTG'=M\Spin\wedge\Sigma^{-3}M\rO(3)$.  Here $w_2(TM)=w_2'(V_{\SO(3)})$, $w_3(TM)+w_1(TM)w_2(TM)=w_3'(V_{\SO(3)})$ and $w_1(TM)$ is nontrivial, $w_1'(V_{\SO(3)})=0$. For $t-s<8$, since there is no odd torsion, the Adams spectral sequence is
\bea
\Ext_{\A_2(1)}^{s,t}(\H^{*+3}(M\rO(3),\Z_2),\Z_2)\Rightarrow\Omega_{t-s}^{(\Pin^+(d)\times \SU(2))/\Z_2}.
\eea
The $\A_2(1)$-module structure of $\H^{*+3}(M\rO(3),\Z_2)$ below degree 5 is shown in Figure \ref{fig:A_2(1)MO3}, from which one can derive the $E_2$ page  shown in Figure \ref{fig:E_2Pin+SU2Z2}. Then one can read off the (co)bordism groups and invariants from the $E_2$ page, as shown in Table \ref{table:Pin+SU2Bordism}. The physical interpretation of cobordism invariants will be discussed in section \ref{secPhyInt}.

\begin{figure}
	\begin{center}
		\begin{tikzpicture}[scale=0.5]
		
		\node[below] at (0,0) {$U$};
		
		\draw[fill] (0,0) circle(.1);
		\draw[fill] (0,1) circle(.1);
		\draw (0,0) -- (0,1);
		\draw[fill] (0,2) circle(.1);
		\draw (0,0) to [out=150,in=150] (0,2);
		\draw[fill] (0,3) circle(.1);
		\draw[fill] (1,3) circle(.1);
		\draw (0,2) -- (0,3);
		\draw (0,1) to [out=30,in=150] (1,3);
		\draw[fill] (1,4) circle(.1);
		\draw (1,3) -- (1,4);
		\draw (0,2) to [out=30,in=150] (1,4);
		\draw[fill] (0,4) circle(.1);
		\draw[fill] (0,5) circle(.1);
		\draw (0,3) to [out=30,in=30] (0,5);
		\draw (0,4) -- (0,5);
		\draw[fill] (0,6) circle(.1);
		\draw (0,4) to [out=150,in=150] (0,6);
		
		\node[below] at (2,2) {$w_1^2U$};
		
		\draw[fill] (2,2) circle(.1);
		\draw[fill] (2,3) circle(.1);
		\draw (2,2) -- (2,3);
		\draw[fill] (2,4) circle(.1);
		\draw (2,2) to [out=150,in=150] (2,4);
		\draw[fill] (2,5) circle(.1);
		\draw[fill] (3,5) circle(.1);
		\draw (2,4) -- (2,5);
		\draw (2,3) to [out=30,in=150] (3,5);
		\draw[fill] (3,6) circle(.1);
		\draw (3,5) -- (3,6);
		\draw (2,4) to [out=30,in=150] (3,6);
		\draw[fill] (3,7) circle(.1);
		\draw (2,5) to [out=30,in=150] (3,7);
		\draw[fill] (3,8) circle(.1);
		\draw (3,7) -- (3,8);
		\draw (3,6) to [out=30,in=30] (3,8);
		
		\node[below] at (4,4) {$w_2^2U$};
		
		\draw[fill] (4,4) circle(.1);
		\draw[fill] (4,5) circle(.1);
		\draw (4,4) -- (4,5);
		\draw[fill] (4,6) circle(.1);
		\draw (4,4) to [out=150,in=150] (4,6);
		\draw[fill] (4,7) circle(.1);
		\draw[fill] (5,7) circle(.1);
		\draw (4,6) -- (4,7);
		\draw (4,5) to [out=30,in=150] (5,7);
		\draw[fill] (5,8) circle(.1);
		\draw (5,7) -- (5,8);
		\draw (4,6) to [out=30,in=150] (5,8);
		\draw[fill] (5,9) circle(.1);
		\draw (4,7) to [out=30,in=150] (5,9);
		\draw[fill] (5,10) circle(.1);
		\draw (5,9) -- (5,10);
		\draw (5,8) to [out=30,in=30] (5,10);
		
		\node[below] at (6,5) {$w_2w_3U$};
		
		\draw[fill] (6,5) circle(.1);
		\draw[fill] (6,6) circle(.1);
		\draw (6,5) -- (6,6);
		\draw[fill] (6,7) circle(.1);
		\draw (6,5) to [out=150,in=150] (6,7);
		\draw[fill] (6,8) circle(.1);
		\draw[fill] (7,8) circle(.1);
		\draw (6,7) -- (6,8);
		\draw (6,6) to [out=30,in=150] (7,8);
		\draw[fill] (7,9) circle(.1);
		\draw (7,8) -- (7,9);
		\draw (6,7) to [out=30,in=150] (7,9);
		\draw[fill] (7,10) circle(.1);
		\draw (6,8) to [out=30,in=150] (7,10);
		\draw[fill] (7,11) circle(.1);
		\draw (7,10) -- (7,11);
		\draw (7,9) to [out=30,in=30] (7,11);

		\end{tikzpicture}
	\end{center}
	\caption{The $\A_2(1)$-module structure of $\H^{*+3}(M\rO(3),\Z_2)$ below degree 5.}
	\label{fig:A_2(1)MO3}
\end{figure}

\begin{figure}
	\begin{center}
		\begin{tikzpicture}
		\node at (0,-1) {0};
		\node at (1,-1) {1};
		\node at (2,-1) {2};
		\node at (3,-1) {3};
		\node at (4,-1) {4};
		\node at (5,-1) {5};
		\node at (6,-1) {$t-s$};
		\node at (-1,0) {0};
		\node at (-1,1) {1};
		\node at (-1,2) {2};
		\node at (-1,3) {3};
		\node at (-1,4) {4};
		\node at (-1,5) {5};
		\node at (-1,6) {$s$};
		
		\draw[->] (-0.5,-0.5) -- (-0.5,6);
		\draw[->] (-0.5,-0.5) -- (6,-0.5);
		
		\draw[fill] (0,0) circle(0.05);
		\draw[fill] (2,0) circle(0.05);
		\draw (4,0) -- (4,1);
		\draw[fill] (4.1,0) circle(0.05);
		\draw[fill] (5,0) circle(0.05);

		\end{tikzpicture}
	\end{center}
	\caption{$E_2$ page of the Adams spectral sequence with symmetry $(\Pin^+(d) \times \SU(2))/\Z_2$. The Bordism group $\Omega_*^{({\Pin^+ \times \SU(2)})/{\Z_2}}$ and the invariants can be read off from the this chart. }
	\label{fig:E_2Pin+SU2Z2}
\end{figure}

\begin{table}[t]
	\centering
	\begin{tabular}{ c c c c }
		\hline
		\multicolumn{4}{c}{Bordism and Cobordism group}\\
		\hline
		$d$ & 
		$\Omega^{(\Pin^+\times \SU(2))/\Z_2}_d$ &
		$\TP^{(\Pin^+\times \SU(2))/\Z_2}_d$ &
		Cobordism Invariant \\
		\hline
		1& $0$& $0$& \\
		2& $\Z_2$ & $\Z_2$&  $w_1^2$\\
		3 & $0$ & $0$ & \\
		4 & $\Z_2\times \Z_4$& $\Z_2\times \Z_4$ & $w_2^2, \eta_{SU(2)}$\\
		5 & $\Z_2$ & $\Z_2$&  $w_2w_3$\\
		\hline
	\end{tabular}
	\caption{
		The bordism and cobordism groups of the symmetry  $(\Pin^+(d)\times \SU(2))/\Z_2$. 
	}
	\label{table:Pin+SU2Bordism}
\end{table}

\subsubsection{(Co)bordism Groups and Invariants of $(\Pin^-(d)\times \SU(2))/\Z_2$}

Let $G'=(\Pin^-(d)\times \SU(2))/\Z_2$, 
then by \cite{Freed2016,1711.11587GPW}, we have
$MTG'=M\Spin\wedge\Sigma^{3}MT\rO(3)$. Here $w_2(TM)+w_1(TM)^2=w_2'(V_{\SO(3)})$, $w_3(TM)+w_2(TM)w_1(TM)=w_3'(V_{\SO(3)})$ and $w_1(TM)$ is nontrivial, $w_1'(V_{\SO(3)})=0$. For $t-s<8$, since there is no odd torsion, we have the Adams spectral sequence
\bea
\Ext_{\A_2(1)}^{s,t}(\H^{*-3}(MT\rO(3),\Z_2),\Z_2)\Rightarrow\Omega_{t-s}^{(\Pin^-(d)\times \SU(2))/\Z_2}.
\eea
The $\A_2(1)$-module structure of $\H^{*-3}(MT\rO(3),\Z_2)$ below degree 5 and the $E_2$ page are shown in Figure \ref{fig:A_2(1)MTO3}, \ref{fig:E_2Pin-SU2Z2}. One can then read off the (co)bordism invariants from the $E_2$ page, as shown in Table \ref{table:Pin-SU2Bordism}. The physical interpretation of cobordism invariants will be discussed in section \ref{secPhyInt}.

\begin{figure}
	\begin{center}
		\begin{tikzpicture}[scale=0.5]
		
		\node[below] at (0,0) {$U$};
		
		\draw[fill] (0,0) circle(.1);
		\draw[fill] (0,1) circle(.1);
		\draw (0,0) -- (0,1);
		\draw[fill] (0,2) circle(.1);
		\draw (0,0) to [out=150,in=150] (0,2);
		\draw[fill] (0,3) circle(.1);
		\draw[fill] (1,3) circle(.1);
		\draw (0,2) -- (0,3);
		\draw (0,1) to [out=30,in=150] (1,3);
		\draw[fill] (1,4) circle(.1);
		\draw (1,3) -- (1,4);
		\draw (0,2) to [out=30,in=150] (1,4);
		\draw[fill] (1,5) circle(.1);
		\draw[fill] (1,6) circle(.1);
		\draw (0,3) to [out=30,in=150] (1,5);
		\draw (1,5) -- (1,6);
		\draw (1,4) to [out=30,in=30] (1,6);
		
		\node[below] at (2,2) {$w_2U$};
		
		\draw[fill] (2,2) circle(.1);
		\draw[fill] (2,3) circle(.1);
		\draw (2,2) -- (2,3);
		\draw[fill] (2,4) circle(.1);
		\draw (2,2) to [out=150,in=150] (2,4);
		\draw[fill] (2,5) circle(.1);
		\draw (2,4) -- (2,5);
		\draw[fill] (2,6) circle(.1);
		\draw[fill] (2,7) circle(.1);
		\draw (2,5) to [out=150,in=150] (2,7);
		\draw (2,6) -- (2,7);
		\draw[fill] (2,8) circle(.1);
		\draw (2,6) to [out=30,in=30] (2,8);
		\draw[fill] (2,9) circle(.1);
		\draw (2,8) -- (2,9);
		
		\node[below] at (4,4) {$w_2^2U$};
		
		\draw[fill] (4,4) circle(.1);
		\draw[fill] (4,5) circle(.1);
		\draw (4,4) -- (4,5);
		\draw[fill] (4,6) circle(.1);
		\draw (4,4) to [out=150,in=150] (4,6);
		\draw[fill] (4,7) circle(.1);
		\draw[fill] (5,7) circle(.1);
		\draw (4,6) -- (4,7);
		\draw (4,5) to [out=30,in=150] (5,7);
		\draw[fill] (5,8) circle(.1);
		\draw (5,7)-- (5,8);
		\draw (4,6) to [out=30,in=150] (5,8);
		\draw[fill] (5,9) circle(.1);
		\draw[fill] (5,10) circle(.1);
		\draw (4,7) to [out=30,in=150] (5,9);
		\draw (5,9) -- (5,10);
		\draw (5,8) to [out=30,in=30] (5,10);

		\node[below] at (6,4) {$w_1^4U$};
		
		\draw[fill] (6,4) circle(.1);
		\draw[fill] (6,5) circle(.1);
		\draw (6,4) -- (6,5);
		\draw[fill] (6,6) circle(.1);
		\draw (6,4) to [out=150,in=150] (6,6);
		\draw[fill] (6,7) circle(.1);
		\draw[fill] (7,7) circle(.1);
		\draw (6,6) -- (6,7);
		\draw (6,5) to [out=30,in=150] (7,7);
		\draw[fill] (7,8) circle(.1);
		\draw (7,7)-- (7,8);
		\draw (6,6) to [out=30,in=150] (7,8);
		\draw[fill] (7,9) circle(.1);
		\draw[fill] (7,10) circle(.1);
		\draw (6,7) to [out=30,in=150] (7,9);
		\draw (7,9) -- (7,10);
		\draw (7,8) to [out=30,in=30] (7,10);
		
		\node[below] at (8,6) {$w_2^3U$};
		
		\draw[fill] (8,6) circle(.1);
		\draw[fill] (8,7) circle(.1);
		\draw (8,6) -- (8,7);
		\draw[fill] (8,8) circle(.1);
		\draw (8,6) to [out=150,in=150] (8,8);
		\draw[fill] (8,9) circle(.1);
		\draw (8,8) -- (8,9);
		\draw[fill] (8,10) circle(.1);
		\draw[fill] (8,11) circle(.1);
		\draw (8,9) to [out=150,in=150] (8,11);
		\draw (8,10) -- (8,11);
		\draw[fill] (8,12) circle(.1);
		\draw (8,10) to [out=30,in=30] (8,12);
		\draw[fill] (8,13) circle(.1);
		\draw (8,12) -- (8,13);
		
		\node[below] at (9,5) {$w_2w_3U$};
		
		\draw[fill] (9,5) circle(.1);
		\draw[fill] (9,6) circle(.1);
		\draw (9,5) -- (9,6);
		\draw (9,5) to [out=150,in=30] (8,7);
		\draw[fill] (9,8) circle(.1);
		\draw[fill] (9,9) circle(.1);
		\draw (9,6) to [out=30,in=30] (9,8);
		\draw (9,8) -- (9,9);
		\draw (8,7) to [out=30,in=150] (9,9);
		
		%

		\end{tikzpicture}
	\end{center}
	\caption{The $\A_2(1)$-module structure of $\H^{*-3}(MT\rO(3),\Z_2)$ below degree 5.}
	\label{fig:A_2(1)MTO3}
\end{figure}

\begin{figure}
	\begin{center}
		\begin{tikzpicture}
		\node at (0,-1) {0};
		\node at (1,-1) {1};
		\node at (2,-1) {2};
		\node at (3,-1) {3};
		\node at (4,-1) {4};
		\node at (5,-1) {5};
		\node at (6,-1) {$t-s$};
		\node at (-1,0) {0};
		\node at (-1,1) {1};
		\node at (-1,2) {2};
		\node at (-1,3) {3};
		\node at (-1,4) {4};
		\node at (-1,5) {5};
		\node at (-1,6) {$s$};
		
		\draw[->] (-0.5,-0.5) -- (-0.5,6);
		\draw[->] (-0.5,-0.5) -- (6,-0.5);
		
		\draw[fill] (0,0) circle(0.05);
		\draw[fill] (2,0) circle(0.05);
		
		\draw[fill] (3.9,0) circle(0.05);
		\draw[fill] (4.1,0) circle(0.05);
		
		\draw (4,1) -- (5,2);
		
		\draw[fill] (5,0) circle(0.05);
		
		
		\end{tikzpicture}
	\end{center}
	\caption{$E_2$ page of the Adams spectral sequence with symmetry $(\Pin^-(d) \times \SU(2))/\Z_2$. The Bordism group $\Omega_*^{({\Pin^- \times \SU(2)})/{\Z_2}}$ and the invariants can be read off from the this chart.}
	\label{fig:E_2Pin-SU2Z2}
\end{figure}

\begin{table}[t]
	\centering
	\begin{tabular}{ c c c c }
		\hline
		\multicolumn{4}{c}{Bordism and Cobordism group}\\
		\hline
		$d$ & 
		$\Omega^{(\Pin^-\times \SU(2))/\Z_2}_d$ &
		$\TP^{(\Pin^-\times \SU(2))/\Z_2}_d$ &
		Cobordism Invariant \\
		\hline
		1& $0$& $0$& \\
		2& $\Z_2$ & $\Z_2$&  $w_1^2$\\
		3 & $0$ & $0$ & \\
		4 & $\Z_2\times \Z_2\times \Z_2$& $\Z_2\times \Z_2\times \Z_2$ & $w_1^2, w_2^2, N_0$\\
		5 & $\Z_2\times \Z_2$ & $\Z_2\times \Z_2$&  $w_2w_3, \widehat{\mathcal{I}}_{1/2}$\\
		\hline
	\end{tabular}
	\caption{
		The bordism and cobordism groups of the symmetry  $(\Pin^-(d)\times \SU(2))/\Z_2$. 
	}
	\label{table:Pin-SU2Bordism}
\end{table}

\subsubsection{Physical Interpretations of the Cobordism Invariants and Classification of Anomalies}
\label{secPhyInt}

The bordism and cobordism groups and  the cobordism invariants are enumerated  in Table \ref{table:OSU2Bordism}, \ref{table:ESU2Bordism}, \ref{table:Pin+SU2Bordism} and \ref{table:Pin-SU2Bordism}. We discuss the physical interpretations of the cobordism invariants in these tables.  We will not discuss the anomalies $w_1^2, w_1^4, w_2^2$ and $w_2w_3$ which have already been discussed in section \ref{subsecO}.

\subsubsection*{Nontrivial Anomalies from $\TP^{\O\times \SU(2)}_d$}

Modulo the cobordism invariants from $\TP^{\O}_d$, the only nontrivial invariant in Table \ref{table:OSU2Bordism} is $c_2^{\SU(2)}\mod 2$.  The physical interpretation of this cobordism invariant as an anomaly is as follows. Suppose $A$ is the $\SU(2)$ background gauge field, and the partition function of an SU(2) invariant system in $2+1$d is $Z[A]$. Under time reversal, the partition function transforms as $Z[A]\to Z[A]\exp(\ii CS_3^{\SU(2)}[A])$. This means that there is a mixed anomaly between time reversal and $\SU(2)$ symmetry. This mixed anomaly can be canceled by attaching a $3+1$d cobordism invariant, so that the total partition function 
\begin{eqnarray}
\exp\left(\ii  w_1 \cup  CS_3^{\SU(2)}[A]\right) =  \exp\left(\ii \frac{\delta}{2} \mathrm{CS}_3^{\SU(2)}[A]\right) = \exp\left(\ii \pi  c_2^{\SU(2)}[A]\right).
\end{eqnarray}
Hence $Z[A]\exp(\ii \pi  c_2^{\SU(2)}[A])$ is time reversal invariant.

\subsubsection*{Nontrivial Anomalies from $\TP^{(\rE\times \SU(2))/\Z_2}_d$}

The cobordism invariants of the total symmetry $(\rE(d)\times \SU(2))/\Z_2$ in Table \ref{table:ESU2Bordism} are the same as the cobordism invariants of the symmetry $\O(d)\times \SU(2)$. This should not be a surprise, because all the invariants already appeared in $\TP^{\O}_{d}$ should persist, and the remaining anomaly $c_2^{\SU(2)}$ mod 2 is linear in the time reversal background field hence it should not depend on the Kramers parity.

\subsubsection*{Nontrivial Anomalies from $\TP^{(\Pin^+\times \SU(2))/\Z_2}_d$}

Comparing with the cobordism invariants in $\TP^O_d$, the only additional invariant $\eta_{SU(2)}$ in $\TP^{\Pin^+\times \SU(2)}_d$ is the eta invariant of the Dirac operator coupled to $\Pin^+$-SU(2) background gauge field $A$. It can be obtained via taking the large negative mass limit of the free $\SU(2)$ invariant Dirac fermion coupled to $A$. The partition function is \cite{1711.11587GPW, Metlitski:2015yqa}
\begin{eqnarray}\label{etaSU2}
\exp\left( 2\pi \ii \nu \eta_{\SU(2)}\right).
\end{eqnarray}
See \cite{1711.11587GPW} for a systematic discussion of $\eta_{\SU(2)}$.  On an unorientable manifold, there is a relation $4\eta_{\SU(2)}= w_1^4 \mod 2$. 
This means  
\begin{enumerate}
	\item $\nu \sim \nu+4$, hence this cobordism invariant generates a $\Z_4$ group, as shown in Table \ref{table:Pin+SU2Bordism}. The four classes are labeled by $\nu=0,1,2,3$ mod 4. 
	\item When $\nu=2\mod 4$, using the above relation, the invariant is \eqref{etaSU2} reduces to $\exp\left( 4\pi \ii  \eta_{\SU(2)}\right)= \exp\left(\ii \pi w_1^4\right)$ \cite{1711.11587GPW} which is the cobordism invariant in $\TP^{\O}_d$. Hence although $w_1^4$ does not appear explicitly in Table \ref{table:Pin+SU2Bordism}, it is implicitly encoded in the invariant $\eta_{\SU(2)}$. This is consistent with the fact that once $\SU(2)$ is explicitly broken but time reversal is still preserved, the invariants in $\TP^{\Pin^+\times \SU(2)}_d$ should reduce to the invariants in $\TP^O_d$. 
	\item The  invariant that does not appear in $\TP^{\O}_d$ correspond to $\nu=1,3$ mod 4. Since the $\nu=1$ mod 4  and $\nu=3$ mod 4 differ by $\mu=2$ mod 4 which belongs to $\TP^{\O}_d$, it suffices to discuss $\nu=1$ mod 4 only. As we mentioned above, the quantum field theory realizing such a SPT is $\nu$ free 4-component Dirac fermions coupled to $\Pin^+$-SU(2) connection, all with a large negative mass. Thus the boundary theory of this SPT phase is $\nu$ massless free 2-component Dirac fermions coupled to $\Pin^+$-SU(2) connection. The $\nu=1$ anomaly even persists when the $\Pin^+$-SU(2) connection is replaced by the $\SU(2)$ connection. After such replacement, the original $\nu=1$ anomaly reduces to $c_2^{\SU(2)}\mod 2$, which is the mixed anomaly between time reversal and the $\SU(2)$ symmetry. 
	\item The $\Z_4$ class from the free Dirac fermion coupled to $\Pin^+$-$\SU(2)$ connection (CI symmetry class) is intimately related to the $\Z_8$ class from the free Dirac fermion coupled to $\Pin^+$-U(1) connection (AIII symmetry class). The two cases are related by restricting the $\SU(2)$ to its $\U(1)$ subgroup. It is also related to the $\Z_{16}$ class from the free Majorana fermion in the $\Pin^+$ class (DIII symmetry class). \cite{1711.11587GPW}
\end{enumerate}

\subsubsection*{Nontrivial Anomalies from $\TP^{(\Pin^-\times \SU(2))/\Z_2}_d$}

Comparing with the cobordism invariants in $\TP^{\O}_d$, the only additional invariants are $N_0$ and $\widehat{\mathcal{I}}_{1/2}$.  $N_0$ is the eta invariant which counts the number of zero modes of the Dirac fermion coupled to $\Pin^-$-SU(2) connection. \cite{1711.11587GPW, Metlitski:2015yqa}
\begin{eqnarray}
Z_\nu[A] = \lim_{m\to -\infty, M\to \infty}\left(\frac{\det(\slashed D_A +m)}{\det(\slashed D_A +M)}\right)^\nu = (-1)^{\nu N_0},
\end{eqnarray}
where $M$ mass in the denominator is the Pauli-Villas regularization. Time reversal requires that the fermion mass $m$ is real. When $m$ is positive and large, the theory is in the trivially gapped, and when $m$ is negative and large, the theory is the nontrivial SPT. 
The number of zero modes is topological only mod 2. \cite{1711.11587GPW} When time reversal is explicitly broken so that the symmetry class reduces to $(\Spin(d)\times \SU(2))/\Z_2$, the fermion mass no longer has to be real, and one can turn on complex mass $m$ of the fermion to connect $m\in \mathbb{R}_+$ and $m\in \mathbb{R}_-$ without encountering the massless point. This is consistent with $\TP_4^{(\Spin(d)\times \SU(2))/\Z_2}$ being trivial. 

$\widehat{\mathcal{I}}_{1/2}$ in $d=5$ is the same invariant in the symmetry class $(\Spin(d)\times \SU(2))/\Z_2$. In fact, the existence of $\widehat{\mathcal{I}}_{1/2}$ is independent of whether time reversal exists. One way to see this is to notice that one can not write down a mass term of the left handed SU(2) invariant Weyl fermion in $3+1$d, hence forbidding SU(2) invariant mass term (which drives the massless fermion to trivially gapped phase) does not require time reversal symmetry.  Therefore $\widehat{\mathcal{I}}_{1/2}$ in the two symmetry classes $(\Spin(d)\times \SU(2))/\Z_2$ and $(\Pin^-(d)\times \SU(2))/\Z_2$ are identical. 

As remarked in section \ref{subsecSpinSU2},  $\widehat{\mathcal{I}}_{1/2}$ can be expressed in terms of a twisted version of Stiefel-Whitney class $w_3'$ and the $\arf$ invariant. Similarly $N_0$ can be written in terms of a twisted version of Stiefel-Whitney class $w_3'$ and the $\tilde\eta$ invariant.  A more precise relation will be discussed in \cite{JuvenWIP}.

\subsection{Brief Comments on $\LG= \Pin^\pm(d)$}

When $\LG= \Pin^\pm(d)$, the spacetime manifold is an unorientable manifold which is equipped with a $\Pin^\pm$ structure. We will denote such manifold as the $\Pin^\pm$ manifold respectively.  In particular, $w_1$ is unconstrained while $w_2$ is trivial for $\Pin^+$ or $w_1^2+w_2$ is trivial for $\Pin^-$. A quantum field theory that can be formulated on the most generic $\Pin^\pm$ manifold must be time reversal symmetric and allows a fermion in the local operator spectrum. The fermion transforms under time reversal symmetry as a Kramers doublet: $\T^2=(-1)^F$ where $F$ measures the fermion number for $\Pin^+$, or as a Kramers singlet: $\T^2=1$ for $\Pin^-$.

\subsubsection{Lorentz Symmetry Extensions for $\LG=\Pin^+(d)$:\\
$\SU(2)\times \Pin^+(d)$ and $(\SU(2)\times \EPin(d))/\Z_2^+$}

Demanding $\LG= \Pin^+(d)$ in the exact sequence \eqref{groupext}, we obtain 
\begin{eqnarray}
1\to \SU(2)\to G' \to \Pin^+(d) \to 1.
\end{eqnarray}
For $\Pin^+(d)$, $w_1, w_1^2$ are both unconstrained, but $w_2=0$. Hence we can identify $w_2(V_{\SO(3)})$ of the SO(3) vector bundle with $0$ or $w_1^2$. This means that there are two choices of extensions, 
\begin{eqnarray}\label{EPin+}
G'= 
\begin{cases}
\SU(2)\times \Pin^+(d), &~~~  w_2(V_{\SO(3)})=0,\\
(\SU(2)\times \EPin(d))/\Z_2^+, &~~~ w_2(V_{\SO(3)})= w_1^2.
\end{cases}
\end{eqnarray}
Here the group $\EPin(d)$ is a double cover of the group $\Pin^+(d)$. Recall that as we discussed in section \ref{secLorentz}, there are  two $\Z_2$ normal subgroups in $\EPin(d)$, i.e., $\Z_2^+\times \Z_2^-$. In the second line in \eqref{EPin+} which corresponds to the nontrivial extension, one is identifying one of the $\Z_2$ subgroup, $\Z_2^+$, in $\EPin(d)$ with the $\Z_2$ subgroup of $\SU(2)$. 


For $G'=\SU(2)\times \Pin^+(d)$, which corresponds to trivial extension, in the local operator spectrum, a fermion should transform as a Kramers doublet under time reversal, and a boson should transform as a Kramers singlet under time reversal. Both the fermion and a boson can carry arbitrary SU(2) isospin $j$. Formally, there are no bundle constraints between the $\SU(2)$ vector bundle and the tangent bundle of the spacetime manifold.

For $G'=(\SU(2)\times \EPin(d))/\Z_2^+$, it corresponds to a nontrivial extension. In the operator spectrum,  there are two transparent fermions $\psi_+$ and $\psi_-$, being Kramers singlet and doublet respectively. The fermion $\psi_+$, being a Kramers singlet,  should also carry $\SU(2)$ isospin $j=\Z+1/2$. However, the other fermion $\psi_-$, being a Kramers doublet, can carry any $\SU(2)$ isospin. Formally, the $\SO(3)$ vector bundle and the tangent bundle of the spacetime manifold are correlated: $w_2(V_{\SO(3)})= w_1^2$ and $w_2=0$. 


\subsubsection{Lorentz Symmetry Extensions for $\LG=\Pin^-(d)$:\\
$\SU(2)\times \Pin^-(d)$ and $(\SU(2)\times \EPin(d))/\Z_2^-$}

Demanding $\LG= \Pin^-(d)$ in the exact sequence \eqref{groupext}, we obtain 
\begin{eqnarray}
1\to \SU(2)\to G' \to \Pin^-(d) \to 1.
\end{eqnarray}
For $\Pin^-(d)$, $w_1$ is unconstrained, but $w_1^2+w_2=0$. Hence we can identify $w_2(V_{\SO(3)})$ of the SO(3) vector bundle with $0$ or $w_1^2$ which is also $w_2$. This means that there are two choices of extensions, 
\begin{eqnarray}\label{EPin-}
G'= 
\begin{cases}
\SU(2)\times \Pin^-(d), &~~~  w_2(V_{\SO(3)}) =0,\\
(\SU(2)\times \EPin(d))/\Z_2^-, &~~~ w_2(V_{\SO(3)})= w_1^2=w_2.
\end{cases}
\end{eqnarray}
Here the group $\EPin(d)$ is a double cover of the group $\Pin^-(d)$, which is the same group that appear  in \eqref{EPin+}. However, notice that here we use a different $\Z_2$ normal subgroup of $\EPin(d)$, i.e. $\Z_2^-$,  to identify with the $\Z_2$ in $\SU(2)$.

For $G'=\SU(2)\times \Pin^-(d)$, which corresponds to trivial extension, in the local operator spectrum, both boson and fermion should transform as a Kramers singlet under time reversal. Both the fermion and a boson can carry arbitrary SU(2) isospin $j$. Formally, there are no bundle constraints between the $\SU(2)$ vector bundle and the tangent bundle of the spacetime manifold. 

For $G'=(\SU(2)\times \EPin(d))/\Z_2^-$, it corresponds to a nontrivial extension. In the operator spectrum,  there are two transparent fermions $\psi_+$ and $\psi_-$, being Kramers singlet and doublet respectively. The fermion $\psi_-$, being a Kramers doublet,  should also carry $\SU(2)$ isospin $j=\Z+1/2$. However, the other fermion $\psi_+$, being a Kramers singlet, can carry any $\SU(2)$ isospin. Formally, the $\SO(3)$ vector bundle and the tangent bundle of the spacetime manifold are correlated: $w_2(V_{\SO(3)})= w_1^2$ and $w_2=w_1^2$.

We will leave  the calculation of (co)bordism invariants and their physical interpretations in a separate paper.

\section{Promoting $\SU(2)$ To Dynamical Gauge Theories 
}
\label{GaugingSU2}

In this section, we try to promoting the $\SU(2)$ internal global symmetry to the dynamical gauge group. We will only consider the $\SU(2)$ gauge theory with the action 
\begin{eqnarray}
S= -\frac{1}{4g^2 } \int_{M} \Tr\left(f \wedge \star f\right) + S_{\text{SPT}},
\end{eqnarray}
where the topological term $S_{\text{SPT}}$ is the cobordism invariants computed in section \ref{secSU2}. 
The topological term can  be either the discrete theta term of the $\SU(2)$ gauge field, or the term with both the $\SU(2)$ gauge field and the spacetime background fields, or the counter term involving on the background field of the spacetime. Moreover, because we only consider the pure gauge theory, there is an emergent $\Z_{2,[1]}$ 1-form center symmetry in the resulting gauge theory. Denote the 2-form background gauge field of the $\Z_{2,[1]}$ 1-form symmetry as $B$. If before gauging $\SU(2)$ the  bundle constraint between $\SU(2)$ and the Lorentz symmetry is
 $w_2(V_{\SO(3)}) = K_1 w_1^2+ K_2 w_2$,  then after coupling to the background field $B$, the gauge bundle constraint is modified to\footnote{This is the sensible modification because taking $w_i=0$ or $B=0$ separately produces the correct gauge bundle constraint in different limits. When taking $B=0$, it reduces $w_2(V_{\SO(3)}) = K_1 w_1^2+ K_2 w_2$ as expected. On the other hand, when taking $w_i=0$, it reduces to $w_2(V_{\SO(3)})=B$, which is expected because the one form symmetry should be irrelevant to whether the spacetime topology is nontrivial or not. }
\begin{eqnarray}
w_2(V_{\SO(3)}) = B+ K_1 w_1^2 + K_2 w_2.
\end{eqnarray}
There are no constraints $B$ and Lorentz background fields, hence after dynamical gauging $\SU(2)$, the global symmetry is $H = \Z_{2,[1]}\times \LG$.  
As we will see, different choices of $(K_1, K_2)$ can lead to different anomalies.


As a preliminary, we comment on the $\SU(2)$ Yang-Mills theory without any topological terms. Because the Yang-Mills action is a functional of the $\SU(2)$ field strength, and the field strength has two antisymmetric indices, the Yang-Mills theory can only be defined in $1+1$ or higher dimensions.  Throughout this section, we will thus not discuss the $0+1$d systems. In $1+1$d, $\SU(2)$ Yang-Mills is exactly solvable \cite{witten1991}, and it can be shown that there is only one ground state on the spatial manifold $\mathbb{R}$. In higher dimensions, Yang-Mills theory in the infrared is strongly coupled, and it is commonly believed that the theory should also have a trivially gapped ground state. Therefore, nontrivial anomalies involving one form symmetry requires nontrivial topological terms. Hence, in the following sections, we will focus on the cases with nontrivial cobordism groups, and see if nontrivial anomaly for the emergent one form symmetry arises.

\subsection{$\LG= \SO(d)$}
\label{subsecSOd}

When the Lorentz symmetry is $\SO(d)$, there are two possibilities of the total symmetries, given in \eqref{gbcSO}. When promoting the $\SU(2)$ global symmetry to dynamical and including the $\Z_{2,[1]}$ one form symmetry background $B_2$, the resulting global symmetries and the gauge bundle constraints for the resulting dynamical $\SU(2)$ gauge theory are
\begin{eqnarray}\label{gbcSOB}
G'= 
\begin{cases}
\SU(2)\times \SO(d)\\
({\SU(2)\times \Spin(d)})/{\Z_2}
\end{cases} \Rightarrow
H= 
\begin{cases}
\Z_{2,[1]}\times \SO(d), &~~~ w_2(V_{\SO(3)})=B_2,\\
\Z_{2,[1]}\times \SO(d), &~~~ w_2(V_{\SO(3)})=B_2+w_2.
\end{cases}
\end{eqnarray}
Here $w_2(V_{\SO(3)})$ is the second Stiefel-Whitney class of the dynamical gauge bundle, which should be distinguished from that appears in \eqref{gbcSO}.

We comment on the physical meanings in these two cases. In both cases, the Lorentz symmetry is $\SO(d)$, which means in both cases the theories should be able to be defined on a most generic non-spin manifold. In particular, in the second case, after gauging, the $\SO(d)$ is no longer lifted to $\Spin(d)$ to identify its center with that of the internal symmetry. This means that in the operator spectrum, there is no transparent fermion. However, after gauging, there is a gauge invariant non-transparent fermion line operator (which does not commute with all other gauge invariant operators). This can be seen by noticing that the Wilson line in the fundamental representation of $\SU(2)$, $W_{1/2}$,  bounds a disk which supports a two dimensional SPT $B+w_2$, and the boundary of the 2d SPT $w_2$ is the world line of a fermion. Furthermore, $W_{1/2}$ carries charge 1 under $\Z_{2,[1]}$.

\subsubsection{(Co)bordism Groups and Invariants of $\SO(d)\times \Z_{2, [1]}$}

We further discuss the anomaly of the emergent symmetry $H$. Because the emergent symmetry is $H= \Z_{2,[1]}\times \SO(d)$, one can compute the bordism and cobordism group and enumerate all possible cobordism invariants as shown in Table \ref{table:Z2SOBordism}. See  \cite{Wan2018bns1812.11967} for the derivation of the (co)bordism calculations. 
We comment on which anomaly can be saturated by gauging the SPTs in Table \ref{table:SOSU2Bordism} and \ref{table:SpinSU2Bordism}.  

\begin{table}[t]
	\centering
	\begin{tabular}{ c c c c }
		\hline
		\multicolumn{4}{c}{Bordism and Cobordism group}\\
		\hline
		$d$ & 
		$\Omega^{\SO\times  \Z_{2,[1]}}_d$ &
		$\TP^{\SO\times  \Z_{2,[1]}}_d$ &
		Cobordism Invariant \\
		\hline
		1& $0$& $0$& \\
		2& $\Z_2$ & $\Z_2$&  $B_2$\\
		3 & $0$ & $\Z$ & $16\CSg$\\
		4 & $\Z\times \Z_{4}$& $\Z_4$ & $\mathcal{P}(B_2)$\\
		5 & $\Z_2\times \Z_2$ & $\Z_2\times \Z_2$& $w_2 w_3, B_2\Sq^1 B_2$ \\
		\hline
	\end{tabular}
	\caption{
		The bordism and cobordism groups of the symmetry  $\SO(d)\times \Z_{2,[1]}$. 
	}
	\label{table:Z2SOBordism}
\end{table}

The $d$ dimensional cobordism invariant in Table \ref{table:Z2SOBordism} can be regarded as the 't Hooft anomaly of $(d-1)$ dimensional gauge theory obtained by gauging $\SU(2)$ in Table \ref{table:SOSU2Bordism} and \ref{table:SpinSU2Bordism}. For the nontrivial cobordism invariants in Table \ref{table:SOSU2Bordism} and \ref{table:SpinSU2Bordism}, we will only focus on those that involve $\SU(2)$ gauge bundle, and also only focus on those in two and three and four dimensions, which are physically relevant.

\subsubsection{Gauging $\SU(2)$ in $\SU(2)\times \SO(d)$}

The only cobordism that involves the $\SU(2)$ bundle  in Table \ref{table:SOSU2Bordism} is $\mathrm{CS}_3^{\SU(2)}$. This means after gauging $\SU(2)$, the theory is the Chern-Simons-Yang-Mills theory in $2+1$d, with the action
\begin{eqnarray}\label{SU2CS}
S= -\frac{1}{4g^2} \int_{M_3} \Tr(f\wedge \star f) + \frac{k}{4\pi}\int \Tr\left(a\dd a - \frac{2\ii}{3}a^3\right),
\end{eqnarray}
where $k\in \Z$ labels which $\SU(2)$ SPT from which we gauge. There are two related ways to see the anomaly of $\Z_{2,[1]}$. One way is to check the topological spin of the $\Z_{2,[1]}$ symmetry generator. The second way to the directly couple the $\Z_{2,[1]}$ background field $B$ to the action. Both of which conclude that the anomaly inflow action for the $\Z_{2,[1]}$ symmetry is given by the invertible TQFT
\begin{eqnarray}\label{SU2CSanomaly}
2\pi \frac{k}{4}\int_{M_4} \mathcal{P}(B_2) , ~~k\in \Z_4.
\end{eqnarray}
Notice that the anomaly only depends on $k\mod 4$. Thus one sees that after gauging $\Z$ classified $\SU(2)$ SPT in $2+1$d, we arrive at a $\Z_4$ classified anomalous $\SU(2)$ pure gauge theory.

\subsubsection{Gauging $\SU(2)$ in $(\SU(2)\times \Spin(d))/\Z_2$}
\label{secsu2gauge}

The only SPT in $d=2,3,4$ that involves the $\SU(2)$ bundle is the $\widehat{\mathrm{CS}}_3^{\SU(2)}$ Chern Simons theory. This theory is almost identical to \eqref{SU2CS}, except that the dynamical $\SU(2)$ gauge field in \eqref{SU2CS} is replaced by the Spin-SU(2) connection, and suitable gravitational Chern SImons term should also be included which will be determined below.\footnote{This is similar to the $\U(1)$ case. If $A$ is a $\U(1)$ gauge field, the theory $\frac{k}{4\pi} A\dd A$ is well defined (for even $k$) on a non-spin manifold. If $A$ is a $\Spin_c$ connection, one needs to append suitable gravitational Chern Simons term $\frac{k}{4\pi} A\dd A+ 2k CS_g$. } The anomaly can be obtained directly by replacing $B_2$ in \eqref{SU2CSanomaly} with $B+ w_2$. The anomaly is
\begin{eqnarray}\label{PB2w2}
2\pi \frac{k}{4}\int_{M_4} \mathcal{P}(B_2+ w_2) = 2\pi \frac{k}{4}\int_{M_4} \mathcal{P}(B_2) + \pi k \int_{M_4}w_2 B_2 + 2\pi \frac{k}{4}\int_{M_4} \mathcal{P}(w_2).
\end{eqnarray}
On the right hand side, the second term can also be written as $\pi k \int_{M_4} \mathcal{P}(B_2)$, by using $\mathcal{P}(B_2) = w_2 B_2$ on an orientable manifold. The last term can be written as 
\begin{eqnarray}\label{Pw2}
2\pi \frac{k}{4}\int_{M_4} \mathcal{P}(w_2)= -2\pi \frac{k}{4}\int_{M_4} p_1= -2\pi \frac{k}{4} \int 8\widehat{A} =  -2\pi \frac{k}{4} \sigma.
\end{eqnarray}
Because on nonspin manifold, $\sigma\in \Z$, \eqref{Pw2} only vanishes for $k\in 4\Z$. This means certain gravitational Chern Simons term $4k \CSg$ should be added to cancel this contribution. Combining \eqref{PB2w2} and \eqref{Pw2}, we find the anomaly
\begin{eqnarray}
2\pi \frac{3k}{4}\int_{M_4} \mathcal{P}(B_2) = -2\pi \frac{k}{4}\int_{M_4} \mathcal{P}(B_2). 
\end{eqnarray}
Furthermore, since there is no nontrivial $\SU(2)$ SPT in 4d,  the 5d cobordism invariant $B_2 \Sq^1 B_2$ can not be saturated by a theory obtained via gauging an $\SU(2)$ SPT in 4d.

\subsection{$\LG= \Spin(d)$}

When the Lorentz symmetry is $\Spin(d)$, there is only one possible total symmetry, given in \eqref{Spintotalsym}. When promoting the $\SU(2)$ global symmetry to dynamical and including the $\Z_{2,[1]}$ one form symmetry background $B_2$, the resulting global symmetries and the gauge bundle constraint for the resulting dynamical $\SU(2)$ gauge theory is 
\begin{eqnarray}
G'= \SU(2)\times \Spin(d) ~~\Rightarrow~~ H= \Z_{2,[1]} \times \Spin(d), ~~~ w_2(V_{\SO(3)}) = B_2.
\end{eqnarray}
After gauging, the resulting $\SU(2)$ gauge theory is still a fermionic theory which depends on the choice of spin structure. This means that there is a transparent fermion line which commutes with every other gauge invariant operators. 

We further discuss the anomaly of the emergent global symmetry $H= \Z_{2,[1]} \times \Spin(d)$. One can compute the bordism and cobordism group and enumerate all possible cobordism invariants, as shown in Table \ref{table:Z2SpinBordism}. See  \cite{Wan2018bns1812.11967} for the derivation of the (co)bordism calculations. 

\begin{table}[t]
	\centering
	\begin{tabular}{ c c c c }
		\hline
		\multicolumn{4}{c}{Bordism and Cobordism group}\\
		\hline
		$d$ & 
		$\Omega^{\Spin\times  \Z_{2,[1]}}_d$ &
		$\TP^{\Spin\times  \Z_{2,[1]}}_d$ &
		Cobordism Invariant \\
		\hline
		1& $\Z_2$& $\Z_2$& $\tilde{\eta}$\\
		2& $\Z_2\times \Z_2$ & $\Z_2\times \Z_2$&  $B_2, \arf$\\
		3 & $0$ & $\Z$ & $\CSg$\\
		4 & $\Z\times \Z_{2}$& $\Z_2$ & $\mathcal{P}(B_2)/2$\\
		5 & $0$ & $0$&  \\
		\hline
	\end{tabular}
	\caption{
		The bordism and cobordism groups of the symmetry  $\Spin(d)\times \Z_{2,[1]}$. 
	}
	\label{table:Z2SpinBordism}
\end{table}

\subsubsection{Gauging $\SU(2)$ in $\SU(2)\times \Spin(d)$}

The only SPT in $d=2,3,4$ that involves the $\SU(2)$ bundle is the $\SU(2)$ Chern Simons theory $\mathrm{CS}_3^{\SU(2)}$. This theory is identical to \eqref{SU2CS}, except that here  $\mathrm{CS}_3^{SU(2)}$ is defined on a spin manifold. The anomaly of the $\Z_{2,[1]}$ global symmetry is the same as \eqref{SU2CSanomaly}. But on a spin manifold, \eqref{SU2CSanomaly} can be further simplified, because when $k=2$, the anomaly is $\pi \int_{M_4} \mathcal{P}(B_2)= \pi \int_{M_4} w_2 B_2= 0$ due to $w_2=0$ on a spin manifold. Thus the anomaly can be rewritten as
\begin{eqnarray}
2\pi \frac{k}{2}\int_{M_4} \frac{\mathcal{P}(B_2) }{2}, ~~k\in \Z_2.
\end{eqnarray}

\subsection{$\LG= \O(d)$}

When the Lorentz symmetry is $\O(d)$, there are four possibilities of the total symmetries, given in  \eqref{Osym}. When promoting the $\SU(2)$ global symmetry to dynamical and including the $\Z_{2,[1]}$ one form symmetry background $B_2$, the resulting global symmetries and the gauge bundle constraints for the resulting dynamical $\SU(2)$ pure gauge theories are 
\begin{eqnarray}
G'=
\begin{cases}
\SU(2)\times \O(d)\\
(\SU(2)\times \rE(d))/\Z_2\\
(\SU(2)\times \Pin^+(d))/\Z_2\\
(\SU(2)\times \Pin^-(d))/\Z_2
\end{cases}\Rightarrow
H= 
\begin{cases}
\Z_{2,[1]}\times \O(d), &~~~ w_2(V_{\SO(3)}) =B_2,\\
\Z_{2,[1]}\times \O(d), &~~~ w_2(V_{\SO(3)}) =B_2+w_1^2,\\
\Z_{2,[1]}\times \O(d), &~~~ w_2(V_{\SO(3)}) =B_2+w_2,\\
\Z_{2,[1]}\times \O(d), &~~~ w_2(V_{\SO(3)}) =B_2+ w_1^2 + w_2.\\
\end{cases}
\end{eqnarray}
In all four cases, the Lorentz symmetry is $\O(d)$, which means that the resulting Yang-Mills theories do not contain a transparent fermion line and is bosonic, despite that the Wilson line in the fundamental representation $W_{1/2}$ are nontransparent and fermionic in the last two cases. Under time reversal, the Wilson line $W_{j}$ with $\SU(2)$ isospin $j$ transforms as 
\begin{eqnarray}
\T^2= 1, (-1)^{2j}, (-1)^{2j}, 1
\end{eqnarray}
in the four cases respectively.

We further compute the bordism group and cobordism group of the emergent global symmetry $H$, and enumerate the cobordism invariants in Table \ref{table:Z2OBordism}. See  \cite{Wan2018bns1812.11967} for the derivation of the (co)bordism calculations. As we will see, only certain linear combinations of the cobordism invariants in 5d in Table \ref{table:Z2OBordism} can be realized as the anomalies  of theories obtained by gauging the SU(2) SPTs in 4d. In particular, $B_2\Sq^1 B_2$ always come together with $\Sq^2\Sq^1B_2$. Moreover, because among Tables \ref{table:OSU2Bordism}, \ref{table:ESU2Bordism}, \ref{table:Pin+SU2Bordism} and \ref{table:Pin-SU2Bordism}, there are no cobordism invariants involving $\SU(2)$ bundle in 3d, hence the 4d cobordism invariants $B_2 w_1^2 $ and $B_2 w_2$ can not be realized via gauging $\SU(2)$ SPT. For the same reason, $w_1B_2$ in 3d can not be realized in the same way either. However, these anomalies can be realized in various systems, for instance the $\U(1)$ gauge theories. For instance, $B_2\Sq^1 B_2$ alone can be realized in Maxwell theory in 4d with $\theta=2\pi$ theta term.

\begin{table}[t]
	\centering
	\begin{tabular}{ c c c c }
		\hline
		\multicolumn{4}{c}{Bordism and Cobordism group}\\
		\hline
		$d$ & 
		$\Omega^{\O\times  \Z_{2,[1]}}_d$ &
		$\TP^{\O\times  \Z_{2,[1]}}_d$ &
		Cobordism Invariant \\
		\hline
		1& $0$& $0$& \\
		2& $\Z_2\times \Z_2$ & $\Z_2\times \Z_2$&  $B_2, w_1^2$\\
		3 & $\Z_2$ & $\Z_2$ & $w_1 B_2$\\
		4 & $\Z_2\times \Z_{2}\times \Z_2\times \Z_2$& $\Z_2\times \Z_{2}\times \Z_2\times \Z_2$ & $w_1^4, w_2^2, B_2 w_1^2, B_2 w_2$\\
		5 & $\Z_2\times \Z_{2}\times \Z_2\times \Z_2$ & $\Z_2\times \Z_{2}\times \Z_2\times \Z_2$& $w_2w_3, B_2 \Sq^1 B_2, \Sq^2\Sq^1B_2, w_1^2 \Sq^1 B_2$ \\
		\hline
	\end{tabular}
	\caption{
		The bordism and cobordism groups of the symmetry  $\O(d)\times \Z_{2,[1]}$. 
	}
	\label{table:Z2OBordism}
\end{table}

\subsubsection{Gauging $\SU(2)$ in $\SU(2)\times \O(d)$}

The only cobordism invariant in $d=2,3,4$ that involves the $\SU(2)$ bundle in Table \ref{table:OSU2Bordism}  is the second Chern class mod 2, i.e., $c_2^{\SU(2)}\mod 2$ in $3+1$d. Written in terms of the $\SU(2)$ gauge field, the topological term is the theta term with $\theta=\pi$. The action is
\begin{eqnarray}\label{SU2YM}
S= -\frac{1}{4g^2} \int_{M_4} \Tr(f\wedge \star f) + \frac{\pi}{8\pi^2} \int_{M_4} \Tr(f\wedge f).
\end{eqnarray} 
The symmetries and the 't Hooft anomalies have been studied extensively in \cite{Gaiotto2017yupZoharTTT1703.00501, Wan2019oyr1904.00994, Wan2018zql1812.11968, Wang:2019obe}. 
Since in this case the gauge bundle constraint is $w_2(V_{\SO(3)}) = B_2$, the Wilson line $W_{1/2}$ is a worldline of a boson, and is a Kramers singlet, $\T^2=1$. The anomaly involving the one form symmetry is
\begin{eqnarray}\label{YManom}
\pi \int_{M_5} B_2 \Sq^1 B_2 + \Sq^2 \Sq^1 B_2,
\end{eqnarray}
which is the combination of  two among the four 5d cobordism invariants in Table \ref{table:Z2OBordism}.

\subsubsection{Gauging $\SU(2)$ in $(\SU(2)\times \rE(d))/\Z_2$}

The only cobordism invariant in $d=2,3,4$ that involves the $\SU(2)$ bundle in Table \ref{table:ESU2Bordism}  is the second Chern class mod 2, i.e., $c_2^{\SU(2)}\mod 2$ in $3+1$d. As remarked below Table \ref{table:ESU2Bordism}, this topological term is the same as that in \eqref{SU2YM}, except the gauge bundle constraint is modified to $w_2(V_{\SO(3)}) = B_2+ w_1^2$. The gauge bundle constraint here means that the Wilson line $W_{1/2}$ is a worldline of a boson, and is a Kramers doublet $\T^2=-1$. The anomaly involving the one form symmetry is 
\begin{eqnarray}\label{YManomw1}
\pi \int_{M_5} B_2 \Sq^1 B_2 + \Sq^2 \Sq^1 B_2 + w_1^2 \Sq^1 B_2,
\end{eqnarray}
which is the combination of  three among the four 5d cobordism invariants in Table \ref{table:Z2OBordism}.

\subsubsection{Gauging $\SU(2)$ in $(\SU(2)\times \Pin^+(d))/\Z_2$}

The only cobordism invariant in $d=2,3,4$ that involves the $\SU(2)$ bundle in Table \ref{table:Pin+SU2Bordism}  is the eta invariant $\eta_{\SU(2)}$. The topological term is in \eqref{SU2YM} should be modified to 
\begin{eqnarray}\label{eta}
2\pi \ii \nu \int_{M_4} \eta_{\SU(2)}, ~~ \nu\in \Z_4,
\end{eqnarray}
where $\nu$ labels the $\Z_4$ class of cobordism invariants. Since for $\nu=2$, $\exp(4\pi \ii \eta_{\SU(2)})= \exp(\ii \pi w_1^4)$ is a purely counter term in terms of the background field $w_1$, we conclude that $\nu=2$ does not produce anomaly. 
The remaining situation is $\nu=1\mod 2$. In this case, \eqref{eta} can be rewritten as the $\theta=\pi$ theta term with $\SU(2)$ gauge field being replaced by the twisted $\SO(3)$ gauge field with the gauge bundle constraint $w_2(V_{\SO(3)}) =B_2+w_2$. The anomaly has been worked out in \cite{Wan2019oyr1904.00994}, which can also be conveniently obtained by replacing $B_2$ with $B_2+w_2$. The anomaly is the same as \eqref{YManom}.


\subsubsection{Gauging $\SU(2)$ in $(\SU(2)\times \Pin^-(d))/\Z_2$}

The only cobordism invariant in $d=2,3,4$ that involves the $\SU(2)$ bundle in Table \ref{table:Pin-SU2Bordism}  is the eta invariant $N_0$ counting the number of zero modes mod 2. The topological term is in \eqref{SU2YM} should be modified to 
\begin{eqnarray}\label{N0}
\pi \ii \nu \int_{M_4} N_0, ~~ \nu\in \Z_2,
\end{eqnarray}
where $\nu$ labels the $\Z_2$ class of cobordism invariants. We will focus on the nontrivial case $\nu=1$. 
In this case, \eqref{N0} can be rewritten as the $\theta=\pi$ theta term with $\SU(2)$ gauge field being replaced by the twisted $\SO(3)$ gauge field with the gauge bundle constraint $w_2(V_{\SO(3)}) =B_2+w_1^2+ w_2$. The anomaly has been worked out in \cite{Wan2019oyr1904.00994}, which can also be conveniently obtained by replacing $B_2$ with $B_2+w_1^2+ w_2$. The anomaly is the same as \eqref{YManomw1}.

%

\section{Comments on the internal symmetry $G= \SU(N), \Spin(N), \Sp(N)$,\\ 
or $G= \SO(3)$ and $\U(1)$}

The majority of this section discusses pure gauge theories by gauging the  $G=\SU(2)$ symmetry, and study their emergent global symmetry. Similar analysis can be generalized to other global symmetries $G$. We will find new features arise if $G$ admits nontrivial magnetic flux characterized by $\H^2(\B G, R)$ for $R=\U(1)$ or $\Z_n$ for some $n$, such as $G= \U(1)$ (for $R=\U(1)$) or $\SO(3)$ (for $R=\Z_2$).

For $G=\SU(2)$,  the emergent symmetry is an electric 1-form center symmetry $\Z_{2,[1]}$ --- because $\SU(2)$ has a nontrivial $\Z_2$ center. Moreover, there is no emergent magnetic symmetry because $\H^2(\B \SU(2), \Z_2)$ which measures the $\SU(2)$ flux is always trivial. Thus we find that the discussion in section \ref{GaugingSU2} can be directly generalized to an arbitrary Lie group $G$ which has nontrivial center and trivial $\H^2(\B G, \Z_n)$ for all $n$.  Such $G$ include $G= \SU(N), \Spin(N), \Sp(N)$ for every $N$.

If $G$ admits nontrivial magnetic flux, then there are emergent magnetic symmetry. As long as there are no monopole operators explicitly in the Hamiltonian, the magnetic symmetry is not broken. In contrast to the electric center global symmetry which is always a 1-form symmetry in arbitrary spacetime dimension, the form of the emergent magnetic global symmetry depends on the dimension. In $d$ spacetime dimensions, the magnetic symmetry is $d-3$ form. Denote the $(d-2)$ form background gauge field as $B_{m,[d-2]}$, then it couples to the background field as
\begin{eqnarray}
\frac{1}{2\pi} \int_{M_d} f\wedge B_{m,[d-2]}
\end{eqnarray}
for $G=\U(1)$ where $f= da$ is the $\U(1)$ field strength ($a$ is the $\U(1)$ gauge field), and $B_{m,[d-2]}$ is the $\U(1)$ gauge field;  and 
\begin{eqnarray}
\pi \int_{M_d} w_2(V_{\SO(3)}) \cup B_{m, [d-2]}
\end{eqnarray}
for $G= \SO(3)$ where {$w_2(V_{\SO(3)}) \in \H^2(\B \SO(3), R)= \H^2(\B \SO(3), \Z_2)$} is the magnetic flux, and $B_{m, [d-2]}$ is the $\Z_2${-valued} $(d-2)$-cocycle. For $G=\SO(3)$, there is no emergent 1-form symmetry, with and without coupling to matter fields in the vector representation of $\SO(3)$.

We will not list the cobordism invariants for $G=\SO(3)$ and $\U(1)$ in this paper.\footnote{{Although we had listed part of the
cobordism groups and invariants for $G=\SO(3)$ and $\U(1)$ in the arXiv's first version of this paper \cite{Wan1912.13504}, we had decided to defer them to a more systematic exploration in a future work.}} But there are a few interesting cases we would like to comment on. Notice that $\SO(3)$ pure gauge theory can be obtained by gauging the $\Z_{2,[1]}$ center 1-form symmetry of the $\SU(2)$ pure gauge theory. If the the $\Z_{2,[1]}$ symmetry of a $d$-dimensional $\SU(2)$ gauge theory is anomalous, the $\SO(3)$ gauge theory obtained after gauging $\Z_{2, [1]}$ can be interpreted as a $d$d and $(d+1)$d combined system, where the $(d+1)$d bulk theory is a non-invertible TQFT. However, 
based on symmetry-extension construction \cite{Wang2017locWWW1705.06728} and related arguments \cite{2018PTEP1801.05416,Prakash2018ugo1804.11236, Guo2018vij1812.11959, Prakash202009},
as explored in \cite{Tachikawa2017gyf, Benini2018reh1803.09336, Cordova2018cvg2group1802.04790}, in certain cases the resulting theory after gauging $\Z_{2, [1]}$ can be a genuine $d$d theory without coupling to the $(d+1)$d bulk {(such as SPTs)}, by modifying the global symmetry bundle to be non-abelian or a higher group. For example, consider an $\SU(2)$ Chern Simons theory by gauging the $k \CS_3^{\SU(2)}$ SPT on a spacetime manifold with Lorentz symmetry $\SO(d)=\SO(3)$ in $3$d. As shown in  \eqref{SU2CSanomaly}, the theory has an anomaly depending on $k\mod 4$. Let us consider the special case $k=2\mod 4$. The anomaly \eqref{SU2CSanomaly} can be simplified to 
\begin{eqnarray}
\pi \int_{M_4} B_2 w_2.
\end{eqnarray}

Next let us gauge $\Z_{2, [1]}$. There is an emergent $0$-form $\Z_2$ global symmetry, whose 1-form background field is denoted as $\widehat{B}_{m,[1]}$. 
The $\widehat{B}_{m,[1]}$ couples to the system via the topological term 
\begin{eqnarray}
\pi \int_{M_3} B_2 \widehat{B}_{m,[1]}.
\end{eqnarray}
Then there are two options to interpret the global symmetry in the resulting $\SO(3)$ {gauge} theory. 
\begin{enumerate}
	\item The first option is to regard the $\SO(3)_{k/2}$ Chern Simons theory as coupled to a dynamical bulk TQFT in 4d. The global symmetry is $\Z_{2} \times \SO(3)$, where $\Z_2$ is the emergent 0-form magnetic symmetry. 
	\item The second option is to regard the $\SO(3)_{k/2}$ Chern Simons theory as a genuine $3$d theory. However, the emergent 0-form $\Z_2$ symmetry bundle and the Lorentz symmetry bundle form nontrivial correlation
	\begin{eqnarray}\label{twist}
	\delta \widehat{B}_{m,[1]} = w_2 \mod 2
	\end{eqnarray}
	\eqref{twist} means the resulting global symmetry is not $\Z_2\times \SO(3)$, but rather 
	\begin{eqnarray}
	\frac{\Z_2\times \Spin(3)}{\Z_2}
	\end{eqnarray}
	where the $\Z_2$ in the denominator is the fermion parity. Physically, it means the $\Z_2$ monopole operator is a fermion. This is consistent with the fact that the topological spin of the monopole operator in $\SO(3)_{k/2}$ theory is $k/4\mod 1$. 
\end{enumerate}

\section{Comments on Symmetry-Extension, Trivialization of Bulk, and Nontrivial Boundary States}

Lorentz symmetry extension provides a guidance for constructing the boundary states that saturate the 't Hooft anomaly for the Lorentz symmetry. In this section, we comment on the examples where the 't Hooft anomaly for Lorentz symmetry found in section \ref{secLG}. 

Let us briefly review how the group extension can be used to trivialize the anomaly discussed in \cite{Wang2017locWWW1705.06728}. Suppose there is an extension 
\begin{eqnarray}
1\to K \to H \overset{r}{\to} G\to 1
\end{eqnarray}
where $G$ is the global symmetry for a quantum field theory $\mathcal{Q}$ in $d$ dimensions and is also the global symmetry for an invertible TQFT(representing the anomaly of $\mathcal{Q}$) in $(d+1)$ dimensions.   $K$ is an emergent symmetry when is only present in $d$ dimensions. $H$ is the total symmetry. We denote the quotient map $H\to G=H/K$ by $r$.  In the setup of the present paper, $K, H, G$ corresponds to $G, G', \LG$ respectively. Suppose the quantum field theory in $d$ dimensions is $G$-anomalous, and the anomaly is captured by the $d+1$ dimensional invertible TQFT (cobordism invariant) $\omega^G_{d+1}$. The anomaly can be trivialized by pulling back from $G$ to $H$ if we can find a cochain $\mu_{d}^{H}$ satisfying 
\begin{eqnarray}\label{triv}
r^*\omega^G_{d+1} = \delta \mu_{d}^{H}
\end{eqnarray}
\eqref{triv} means when we use $r$ to pull back the anomaly $\omega^G_{d+1}$ from $G$ to $H$, the anomaly becomes exact, i.e., trivial. This method has been extensively explored to construct the boundary states of SPTs \cite{Wang2017locWWW1705.06728, 2018PTEP1801.05416,Prakash2018ugo1804.11236, Guo2018vij1812.11959, Kobayashi1905.05391, Prakash202009}, has been generalized to higher form symmetries\cite{Wan2018bns1812.11967}, and construct quantum field theory examples with exotic 2-group symmetries and non-invertible symmetries \cite{Tachikawa:2017gyf, Hsin2007.05915}.
In the rest of this section, we will comment on various group extensions involving the Lorentz groups.

\subsection{Trivializing $w_1^2$ in $\TP_2^{\O}$}

We can consider the group extension
\begin{eqnarray}
1\to \Z_2 \to \rE(2)\overset{r}{\to} \O(2)\to 1
\end{eqnarray}
where $\rE(2)$ is also an Lorentz symmetry group discussed in the introduction. The only nontrivial cobordism invariant in $\TP_2^{\O}$ is $w_1^2$. Using the projection map $r$, pulling back $w_1^2$ from $\O(2)$ to $\rE(2)$ also gives us $w_1^2$. However, from table \ref{table:SWLG}, we find that $w_1^2$ is trivial for $\rE(2)$. This means that after pulling back $w_1^2$ from $\O(2)$ to $\rE(2)$, it can be written as a total derivative, 
\begin{eqnarray}\label{w1}
w_1^2 =\delta \mu_1\mod 2
\end{eqnarray}
where $\mu_1$ is a $1$-cochain, representing the $\rE$-structure of the underlying spacetime manifold. 


\subsection{Trivializing $w_1^4$ in $\TP_4^{\O}$} 
\label{secw14}

Since $w_1^2$ in $\O(2)$ can be trivialized by pulling back to $\rE(2)$, it immediately follows that $w_1^4$ in $\O(4)$ can also be trivialized when pulled back to $\rE(4)$, using
\begin{eqnarray}
w_1^4= \delta (w_1^2 \mu_1)\mod 2
\end{eqnarray}
where $\mu_1$ is a $1$-cochain representing the $\rE$-structure. Indeed, one can check, using the chain rule, that $\delta (w_1^2 \mu_1) = \delta (w_1^2) \mu_1 + w_1^2 \delta \mu_1$. The first term  vanishes because $\delta w_1=0\mod 2$. The second term survives, and by using \eqref{w1}, we obtain $w_1^4$ as desired. 


\subsection{Trivializing $w_2 w_3$ and $w_2^2$ in $\TP_5^{\O}$ and $\TP_4^{\O}$} 

We consider the group extension
\begin{eqnarray}
1\to \Z_2 \to \Pin^+(5)\to\O(5)\to 1
\end{eqnarray}
The only nontrivial cobordism invariant in $\TP_5^{\O}$ is $w_2 w_3$. When we pulled back  $w_2 w_3$ from $\O(5)$ to $\Pin^+(5)$, we find that $w_2$ is trivialized (see table \ref{table:SWLG}). Hence $w_2= \delta \rho_+$ where $\rho_+$ is actually the $\Pin^+$ structure. Then it follows that 
\begin{eqnarray}
w_2 w_3 = \delta( \rho_+ w_3)\mod 2
\end{eqnarray}

Using the same group extension for $d=4$
\begin{eqnarray}
1\to \Z_2 \to \Pin^+(4)\to\O(4)\to 1
\end{eqnarray}
we find that $w_2^2$ in $\TP_4^{\O}$ can be trivialized by pulling back from $\O(4)$ to $\Pin^+(4)$ in a similar way, 
\begin{eqnarray}
w_2^2= \delta(\rho_+ w_2)\mod 2
\end{eqnarray}

\subsection{Trivializing $8\eta$ in $\TP_4^{\Pin^+}$} 
\label{8eta}

We consider the group extension 
\begin{eqnarray}\label{triv1}
1\to \Z_2 \to \EPin(4)\to\Pin^+(4)\to 1
\end{eqnarray}
The only cobordism invariant in $\TP_4^{\Pin^+}$ is the eta invariant $\eta$. We claim that 8 copies of $\eta$ can be trivialized using \eqref{triv1}. First we notice the identity
$8\eta = w_1^4 \mod 2$. Hence trivializing $8\eta $ really amounts to trivializing $w_1^4$. Since when pulling backing $w_1^2$ from $\Pin^+(4)$ to $\EPin(4)$,   $w_1^2$ is trivialized (see table \ref{table:SWLG}), it follows that $w_1^4$ is also trivialized. Thus we can write 
\begin{eqnarray}\label{EPinst}
w_1^2= \delta \nu_1 \mod 2 
\end{eqnarray}
where we denote $\nu_1$ as the $\EPin$ structure. This means $w_1^2$ is trivialized by the $\EPin$ structure. Then using the same analysis in section \ref{secw14}, we conclude that $w_1^4$ can be trivialized as follows
\begin{eqnarray}
w_1^4= \delta (w_1^2 \nu_1) \mod 2 
\end{eqnarray}

\subsection{Trivializing $4\abk$ in $\TP_2^{\Pin^-}$} 
\label{abk}

Let us consider the group extension
\begin{eqnarray}
1\to \Z_2 \to \EPin(2)\to\Pin^-(2)\to 1
\end{eqnarray}
The generator of $\TP_2^{\Pin^-}= \Z_8$ is the $\abk$ invariant. Notice that the 4 copies of $\abk$ is $w_1^2$.  One can trivialize $w_1^2$ using the $\EPin$ structure as shown in \eqref{EPinst}, 
\begin{eqnarray}
4\abk = w_1^2 = \delta \nu_1 \mod 2 
\end{eqnarray}


\subsection{Trivializing $4 a\cup \abk$ in $\TP_3^{\Z_2\times \Spin}$} 
\label{aabk}

It is known that the $\Z_2$ fermionic SPT in $2+1$d is classified by $\Z_8$. Concretely, this means $\TP_3^{\Z_2\times \Spin}= \Z_8$, and the generator is $a\cup \abk$, where $a$ is the generator for $\Z_2$ and $\abk$ is the ABK-invariant generating $\TP_2^{\Pin^-}$. We will show that $4 a\cup \abk$ can be trivialized by pulling back from $\Z_2\times \Spin(3)$ to $\Z_4\times \Spin(3)$, via 
\begin{eqnarray}\label{ext}
1\to \Z_2 \to \Z_4\times \Spin(3)\to\Z_2\times \Spin(3)\to 1
\end{eqnarray}
To see this, we first use the identity 
\begin{eqnarray}
4 a\cup \abk= a^3\mod 2
\end{eqnarray}
The right hand side is just the $\Z_2$ bosonic SPT in $2+1$d. To trivialize $a^3$, we can extend $\Z_2$ to $\Z_4$, where  the generator $a$ in $\Z_2$ is extended to generator $2b+a$ in $\Z_4$, with the constraint 
\begin{eqnarray}\label{con}
\delta b = a^2 \mod 2
\end{eqnarray}
Notice that $a^2\in \H^2(\B\Z_2, \Z_2)$ classifies the extension $1\to \Z_2 \to \Z_4\to\Z_2\to 1$, which also consequently classifies the extension \eqref{ext}. Using the bundle constraint \eqref{con}, it follows that
\begin{eqnarray}
4 a\cup \abk= a^3 = \delta (a b)\mod 2
\end{eqnarray}



%
%

\subsection{Trivializing the Cobordism Invariants in the Smith Homomorphisms:
\\ $\dots \to\Omega_{d}^{\Spin\times \Z_2}\to \Omega_{d-1}^{\Pin^-} \to \Omega_{d-2}^{\Spin\times_{\Z_2} \Z_4}\to \Omega_{d-3}^{\Pin^+}  \to \dots$
}


Smith homomorphism provides a chain of bordism invariants in any dimension (see 
\cite{2018arXiv180502772T} and also \cite{Kapustin:2014dxa, Guo2018vij1812.11959, Hason2019akwRyanThorngren1910.14039, JW2006.16996}):
\begin{eqnarray}\label{bor}
\cdots \to \Omega_{d}^{\Spin\times \Z_2}\to \Omega_{d-1}^{\Pin^-} \to \Omega_{d-2}^{\Spin\times_{\Z_2} \Z_4}\to \Omega_{d-3}^{\Pin^+} \to \Omega_{d-4}^{\Spin\times \Z_2} \to \cdots
\end{eqnarray}
There is correspondingly a similar Smith homomorphism chain of cobordism invariants, 
\begin{eqnarray}
\cdots \leftarrow \TP_{d}^{\Spin\times \Z_2}\leftarrow \TP_{d-1}^{\Pin^-} \leftarrow \TP_{d-2}^{\Spin\times_{\Z_2} \Z_4}\leftarrow \TP_{d-3}^{\Pin^+} \leftarrow \TP_{d-4}^{\Spin\times \Z_2} \leftarrow \cdots
\end{eqnarray}
In this subsection, we will focus on the sequence \eqref{bor}, and in particular we will discuss $d=7$. We enlist the bordism classifications, the manifold generators, and the bordism invariants in table \ref{tab:SmithSPTs}. 
\begin{table}
	\centering
	\begin{tabular}{ | c c  c  c |}
		\hline
		Bordism Group & Classification & Manifold Generator & Invariants\\
		\hline
		$\Omega_8^{\Pin^+}$ & $\Z_{32} \times \Z_2$ & $\RP^8$ & $\exp\left(\frac{2\pi \ii}{32}\nu \eta_8\right)$\\
		$\Omega_7^{\Spin \times {\Z_2}}$ & $\Z_{16}$ & $\RP^7$ & $\exp\left(\frac{2\pi \ii}{16}\nu \eta(\text{PD}(a^3))\right)$\\
		$\Omega_6^{\Pin^-}$ & $\Z_{16}$ & $\RP^6$ & $\exp\left(\frac{2\pi \ii}{16}\nu \eta(\text{PD}(w_1^2))\right)$\\
		$\Omega_5^{\Spin \times_{\Z_2} \Z_4}$ & $\Z_{16}$ & $\RP^5$ & $\exp\left(\frac{2\pi \ii}{16}\nu \eta(\text{PD}(a))\right)$\\
		$\Omega_4^{\Pin^+} $ & $\Z_{16}$ & $\RP^4$ & $\exp\left(\frac{2\pi \ii}{16}\nu \eta\right)$\\
		$\Omega_3^{\Spin \times {\Z_2}}$ & $\Z_{8}$ & $\RP^3$ & $\exp\left(\frac{2\pi \ii}{8}\nu \abk(\text{PD}(a))\right)$\\
		$\Omega_2^{\Pin^-} $ & $\Z_{8}$ & $\RP^2$ & $\exp\left(\frac{2\pi \ii}{8}\nu \abk\right)$\\
		$\Omega_1^{\Spin \times_{\Z_2} \Z_4}$ & $\Z_{4}$ & $\RP^1$ & $\exp\left(\frac{2\pi \ii}{4}\nu \eta'\right)$\\
		\hline
	\end{tabular}
    \caption{Bordism classifications, the manifold generators, and the bordism invariants for groups in the Smith homomorphisms. In odd dimensions, $a$ is $\Z_2$ 1-cocycle, generating $\Z_2$ (for $d=4k+3$) and $\Z_4/\Z_2$ (for $d=4k+1$). PD is the Poincare dual of the cocycles. In $d=8$, we just list the manifold generator and bordism invariant for $\Z_{32}$, 
    which is generated by the eta invariant $\eta_8$ in 8d. }
    \label{tab:SmithSPTs}
\end{table}

The trivialization of the middle class of the bordism invariants for $d=2,3,4$ have been studied in previous subsections. In section \ref{abk}, we trivialized the middle class $\nu=4\in \Z_8$, i.e. $4 \abk$, by pulling back to $\EPin(2)$. In section \ref{aabk}, we trivialized the middle class $\nu=4\in \Z_8$, i.e. $4 a\cup\abk = 4 \abk(\text{PD}(a))$, by pulling back to $\Spin(3)\times \Z_2$. In section \ref{8eta}, we trivialized the middle class $\nu=8\in \Z_{16}$, i.e. $8 \eta$, by pulling back to $\EPin(4)$.  

In general, we propose that the given a $\Z_{2^n}$ class, the $\nu=2^{n-1} \in \Z_{2^n}$ class can be trivialized by the following extensions:
\bea
\begin{array}{rcl}
	d=4k &:&1 \to \Z_2\to \EPin(d) \to \Pin^+(d) \to1.\\
	d=4k+1 &:&1 \to \Z_2\to {\Spin(d) \times \Z_4} \to {\Spin(d) \times_{\Z_2} \Z_4} \to1.\\
	d=4k+2 &:&1 \to \Z_2\to \EPin(d) \to \Pin^-(d) \to1.\\
	d=4k+3 &:&1 \to \Z_2\to {\Spin(d) \times \Z_4} \to {\Spin(d) \times \Z_2} \to 1.
\end{array}
\eea
The key feature is that in even dimensions, the middle class of the bordism invariants in the Smith homomorphism chain can be trivialized by pulling back to $\EPin(d)$, while in odd dimensions, they can be trivialized by pulling back to $\Spin(d)\times \Z_4$. 
Repeating similar calculations, we can check the extensions for cases in $d=5,6,7,8$.  We enumerate the trivializations below: 
\begin{enumerate}
	\item For $d=5$, we have the identity $8 \eta(\text{PD}(a))= a^5 \mod 2$. Hence one can trivialize $a^5$ in $\Z_4\times \Spin(5)$ via $a^5 = \delta (a^3 b)$. where $b$ satisfies \eqref{con}. 
	\item For $d=6$, we have the identity $8 \eta(\text{PD}(w_1^2))= w_1^6 \mod 2$. Hence one can trivialize $w_1^6$ in $\EPin(6)$ via $w_1^6 = \delta (w_1^4 \mu_1)$ where $\mu_1$ is defined in \eqref{w1}. 
	\item For $d=7$, we have the identity $8 \eta(\text{PD}(a^3))= a^7 \mod 2$. Hence one can trivialize $a^7$ in $\Z_4\times \Spin(5)$ via $a^7 = \delta (a^5 b)$. where $b$ satisfies \eqref{con}. 
	\item For $d=8$, we have the identity $16 \eta_8= w_1^8 \mod 2$. Hence one can trivialize $w_1^8$ in $\EPin(8)$ via $w_1^8 = \delta (w_1^6 \mu_1)$ where $\mu_1$ is defined in \eqref{w1}. 
\end{enumerate}
Other ways to do symmetry extension for fermionic systems are also studied \cite{Guo2018vij1812.11959, Kobayashi1905.05391} and \cite{Prakash202009}.

\section{Acknowledgement}

JW thanks Kantaro Ohmori for email correspondences.
YZ thanks Shu-Heng Shao and Ryan Thorngren for email correspondences.  
JW and YZ thank Julio Parra-Martinez for discussions and a collaboration for a upcoming work.  ZW  acknowledges previous supports from NSFC grants 11431010 and 11571329. 
ZW is supported by the Shuimu Tsinghua Scholar Program.  JW is supported by Center for Mathematical Sciences and Applications at Harvard University.  YZ is supported by department of physics in Princeton University.

\appendix

\section{Bordism and Cobordism Groups involving $\rE(d)$}
\label{app}

\subsection{Bordism and Cobordism Groups of $\rE(d)$}

In this subsection, we will compute the cobordism group of $\rE(d)$. Recall that $\rE$ is defined to be the subgroup of $\rO\times\Z_4$ consisting of the pairs $(A,j)$ such that $\det A=j^2$, there is a fibration $\B\rE\to\B\rO\xrightarrow{w_1^2}\B^2\Z_2$. 

We can also think of the space $\B\rE$ as the fiber of $w_1 + x: \B\rO \times \B\Z_4 \to \B\Z_2$, where $x$ is the generator of $\H^1(\B\Z_4, \Z_2)$.  
Take $W$ to be the line bundle on $\B\Z_4$ determined by $\B\Z_4 \to \B\Z_2 = \B\rO(1)$.    Define a map $f: \B\rO \times \B\Z_4 \to \B\rO \times \B\Z_4$ by $(V, V') \to (V + W - 1, V')$, with inverse $(V, V') \to (V - W + 1, V')$.  
Observe $f^*(w_1) = w_1 + x$, so that $\B\rE$ is homotopy equivalent to $\B\SO \times \B\Z_4$.  
The canonical bundle $\B\rE \to \B\rO$ corresponds to $V - W + 1$ on $\B\SO \times \B\Z_4$.  
So $MT\rE = MT\SO \wedge \text{Thom}(\B\Z_4, W-1) 
=M\SO\wedge\Sigma^{-1}M\Z_4$.  

The localization of the Thom spectrum $M\SO$ at the prime 2 is 
\bea
M\SO_{(2)}=H\Z_{(2)}\vee\Sigma^4H\Z_{(2)}\vee\Sigma^5H\Z_2\vee\cdots.
\eea

The mod 2 cohomology of $H\Z$ is
\bea
\H^*(H\Z,\Z_2)=\A_2\otimes_{\A_2(0)}\Z_2
\eea
where $\A_2(0)$ is the subalgebra of $\A_2$ generated by $\Sq^1$.

By K\"unneth formula, we have
\bea
\H^*(M\SO\wedge\Sigma^{-1}M\Z_4,\Z_2)=\H^*(M\SO,\Z_2)\otimes\H^{*+1}(M\Z_4,\Z_2).
\eea

By the generalized Pontryagin-Thom isomorphism,
\bea
\pi_d(MT\rE)=\Omega_d^{\rE}.
\eea

Since there is no odd torsion, the Adams spectral sequence shows:
\bea
\Ext_{\A_2}^{s,t}(\H^*(MT\rE,\Z_2),\Z_2)\Rightarrow\Omega_{t-s}^{\rE}.
\eea

We have
\bea
\H^*(\B\Z_4,\Z_2)=\Z_2[y]\otimes\Lambda_{\Z_2}(x)
\eea
where $\Lambda_{\Z_2}$ is the exterior algebra, $x$ is the generator of $\H^1(\B\Z_4, \Z_2)$, $y$ is the generator of $\H^2(\B\Z_4, \Z_2)$, with $\Sq^1x=\Sq^1y=0$.

On the other hand, by Thom isomorphism,
\bea
\H^{*+1}(M\Z_4,\Z_2)=(\Z_2[y]\otimes\Lambda_{\Z_2}(x))U
\eea
where $U$ is the Thom class of the line bundle $W$ with $\Sq^1U=xU=w_1U$.

The $\A_2(0)$ module structure of $\H^{*+1}(M\Z_4,\Z_2)$ is shown in Figure \ref{fig:A_2(0)MZ4}.

The $E_2$ page of the Adams spectral sequence is shown in Figure \ref{fig:E_2E}.

\begin{figure}[H]
	\begin{center}
		\begin{tikzpicture}[scale=0.5]
		
		\node[right] at (0,0) {$U$};
		\draw[fill] (0,0) circle(.1);
		
		\node[right] at (0,1) {$xU$};
		\draw[fill] (0,1) circle(.1);
		
		\draw (0,0) -- (0,1);
		
		\node[right] at (0,2) {$yU$};
		\draw[fill] (0,2) circle(.1);
		
		\node[right] at (0,3) {$xyU$};
		\draw[fill] (0,3) circle(.1);
		
		\draw (0,2) -- (0,3);
		
		\node[right] at (0,4) {$y^2U$};
		\draw[fill] (0,4) circle(.1);
		
		\node[right] at (0,5) {$xy^2U$};
		\draw[fill] (0,5) circle(.1);
		
		\draw (0,4) -- (0,5);
		
		\end{tikzpicture}
	\end{center}
	\caption{The $\A_2(0)$ module structure of $\H^{*+1}(M\Z_4,\Z_2)$.}
	\label{fig:A_2(0)MZ4}
\end{figure}

\begin{figure}[H]
	\begin{center}
		\begin{tikzpicture}
		\node at (0,-1) {0};
		\node at (1,-1) {1};
		\node at (2,-1) {2};
		\node at (3,-1) {3};
		\node at (4,-1) {4};
		\node at (5,-1) {5};
		\node at (6,-1) {$t-s$};
		\node at (-1,0) {0};
		\node at (-1,1) {1};
		\node at (-1,2) {2};
		\node at (-1,3) {3};
		\node at (-1,4) {4};
		\node at (-1,5) {5};
		\node at (-1,6) {$s$};
		
		\draw[->] (-0.5,-0.5) -- (-0.5,6);
		\draw[->] (-0.5,-0.5) -- (6,-0.5);

		\draw[fill] (0,0) circle(.05);		
		\draw[fill] (2,0) circle(.05);	
		\draw[fill] (4,0) circle(.05);			
		\draw[fill] (4.1,0) circle(.05);
		\draw[fill] (5,0) circle(.05);		
		\end{tikzpicture}
	\end{center}
	\caption{$\Omega_*^{\rE}$}
	\label{fig:E_2E}
\end{figure}

\begin{table}[H]
	\centering
	\begin{tabular}{ c c c c}
		\hline
		\multicolumn{4}{c}{Bordism and Cobordism group}\\
		\hline
		$d$ & 
		$\Omega^{\rE}_d$ & $\TP^{\rE}_d$
		& Cobordism Invariant \\
		\hline
		0& $\Z_2$ & $\Z_2$ &\\
		1& 0 & 0 &\\
		2& $\Z_2$ & $\Z_2$ & $y$ \\
		3 & $0$ & 0 &\\
		4 & $\Z_2^2$ & $\Z_2^2$ &  $y^2,w_2^2$ \\
		5 & $\Z_2$ & $\Z_2$&  $w_2w_3$\\
		\hline
	\end{tabular}
	\caption{Bordism group. Here $y$ is the generator of $\H^2(\B\Z_4,\Z_2)$, $w_i$ is the Stiefel-Whitney class of the tangent bundle. Here the cohomology classes of $\B\Z_4$ are pulled back to the manifold $M$ via the map $M\to\B\rE\to\B\Z_4$.
	}
	\label{table:EBordism}
\end{table}

\subsection{Bordism and Cobordism Groups  of $\rE(d)\times \SU(2)$}

Since there is no odd torsion, the Adams spectral sequence shows:
\bea
\Ext_{\A_2}^{s,t}(\H^*(MT(\rE\times\SU(2)),\Z_2),\Z_2)\Rightarrow\Omega_{t-s}^{\rE\times\SU(2)}.
\eea

We have $MT(\rE\times\SU(2))=MT\rE\wedge(\B\SU(2))_+=M\SO\wedge\Sigma^{-1}M\Z_4\wedge(\B\SU(2))_+$.

The $E_2$ page of the Adams spectral sequence is shown in Figure \ref{fig:E_2ESU2}.

\begin{figure}[H]
	\begin{center}
		\begin{tikzpicture}
		\node at (0,-1) {0};
		\node at (1,-1) {1};
		\node at (2,-1) {2};
		\node at (3,-1) {3};
		\node at (4,-1) {4};
		\node at (5,-1) {5};
		\node at (6,-1) {$t-s$};
		\node at (-1,0) {0};
		\node at (-1,1) {1};
		\node at (-1,2) {2};
		\node at (-1,3) {3};
		\node at (-1,4) {4};
		\node at (-1,5) {5};
		\node at (-1,6) {$s$};
		
		\draw[->] (-0.5,-0.5) -- (-0.5,6);
		\draw[->] (-0.5,-0.5) -- (6,-0.5);
		
		\draw[fill] (0,0) circle(0.05);	
		\draw[fill] (2,0) circle(0.05);	
		\draw[fill] (4,0) circle(0.05);	
		\draw[fill] (4.1,0) circle(0.05);
		\draw[fill] (4.2,0) circle(0.05);
		\draw[fill] (5,0) circle(0.05);

		\end{tikzpicture}
	\end{center}
	\caption{$\Omega_*^{\rE\times\SU(2)}$}
	\label{fig:E_2ESU2}
\end{figure}

\begin{table}[H]
	\centering
	\begin{tabular}{ c c c c}
		\hline
		\multicolumn{4}{c}{Bordism and Cobordism group}\\
		\hline
		$d$ & 
		$\Omega^{\rE\times \SU(2)}_d$ & $\TP^{\rE\times \SU(2)}_d$
		& Cobordism Invariant \\
		\hline
		0& $\Z_2$ & $\Z_2$ &\\
		1& 0 & 0 &\\
		2& $\Z_2$ & $\Z_2$ & $y$ \\
		3 & $0$ & 0 &\\
		4 & $\Z_2^3$ & $\Z_2^3$ &  $y^2,w_2^2, c_2\mod2$ \\
		5 & $\Z_2$ & $\Z_2$&  $w_2w_3$\\
		\hline
	\end{tabular}
	\caption{Bordism group. Here $y$ is the generator of $\H^2(\B\Z_4,\Z_2)$, $w_i$ is the Stiefel-Whitney class of the tangent bundle.  Here the cohomology classes of $\B\Z_4$ are pulled back to the manifold $M$ via the map $M\to\B\rE\to\B\Z_4$.
	}
	\label{table:ESU2Bordism1}
\end{table}

\subsection{Bordism and Cobordism Groups  of $\rE(d)\times \B\Z_2$}

We have $MT(\rE\times\B\Z_2)=MT\rE\wedge(\B^2\Z_2)_+=M\SO\wedge\Sigma^{-1}M\Z_4\wedge(\B^2\Z_2)_+$.

By K\"unneth formula,
\bea
\H^*(M\SO\wedge\Sigma^{-1}M\Z_4\wedge(\B^2\Z_2)_+,\Z_2)=\H^*(M\SO,\Z_2)\otimes\H^{*+1}(M\Z_4,\Z_2)\otimes\H^*(\B^2\Z_2,\Z_2).
\eea

Since there is no odd torsion, we have the Adams spectral sequence
\bea
\Ext_{\A_2}^{s,t}(\H^*(M\SO,\Z_2)\otimes\H^{*+1}(M\Z_4,\Z_2)\otimes\H^*(\B^2\Z_2,\Z_2),\Z_2)\Rightarrow\Omega_{t-s}^{\rE\times\B\Z_2}.
\eea

We have 
\bea
\H^*(\B^2\Z_2,\Z_2)=\Z_2[x_2,x_3,x_5,\dots]
\eea
where $x_2$ is the generator of $\H^2(\B^2\Z_2,\Z_2)$, $x_3=\Sq^1x_2$, $x_5=\Sq^2x_3$, $\dots$.

We also have 
\bea
\H^{*+1}(M\Z_4,\Z_2)=(\Z_2[y]\otimes\Lambda_{\Z_2}(x))U
\eea
where $x$ is the generator of $\H^1(\B\Z_4,\Z_2)$, $y$ is the generator of $\H^2(\B\Z_4,\Z_2)$, $\Lambda_{\Z_2}$ is the exterior algebra, $U$ is the Thom class of the line bundle determined by $\B\Z_4\to\B\Z_2=\B\O(1)$.

We list the elements of $\H^{*+1}(M\Z_4,\Z_2)\otimes\H^*(\B^2\Z_2,\Z_2)$ below degree 5 as follows:
\bea
\begin{array}{ll}0&U\\1&xU\\2&yU,x_2U\\3&xyU,xx_2U,x_3U\\4&y^2U,yx_2U,xx_3U,x_2^2U\\5&xy^2U,xyx_2U,yx_3U,xx_2^2U,x_2x_3U,x_5U.\end{array}
\eea
They satisfy $\Sq^1U=xU$, $\Sq^1yU=xyU$, $\Sq^1y^2U=xy^2U$, $\Sq^1(x_2U)=(xx_2+x_3)U$, $\Sq^1x_2^2U=xx_2^2U$, $\Sq^1(x_5U)=(xx_5+x_3^2)U$,
$\Sq^1(x_2x_3U)=(xx_2x_3+x_3^2)U$, $\Sq^1(xx_2U)=\Sq^1(x_3U)=xx_3U$, $\Sq^1(yx_2U)=(xyx_2+yx_3)U$, and  $\Sq^1(xyx_2U)=\Sq^1(yx_3U)=xyx_3U$.

The differentials $d_1$ are induced by $\Sq^1$.

The $E_2$ page is shown in Figure \ref{fig:E_2EBZ2}.

\begin{figure}[H]
	\begin{center}
		\begin{tikzpicture}
		\node at (0,-1) {0};
		\node at (1,-1) {1};
		\node at (2,-1) {2};
		\node at (3,-1) {3};
		\node at (4,-1) {4};
		\node at (5,-1) {5};
		\node at (6,-1) {$t-s$};
		\node at (-1,0) {0};
		\node at (-1,1) {1};
		\node at (-1,2) {2};
		\node at (-1,3) {3};
		\node at (-1,4) {4};
		\node at (-1,5) {5};
		\node at (-1,6) {$s$};
		
		\draw[->] (-0.5,-0.5) -- (-0.5,6);
		\draw[->] (-0.5,-0.5) -- (6,-0.5);
		
		\draw[fill] (0,0) circle(0.05);
		\draw[fill] (2,0) circle(0.05);
		\draw[fill] (2.1,0) circle(0.05);
		\draw[fill] (3,0) circle(0.05);
		\draw[fill] (3.9,0) circle(0.05);
		\draw[fill] (4,0) circle(0.05);
		\draw[fill] (4.1,0) circle(0.05);
		\draw[fill] (4.2,0) circle(0.05);
		
		\draw[fill] (5,0) circle(0.05);
		\draw[fill] (5.1,0) circle(0.05);
		\draw[fill] (4.9,0) circle(0.05);
		\draw[fill] (5.2,0) circle(0.05);

		\end{tikzpicture}
	\end{center}
	\caption{$\Omega_*^{\rE\times\B\Z_2}$}
	\label{fig:E_2EBZ2}
\end{figure}

\begin{table}[H]
	\centering
	\begin{tabular}{ c c c c}
		\hline
		\multicolumn{4}{c}{Bordism and Cobordism group}\\
		\hline
		$d$ & 
		$\Omega^{\rE\times\Z_{2,[1]}}_d$ & $\TP^{\rE\times\Z_{2,[1]}}_d$
		& Cobordism Invariant \\
		\hline
		0& $\Z_2$ & $\Z_2$ &\\
		1& 0 & 0 &\\
		2& $\Z_2^2$ & $\Z_2^2$ & $x_2,y$ \\
		3 & $\Z_2$ & $\Z_2$ & $xx_2=x_3$\\
		4 & $\Z_2^4$ & $\Z_2^4$ &  $w_2^2,y^2,x_2^2,yx_2$ \\
		5 & $\Z_2^4$ & $\Z_2^4$&  $w_2w_3,xyx_2=yx_3,x_2x_3,x_5$\\
		\hline
	\end{tabular}
	\caption{Bordism group. Here $x$ is the generator of $\H^1(\B\Z_4,\Z_2)$,
		$y$ is the generator of $\H^2(\B\Z_4,\Z_2)$, 
		$x_2$ is the generator of $\H^2(\B^2\Z_2,\Z_2)$, $x_3=\Sq^1x_2$, $x_5=\Sq^2x_3$,
		$w_i$ is the Stiefel-Whitney class of the tangent bundle.
	}
	\label{table:EBZ2Bordism}
\end{table}

\section{Bordism and Cobordism Groups involving $\EPin(d)$}
\label{app2}

\subsection{Bordism and Cobordism Groups of $\EPin(d)$}

In this subsection, we will compute the cobordism group of $\EPin(d)$. Recall that $\EPin$ is a group extension:
\bea
1\to\Z_2\times\Z_2\to\EPin\to\rO\to1
\eea
such that $\B\EPin$ is the fiber of $w_2$ and $w_1^2$ of $\B\rO$. We can also think of the space $\B\EPin$ as the fiber of $w_2$ of $\B\rE$.

Also recall that 
we can think of the space $\B\rE$ as the fiber of $w_1 + x: \B\rO \times \B\Z_4 \to \B\Z_2$, where $x$ is the generator of $\H^1(\B\Z_4, \Z_2)$.  
Take $W$ to be the line bundle on $\B\Z_4$ determined by $\B\Z_4 \to \B\Z_2 = \B\rO(1)$.    Define a map $f: \B\rO \times \B\Z_4 \to \B\rO \times \B\Z_4$ by $(V, V') \to (V + W - 1, V')$, with inverse $(V, V') \to (V - W + 1, V')$.  
Observe $f^*(w_1) = w_1 + x$, so that $\B\rE$ is homotopy equivalent to $\B\SO \times \B\Z_4$.  
The canonical bundle $\B\rE \to \B\rO$ corresponds to $V - W + 1$ on $\B\SO \times \B\Z_4$.  
So $\B\EPin$ is homotopy equivalent to $\B\Spin\times\B\Z_4$. The canonical bundle $\B\EPin\to\B\rO$ corresponds to $V-W+1$ on $\B\Spin\times\B\Z_4$.
So $MT\EPin = MT\Spin \wedge \text{Thom}(\B\Z_4, W-1) 
=M\Spin\wedge\Sigma^{-1}M\Z_4$.

Since there is no odd torsion, the Adams spectral sequence shows: for $t-s<8$,
\bea
\Ext_{\A_2(1)}^{s,t}(\H^{*+1}(M\Z_4,\Z_2),\Z_2)\Rightarrow\Omega_{t-s}^{\EPin}.
\eea

We have
\bea
\H^*(\B\Z_4,\Z_2)=\Z_2[y]\otimes\Lambda_{\Z_2}(x)
\eea
where $\Lambda_{\Z_2}$ is the exterior algebra, $x$ is the generator of $\H^1(\B\Z_4, \Z_2)$, $y$ is the generator of $\H^2(\B\Z_4, \Z_2)$, with $\Sq^1x=\Sq^1y=0$.

On the other hand, by Thom isomorphism,
\bea
\H^{*+1}(M\Z_4,\Z_2)=(\Z_2[y]\otimes\Lambda_{\Z_2}(x))U
\eea
where $U$ is the Thom class of the line bundle $W$ with $\Sq^1U=xU=w_1U$ and $\Sq^2U=0$.

The $\A_2(1)$ module structure of $\H^{*+1}(M\Z_4,\Z_2)$ is shown in Figure \ref{fig:A_2(1)MZ4}.

The $E_2$ page of the Adams spectral sequence is shown in Figure \ref{fig:E_2EPin}.

\begin{figure}[H]
	\begin{center}
		\begin{tikzpicture}[scale=0.5]
		
		\node[right] at (0,0) {$U$};
		\draw[fill] (0,0) circle(.1);
		
		\node[right] at (0,1) {$xU$};
		\draw[fill] (0,1) circle(.1);
		
		\draw (0,0) -- (0,1);
		
		\node[right] at (0,2) {$yU$};
		\draw[fill] (0,2) circle(.1);
		
		\node[right] at (0,3) {$xyU$};
		\draw[fill] (0,3) circle(.1);
		
		\draw (0,2) -- (0,3);
		
		\node[right] at (0,4) {$y^2U$};
		\draw[fill] (0,4) circle(.1);
		
		\node[right] at (0,5) {$xy^2U$};
		\draw[fill] (0,5) circle(.1);
		
		\draw (0,4) -- (0,5);
		
		\draw (0,2) to[out=150,in=150] (0,4);
		
		\draw (0,3) to[out=30,in=30] (0,5);
		
		\end{tikzpicture}
	\end{center}
	\caption{The $\A_2(1)$ module structure of $\H^{*+1}(M\Z_4,\Z_2)$.}
	\label{fig:A_2(1)MZ4}
\end{figure}

\begin{figure}[H]
	\begin{center}
		\begin{tikzpicture}
		\node at (0,-1) {0};
		\node at (1,-1) {1};
		\node at (2,-1) {2};
		\node at (3,-1) {3};
		\node at (4,-1) {4};
		\node at (5,-1) {5};
		\node at (6,-1) {$t-s$};
		\node at (-1,0) {0};
		\node at (-1,1) {1};
		\node at (-1,2) {2};
		\node at (-1,3) {3};
		\node at (-1,4) {4};
		\node at (-1,5) {5};
		\node at (-1,6) {$s$};
		
		\draw[->] (-0.5,-0.5) -- (-0.5,6);
		\draw[->] (-0.5,-0.5) -- (6,-0.5);

		\draw (0,0) -- (2,2);
		\draw (2,1) -- (2,2);
		\draw (2,1) -- (4,3);
		\draw[fill] (2,0) circle(.05);
		\draw[fill] (4,1) circle(.05);
			
		\end{tikzpicture}
	\end{center}
	\caption{$\Omega_*^{\EPin}$}
	\label{fig:E_2EPin}
\end{figure}

\begin{table}[H]
	\centering
	\begin{tabular}{ c c c c}
		\hline
		\multicolumn{4}{c}{Bordism and Cobordism group}\\
		\hline
		$d$ & 
		$\Omega^{\EPin}_d$ & $\TP^{\EPin}_d$
		& Cobordism Invariant \\
		\hline
		0& $\Z_2$ & $\Z_2$ &\\
		1& $\Z_2$ & $\Z_2$ & $\tilde\eta$\\
		2& $\Z_2\times\Z_4$ & $\Z_2\times\Z_4$ & $y,\frac{\text{ABK}}{2}$ \\
		3 & $\Z_2$ & $\Z_2$ & $w_1\text{Arf}$\\
		4 & $\Z_2^2$ & $\Z_2^2$ & $w_1y\tilde\eta, \frac{\eta}{8}$  \\
		5 & $0$ & $0$&  \\
		\hline
	\end{tabular}
	\caption{Bordism group. Here $y$ is the generator of $\H^2(\B\Z_4,\Z_2)$, $w_i$ is the Stiefel-Whitney class of the tangent bundle. $\tilde\eta$ is the mod 2 index of 1d Dirac operator, Arf is the 2d Arf invariant, ABK is the 2d Arf-Brown-Kervaire invariant, $\eta$ is the 4d eta invariant. 
	Here the cohomology classes of $\B\Z_4$ are pulled back to the manifold $M$ via the map $M\to\B\EPin\to\B\rE\to\B\Z_4$.
	}
	\label{table:EPinBordism}
\end{table}

\subsection{Bordism and Cobordism Groups  of $\EPin(d)\times \B\Z_2$}
In this subsection, we will compute the cobordism group of $\EPin(d)\times \B\Z_2$.

Recall that $MT\EPin=M\Spin\wedge\Sigma^{-1}M\Z_4$. So $MT(\EPin\times\B\Z_2)=MT\EPin\wedge(\B^2\Z_2)_+=M\Spin\wedge\Sigma^{-1}M\Z_4\wedge(\B^2\Z_2)_+$.

Since there is no odd torsion, the Adams spectral sequence shows: for $t-s<8$,
\bea
\Ext_{\A_2(1)}^{s,t}(\H^{*+1}(M\Z_4,\Z_2)\otimes\H^*(\B^2\Z_2,\Z_2),\Z_2)\Rightarrow\Omega_{t-s}^{\EPin\times\B\Z_2}.
\eea

Recall that 
\bea
\H^*(\B^2\Z_2,\Z_2)=\Z_2[x_2,x_3,x_5,\dots]
\eea
where $x_2$ is the generator of $\H^2(\B^2\Z_2,\Z_2)$, $x_3=\Sq^1x_2$, $x_5=\Sq^2x_3$, $\dots$.

The $\A_2(1)$ module structure of $\H^{*+1}(M\Z_4,\Z_2)\otimes\H^*(\B^2\Z_2,\Z_2)$ is shown in Figure \ref{fig:A_2(1)MZ4B2Z2}.

The $E_2$ page of the Adams spectral sequence is shown in Figure \ref{fig:E_2EPinBZ2}.

\begin{figure}[H]
	\begin{center}
		\begin{tikzpicture}[scale=0.5]
		
		\node[right] at (0,0) {$U$};
		\draw[fill] (0,0) circle(.1);
		
		\node[right] at (0,1) {$xU$};
		\draw[fill] (0,1) circle(.1);
		
		\draw (0,0) -- (0,1);
		
		\node[right] at (0,2) {$yU$};
		\draw[fill] (0,2) circle(.1);
		
		\node[right] at (0,3) {$xyU$};
		\draw[fill] (0,3) circle(.1);
		
		\draw (0,2) -- (0,3);
		
		\node[right] at (0,4) {$y^2U$};
		\draw[fill] (0,4) circle(.1);
		
		\node[right] at (0,5) {$xy^2U$};
		\draw[fill] (0,5) circle(.1);
		
		\draw (0,4) -- (0,5);
		
		\draw (0,2) to[out=150,in=150] (0,4);
		
		\draw (0,3) to[out=30,in=30] (0,5);
		
		\node at (3,3) {$\bigotimes$};
		
		\node[below] at (6,0) {$1$};

\draw[fill] (6,0) circle(.1);

\node[below] at (6,2) {$x_2$};

\draw[fill] (6,2) circle(.1);
\draw[fill] (6,3) circle(.1);
\draw[fill] (6,4) circle(.1);
\draw (6,2) -- (6,3);
\draw (6,2) to [out=150,in=150] (6,4);
\draw[fill] (7,5) circle(.1);
\draw[fill] (7,6) circle(.1);
\draw (7,5) -- (7,6);
\draw (6,3) to [out=30,in=150] (7,5);
\draw (6,4) to [out=30,in=150] (7,6);
\draw[fill] (8,5) circle(.1);

\node[below] at (8,5) {$x_2x_3$};

\draw (8,5) -- (7,6);
\draw[fill] (8,7) circle(.1);
\draw (8,5) to [out=30,in=30] (8,7);
\draw[fill] (8,8) circle(.1);
\draw (8,7) -- (8,8);
\draw[fill] (8,10) circle(.1);
\draw (8,8) to [out=150,in=150] (8,10);

\node at (-2,-10) {$=$};

\node[right] at (0,-15) {$U$};
		\draw[fill] (0,-15) circle(.1);
		
		\node[right] at (0,-14) {$xU$};
		\draw[fill] (0,-14) circle(.1);
		
		\draw (0,-15) -- (0,-14);
		
		\node[right] at (0,-13) {$yU$};
		\draw[fill] (0,-13) circle(.1);
		
		\node[right] at (0,-12) {$xyU$};
		\draw[fill] (0,-12) circle(.1);
		
		\draw (0,-13) -- (0,-12);
		
		\node[right] at (0,-11) {$y^2U$};
		\draw[fill] (0,-11) circle(.1);
		
		\node[right] at (0,-10) {$xy^2U$};
		\draw[fill] (0,-10) circle(.1);
		
		\draw (0,-11) -- (0,-10);
		
		\draw (0,-13) to[out=150,in=150] (0,-11);
		
		\draw (0,-12) to[out=30,in=30] (0,-10);
		
		\node[below] at (6,-13) {$x_2U$};
		
		\draw[fill] (6,-13) circle(.1);
		
		 \draw[fill] (6,-12) circle(.1);
		 
		 \draw (6,-13) -- (6,-12);
		 
		 \draw[fill] (5,-11) circle(.1);
		\draw (6,-13) to[out=150,in=30] (5,-11);
		
		\draw[fill] (6,-10) circle(.1);
		\draw (6,-12) to[out=30,in=30] (6,-10);
		\draw[fill] (6,-9) circle(.1);
		\draw (6,-10) -- (6,-9);
		\draw (5,-11) to[out=30,in=150] (6,-9);
		\draw[fill] (5,-10) circle(.1);
		\draw (5,-11) -- (5,-10);
		\draw[fill] (4,-12) circle(.1);
		
		\node[below] at (4,-12) {$xx_2U$};
		
		\draw (4,-12) to[out=30,in=150] (5,-10);
		\draw[fill] (4,-11) circle(.1);
		\draw (4,-12) -- (4,-11);
		\draw[fill] (4,-9) circle(.1);
		\draw (4,-11) to[out=150,in=150] (4,-9);
		\draw[fill] (4,-8) circle(.1);
		\draw (4,-9) -- (4,-8);
		\draw (5,-10) to[out=150,in=30] (4,-8);
		
		\node[below] at (8,-10) {$x_2x_3U$};
		
		\draw[fill] (8,-10) circle(.1);
		\draw[fill] (8,-9) circle(.1);
		\draw (8,-10) -- (8,-9);
		\draw[fill] (8,-8) circle(.1);
		\draw (8,-10) to [out=150,in=150] (8,-8);
		\draw[fill] (8,-7) circle(.1);
		\draw[fill] (9,-7) circle(.1);
		\draw (8,-8) -- (8,-7);
		\draw (8,-9) to [out=30,in=150] (9,-7);
		\draw[fill] (9,-6) circle(.1);
		\draw (9,-7) -- (9,-6);
		\draw (8,-8) to [out=30,in=150] (9,-6);
		\draw[fill] (9,-5) circle(.1);
		\draw[fill] (9,-4) circle(.1);
		\draw (8,-7) to [out=30,in=150] (9,-5);
		\draw (9,-5) -- (9,-4);
		\draw (9,-6) to [out=30,in=30] (9,-4);
		
		\node[below] at (10,-11) {$yx_2U$};
		
		\draw[fill] (10,-11) circle(.1);
		\draw[fill] (10,-10) circle(.1);
		\draw (10,-11) -- (10,-10);
		\draw[fill] (10,-9) circle(.1);
		\draw (10,-11) to [out=150,in=150] (10,-9);
		\draw[fill] (10,-8) circle(.1);
		\draw[fill] (11,-8) circle(.1);
		\draw (10,-9) -- (10,-8);
		\draw (10,-10) to [out=30,in=150] (11,-8);
		\draw[fill] (11,-7) circle(.1);
		\draw (11,-8) -- (11,-7);
		\draw (10,-9) to [out=30,in=150] (11,-7);
		\draw[fill] (11,-6) circle(.1);
		\draw[fill] (11,-5) circle(.1);
		\draw (10,-8) to [out=30,in=150] (11,-6);
		\draw (11,-6) -- (11,-5);
		\draw (11,-7) to [out=30,in=30] (11,-5);

		\node[below] at (12,-10) {$xyx_2U$};
		
		\draw[fill] (12,-10) circle(.1);
		\draw[fill] (12,-9) circle(.1);
		\draw (12,-10) -- (12,-9);
		\draw[fill] (12,-8) circle(.1);
		\draw (12,-10) to [out=150,in=150] (12,-8);
		\draw[fill] (12,-7) circle(.1);
		\draw[fill] (13,-7) circle(.1);
		\draw (12,-8) -- (12,-7);
		\draw (12,-9) to [out=30,in=150] (13,-7);
		\draw[fill] (13,-6) circle(.1);
		\draw (13,-7) -- (13,-6);
		\draw (12,-8) to [out=30,in=150] (13,-6);
		\draw[fill] (13,-5) circle(.1);
		\draw[fill] (13,-4) circle(.1);
		\draw (12,-7) to [out=30,in=150] (13,-5);
		\draw (13,-5) -- (13,-4);
		\draw (13,-6) to [out=30,in=30] (13,-4);
		
		\end{tikzpicture}
	\end{center}
	\caption{The $\A_2(1)$ module structure of $\H^{*+1}(M\Z_4,\Z_2)\otimes\H^*(\B^2\Z_2,\Z_2)$.}
	\label{fig:A_2(1)MZ4B2Z2}
\end{figure}
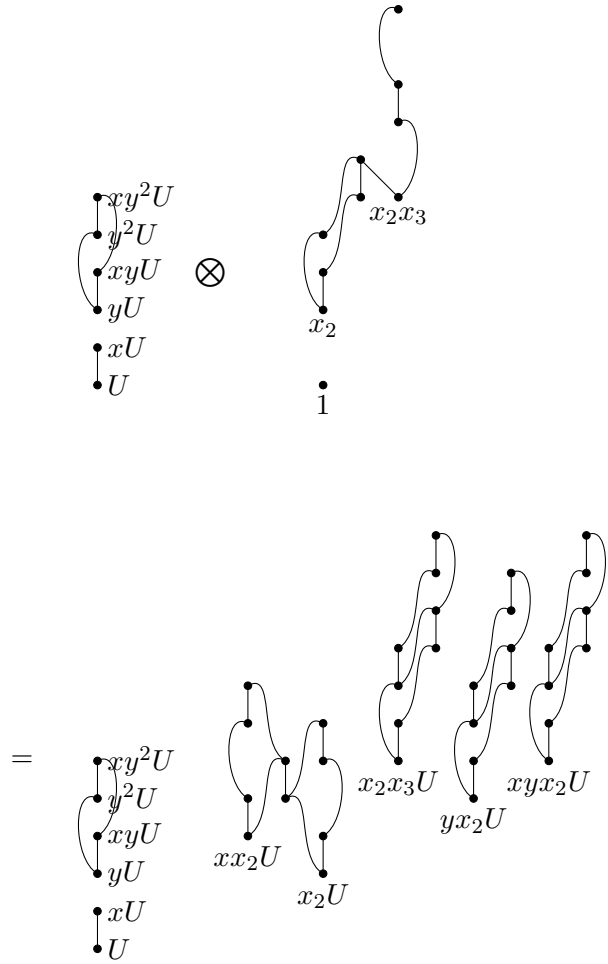

\begin{figure}[H]
	\begin{center}
		\begin{tikzpicture}
		\node at (0,-1) {0};
		\node at (1,-1) {1};
		\node at (2,-1) {2};
		\node at (3,-1) {3};
		\node at (4,-1) {4};
		\node at (5,-1) {5};
		\node at (6,-1) {$t-s$};
		\node at (-1,0) {0};
		\node at (-1,1) {1};
		\node at (-1,2) {2};
		\node at (-1,3) {3};
		\node at (-1,4) {4};
		\node at (-1,5) {5};
		\node at (-1,6) {$s$};
		
		\draw[->] (-0.5,-0.5) -- (-0.5,6);
		\draw[->] (-0.5,-0.5) -- (6,-0.5);

		\draw (0,0) -- (2,2);
		\draw (2,1) -- (2,2);
		\draw (2,1) -- (4,3);
		\draw[fill] (2,0) circle(.05);
		\draw[fill] (4,1) circle(.05);
		\draw[fill] (2.1,0) circle(.05);
		\draw[fill] (4.1,1) circle(.05);
		\draw[fill] (3,0) circle(.05);
		\draw[fill] (4,0) circle(.05);
		\draw[fill] (5,0) circle(.05);
		\draw[fill] (5.1,0) circle(.05);
		\end{tikzpicture}
	\end{center}
	\caption{$\Omega_*^{\EPin\times\B\Z_2}$}
	\label{fig:E_2EPinBZ2}
\end{figure}
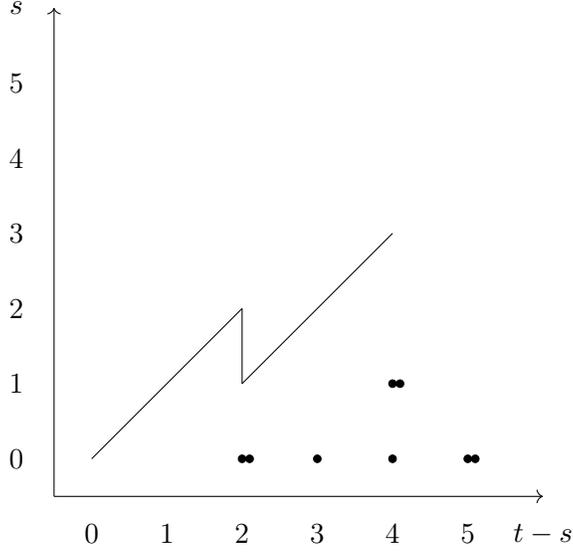

\begin{table}[H]
	\centering
	\begin{tabular}{ c c c c}
		\hline
		\multicolumn{4}{c}{Bordism and Cobordism group}\\
		\hline
		$d$ & 
		$\Omega^{\EPin\times\B\Z_2}_d$ & $\TP^{\EPin\times\B\Z_2}_d$
		& Cobordism Invariant \\
		\hline
		0& $\Z_2$ & $\Z_2$ &\\
		1& $\Z_2$ & $\Z_2$ & $\tilde\eta$\\
		2& $\Z_2^2\times\Z_4$ & $\Z_2^2\times\Z_4$ & $x_2,y,\frac{\text{ABK}}{2}$ \\
		3 & $\Z_2^2$ & $\Z_2^2$ & $w_1x_2,w_1\text{Arf}$\\
		4 & $\Z_2^4$ & $\Z_2^4$ & $yx_2,\frac{\cP_2(x_2)}{2},w_1y\tilde\eta, \frac{\eta}{8}$  \\
		5 & $\Z_2^2$ & $\Z_2^2$&  $x_2x_3,w_1yx_2$ \\
		\hline
	\end{tabular}
	\caption{Bordism group. Here $y$ is the generator of $\H^2(\B\Z_4,\Z_2)$, $w_i$ is the Stiefel-Whitney class of the tangent bundle. $\tilde\eta$ is the mod 2 index of 1d Dirac operator, Arf is the 2d Arf invariant, ABK is the 2d Arf-Brown-Kervaire invariant, $\eta$ is the 4d eta invariant. $x_2$ is the generator of $\H^2(\B^2\Z_2,\Z_2)$, $x_3=\Sq^1x_2$, $\cP_2(x_2)$ is the Pontryagin square of $x_2$. By Wu formula, $\cP_2(x_2)=x_2^2=\Sq^2(x_2)=(w_2+w_1^2)x_2=0\mod2$.
	Here the cohomology classes of $\B\Z_4$ are pulled back to the manifold $M$ via the map $M\to\B\EPin\to\B\rE\to\B\Z_4$.
	}
	\label{table:EPinBZ2Bordism}
\end{table}

\section{Bordism and Cobordism Groups involving $\DPin(d)$}
\label{app3}

\subsection{Bordism and Cobordism Groups of $\DPin(d)$}

In \cite{Kaidi:2019tyf}, the authors consider another group extension 
\bea
1\to\Z_2^+\times\Z_2^-\to \DPin(d)\to\O(d)\to1
\eea
where the orientation reversal in $\O(d)$ acts on $\Z_2^+\times \Z_2^-$ by exchanging the two $\Z_2$ factors.
The bordism groups $\Omega_d^{\DPin}$ for $d\le6$ are also computed in \cite{Kaidi:2019tyf}.

According to \cite{Kaidi:2019tyf}, $MT\DPin=M\Spin\wedge\Sigma^1MT\O(1)\wedge \Sigma^{-1}M\O(1)$. In this subsection, we will reproduce the result of bordism groups $\Omega_d^{\DPin}$ in \cite{Kaidi:2019tyf}.

Since there is no odd torsion, the Adams spectral sequence shows: for $t-s<8$,
\bea
\Ext_{\A_2(1)}^{s,t}(\H^{*-1}(MT\O(1),\Z_2)\otimes\H^{*+1}(M\O(1),\Z_2),\Z_2)\Rightarrow\Omega_{t-s}^{\DPin}.
\eea

By Thom's isomorphism, 
\bea
\H^{*-1}(MT\O(1),\Z_2)=\Z_2[a]U
\eea
where $U$ is the Thom class of the virtual bundle $-E_1$ over $\B \O(1)$, $E_1$ is the universal 1-bundle over $\B \O(1)$ and $a$ is the 1st Stiefel-Whitney class of $E_1$ over $\B \O(1)$.

Also by Thom's isomorphism, 
\bea
\H^{*+1}(M\O(1),\Z_2)=\Z_2[w]V
\eea
where $V$ is the Thom class of the universal 1-bundle $E_1'$ over $\B \O(1)$ and $w$ is the 1st Stiefel-Whitney class of $E_1'$ over $\B \O(1)$.

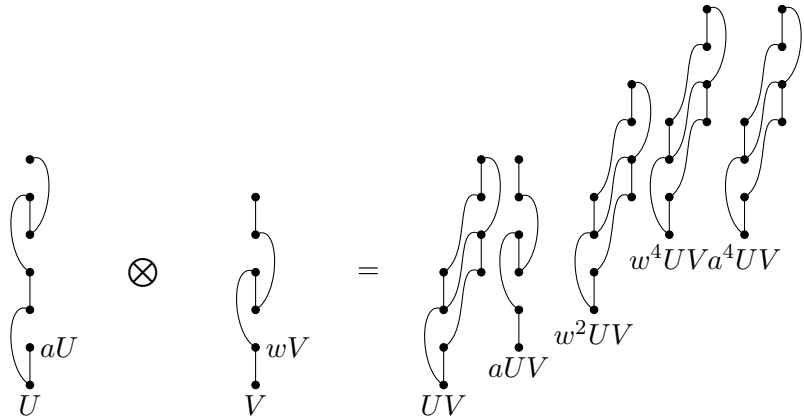
\begin{figure}[H]
	\begin{center}
		\begin{tikzpicture}[scale=0.5]
		
		\node[below] at (0,0) {$U$};
		\node[right] at (0,1) {$aU$};

\draw[fill] (0,0) circle(.1);
\draw[fill] (0,1) circle(.1);
\draw (0,0) -- (0,1);
\draw[fill] (0,2) circle(.1);
\draw (0,0) to [out=150,in=150] (0,2);
\draw[fill] (0,3) circle(.1);
\draw (0,2) -- (0,3);
\draw[fill] (0,4) circle(.1);
\draw[fill] (0,5) circle(.1);
\draw (0,4) -- (0,5);
\draw (0,3) to [out=150,in=150] (0,5);
\draw[fill] (0,6) circle(.1);
\draw (0,4) to [out=30,in=30] (0,6);

\node at (3,3) {$\bigotimes$};

\node[below] at (6,0) {$V$};
\node[right] at (6,1) {$wV$};

\draw[fill] (6,0) circle(.1);
\draw[fill] (6,1) circle(.1);
\draw (6,0) -- (6,1);
\draw[fill] (6,2) circle(.1);
\draw[fill] (6,3) circle(.1);
\draw (6,2) -- (6,3);
\draw (6,1) to [out=150,in=150] (6,3);
\draw[fill] (6,4) circle(.1);
\draw (6,2) to [out=30,in=30] (6,4);
\draw[fill] (6,5) circle(.1);
\draw (6,4) -- (6,5);

\def\x{11}
\def\y{15}

\node at (-2+\x,-10+13) {$=$};

\node[below] at (0+\x,-15+\y) {$UV$};
		
		\draw[fill] (0+\x,-15+\y) circle(.1);
		\draw[fill] (0+\x,-14+\y) circle(.1);
		\draw (0+\x,-15+\y) -- (0+\x,-14+\y);
		\draw[fill] (0+\x,-13+\y) circle(.1);
		\draw (0+\x,-15+\y) to [out=150,in=150] (0+\x,-13+\y);
		\draw[fill] (0+\x,-12+\y) circle(.1);
		\draw[fill] (1+\x,-12+\y) circle(.1);
		\draw (0+\x,-13+\y) -- (0+\x,-12+\y);
		\draw (0+\x,-14+\y) to [out=30,in=150] (1+\x,-12+\y);
		\draw[fill] (1+\x,-11+\y) circle(.1);
		\draw (1+\x,-12+\y) -- (1+\x,-11+\y);
		\draw (0+\x,-13+\y) to [out=30,in=150] (1+\x,-11+\y);
		\draw[fill] (1+\x,-10+\y) circle(.1);
		\draw[fill] (1+\x,-9+\y) circle(.1);
		\draw (0+\x,-12+\y) to [out=30,in=150] (1+\x,-10+\y);
		\draw (1+\x,-10+\y) -- (1+\x,-9+\y);
		\draw (1+\x,-11+\y) to [out=30,in=30] (1+\x,-9+\y);

\node[below] at (2+\x,-14+\y) {$aUV$};

\draw[fill] (2+\x,-14+\y) circle(.1);
\draw[fill] (2+\x,-13+\y) circle(.1);
\draw (2+\x,-14+\y) -- (2+\x,-13+\y);
\draw[fill] (2+\x,-12+\y) circle(.1);
\draw[fill] (2+\x,-11+\y) circle(.1);
\draw (2+\x,-12+\y) -- (2+\x,-11+\y);
\draw (2+\x,-13+\y) to [out=150,in=150] (2+\x,-11+\y);
\draw[fill] (2+\x,-10+\y) circle(.1);
\draw (2+\x,-12+\y) to [out=30,in=30] (2+\x,-10+\y);
\draw[fill] (2+\x,-9+\y) circle(.1);
\draw (2+\x,-10+\y) -- (2+\x,-9+\y);

\node[below] at (4+\x,-13+\y) {$w^2UV$};
		
		\draw[fill] (4+\x,-13+\y) circle(.1);
		\draw[fill] (4+\x,-12+\y) circle(.1);
		\draw (4+\x,-13+\y) -- (4+\x,-12+\y);
		\draw[fill] (4+\x,-11+\y) circle(.1);
		\draw (4+\x,-13+\y) to [out=150,in=150] (4+\x,-11+\y);
		\draw[fill] (4+\x,-10+\y) circle(.1);
		\draw[fill] (5+\x,-10+\y) circle(.1);
		\draw (4+\x,-11+\y) -- (4+\x,-10+\y);
		\draw (4+\x,-12+\y) to [out=30,in=150] (5+\x,-10+\y);
		\draw[fill] (5+\x,-9+\y) circle(.1);
		\draw (5+\x,-10+\y) -- (5+\x,-9+\y);
		\draw (4+\x,-11+\y) to [out=30,in=150] (5+\x,-9+\y);
		\draw[fill] (5+\x,-8+\y) circle(.1);
		\draw[fill] (5+\x,-7+\y) circle(.1);
		\draw (4+\x,-10+\y) to [out=30,in=150] (5+\x,-8+\y);
		\draw (5+\x,-8+\y) -- (5+\x,-7+\y);
		\draw (5+\x,-9+\y) to [out=30,in=30] (5+\x,-7+\y);

\node[below] at (6+\x,-11+\y) {$w^4UV$};
		
		\draw[fill] (6+\x,-11+\y) circle(.1);
		\draw[fill] (6+\x,-10+\y) circle(.1);
		\draw (6+\x,-11+\y) -- (6+\x,-10+\y);
		\draw[fill] (6+\x,-9+\y) circle(.1);
		\draw (6+\x,-11+\y) to [out=150,in=150] (6+\x,-9+\y);
		\draw[fill] (6+\x,-8+\y) circle(.1);
		\draw[fill] (7+\x,-8+\y) circle(.1);
		\draw (6+\x,-9+\y) -- (6+\x,-8+\y);
		\draw (6+\x,-10+\y) to [out=30,in=150] (7+\x,-8+\y);
		\draw[fill] (7+\x,-7+\y) circle(.1);
		\draw (7+\x,-8+\y) -- (7+\x,-7+\y);
		\draw (6+\x,-9+\y) to [out=30,in=150] (7+\x,-7+\y);
		\draw[fill] (7+\x,-6+\y) circle(.1);
		\draw[fill] (7+\x,-5+\y) circle(.1);
		\draw (6+\x,-8+\y) to [out=30,in=150] (7+\x,-6+\y);
		\draw (7+\x,-6+\y) -- (7+\x,-5+\y);
		\draw (7+\x,-7+\y) to [out=30,in=30] (7+\x,-5+\y);

		\node[below] at (8+\x,-11+\y) {$a^4UV$};
		
		\draw[fill] (8+\x,-11+\y) circle(.1);
		\draw[fill] (8+\x,-10+\y) circle(.1);
		\draw (8+\x,-11+\y) -- (8+\x,-10+\y);
		\draw[fill] (8+\x,-9+\y) circle(.1);
		\draw (8+\x,-11+\y) to [out=150,in=150] (8+\x,-9+\y);
		\draw[fill] (8+\x,-8+\y) circle(.1);
		\draw[fill] (9+\x,-8+\y) circle(.1);
		\draw (8+\x,-9+\y) -- (8+\x,-8+\y);
		\draw (8+\x,-10+\y) to [out=30,in=150] (9+\x,-8+\y);
		\draw[fill] (9+\x,-7+\y) circle(.1);
		\draw (9+\x,-8+\y) -- (9+\x,-7+\y);
		\draw (8+\x,-9+\y) to [out=30,in=150] (9+\x,-7+\y);
		\draw[fill] (9+\x,-6+\y) circle(.1);
		\draw[fill] (9+\x,-5+\y) circle(.1);
		\draw (8+\x,-8+\y) to [out=30,in=150] (9+\x,-6+\y);
		\draw (9+\x,-6+\y) -- (9+\x,-5+\y);
		\draw (9+\x,-7+\y) to [out=30,in=30] (9+\x,-5+\y);

		\end{tikzpicture}
	\end{center}
	\caption{The $\A_2(1)$ module structure of $\H^{*-1}(MT\O(1),\Z_2)\otimes\H^{*+1}(M\O(1),\Z_2)$.}
	\label{fig:A_2(1)MTO1MO1}
\end{figure}

The $\A_2(1)$ module structure of $\H^{*-1}(MT\O(1),\Z_2)\otimes\H^{*+1}(M\O(1),\Z_2)$ is shown in Figure \ref{fig:A_2(1)MTO1MO1}, which agrees with the Figure 6 in \cite{Kaidi:2019tyf}.

\begin{figure}[H]
	\begin{center}
		\begin{tikzpicture}
		\node at (0,-1) {0};
		\node at (1,-1) {1};
		\node at (2,-1) {2};
		\node at (3,-1) {3};
		\node at (4,-1) {4};
		\node at (5,-1) {5};
		\node at (6,-1) {$t-s$};
		\node at (-1,0) {0};
		\node at (-1,1) {1};
		\node at (-1,2) {2};
		\node at (-1,3) {3};
		\node at (-1,4) {4};
		\node at (-1,5) {5};
		\node at (-1,6) {$s$};
		
		\draw[->] (-0.5,-0.5) -- (-0.5,6);
		\draw[->] (-0.5,-0.5) -- (6,-0.5);
		
		\draw[fill] (0,0) circle(.05);
	\draw (1,0) -- (3,2);
	\draw (3,0) -- (3,2);
	\draw[fill] (2,0) circle(.05);
	\draw[fill] (4,0) circle(.05);	
	\draw[fill] (4.1,0) circle(.05);	
		\end{tikzpicture}
	\end{center}
	\caption{$\Omega_*^{\DPin}$}
	\label{fig:E_2DPin}
\end{figure}
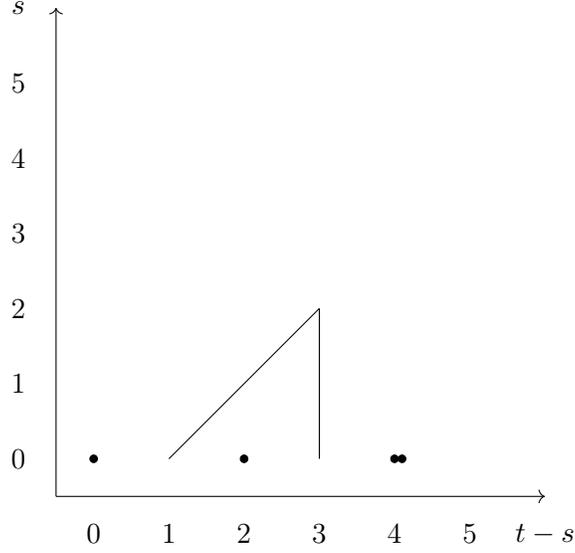

The $E_2$ page is shown in Figure \ref{fig:E_2DPin}.

\begin{table}[H]
	\centering
	\begin{tabular}{ c c c c}
		\hline
		\multicolumn{4}{c}{Bordism and Cobordism group}\\
		\hline
		$d$ & 
		$\Omega^{\DPin}_d$ & $\TP^{\DPin}_d$
		& Cobordism Invariant \\
		\hline
		0& $\Z_2$ & $\Z_2$ &\\
		1& $\Z_2$ & $\Z_2$ & $a$\\
		2& $\Z_2^2$ & $\Z_2^2$ & $w^2,a\tilde\eta$ \\
		3 & $\Z_8$ & $\Z_8$ & $a\text{ABK}$\\
		4 & $\Z_2^2$ & $\Z_2^2$ & $w^4,a^4$  \\
		5 & $0$ & $0$&   \\
		\hline
	\end{tabular}
	\caption{Bordism group. Here $a$ and $w$ are explained before. $\tilde\eta$ is the mod 2 index of 1d Dirac operator, ABK is the 2d Arf-Brown-Kervaire invariant.
	}
	\label{table:DPinBordism}
\end{table}

\subsection{Bordism and Cobordism Groups of $\DPin(d)\times\B\Z_2$}

In this subsection, we will compute the cobordism group of $\DPin(d)\times \B\Z_2$.

Recall that $MT\DPin=M\Spin\wedge\Sigma^{1}MT\O(1)\wedge\Sigma^{-1}M\O(1)$. So $MT(\DPin\times\B\Z_2)=MT\DPin\wedge(\B^2\Z_2)_+=M\Spin\wedge\Sigma^{1}MT\O(1)\wedge\Sigma^{-1}M\O(1)\wedge(\B^2\Z_2)_+$.

Since there is no odd torsion, the Adams spectral sequence shows: for $t-s<8$,
\bea
\Ext_{\A_2(1)}^{s,t}(\H^{*-1}(MT\O(1),\Z_2)\otimes\H^{*+1}(M\O(1),\Z_2)\otimes\H^*(\B^2\Z_2,\Z_2),\Z_2)\Rightarrow\Omega_{t-s}^{\DPin\times\B\Z_2}.
\eea

Recall that 
\bea
\H^*(\B^2\Z_2,\Z_2)=\Z_2[x_2,x_3,x_5,\dots]
\eea
where $x_2$ is the generator of $\H^2(\B^2\Z_2,\Z_2)$, $x_3=\Sq^1x_2$, $x_5=\Sq^2x_3$, $\dots$.

\begin{figure}[H]
	\begin{center}
		\begin{tikzpicture}[scale=0.5]
		
		\node[below] at (0,0) {$UV$};
		
		\draw[fill] (0,0) circle(.1);
		\draw[fill] (0,1) circle(.1);
		\draw (0,0) -- (0,1);
		\draw[fill] (0,2) circle(.1);
		\draw (0,0) to [out=150,in=150] (0,2);
		\draw[fill] (0,3) circle(.1);
		\draw[fill] (1,3) circle(.1);
		\draw (0,2) -- (0,3);
		\draw (0,1) to [out=30,in=150] (1,3);
		\draw[fill] (1,4) circle(.1);
		\draw (1,3) -- (1,4);
		\draw (0,2) to [out=30,in=150] (1,4);
		\draw[fill] (1,5) circle(.1);
		\draw[fill] (1,6) circle(.1);
		\draw (0,3) to [out=30,in=150] (1,5);
		\draw (1,5) -- (1,6);
		\draw (1,4) to [out=30,in=30] (1,6);

\node[below] at (2,1) {$aUV$};

\draw[fill] (2,1) circle(.1);
\draw[fill] (2,2) circle(.1);
\draw (2,1) -- (2,2);
\draw[fill] (2,3) circle(.1);
\draw[fill] (2,4) circle(.1);
\draw (2,3) -- (2,4);
\draw (2,2) to [out=150,in=150] (2,4);
\draw[fill] (2,5) circle(.1);
\draw (2,3) to [out=30,in=30] (2,5);
\draw[fill] (2,6) circle(.1);
\draw (2,5) -- (2,6);

\node[below] at (4,2) {$w^2UV$};
		
		\draw[fill] (4,2) circle(.1);
		\draw[fill] (4,3) circle(.1);
		\draw (4,2) -- (4,3);
		\draw[fill] (4,4) circle(.1);
		\draw (4,2) to [out=150,in=150] (4,4);
		\draw[fill] (4,5) circle(.1);
		\draw[fill] (5,5) circle(.1);
		\draw (4,4) -- (4,5);
		\draw (4,3) to [out=30,in=150] (5,5);
		\draw[fill] (5,6) circle(.1);
		\draw (5,5) -- (5,6);
		\draw (4,4) to [out=30,in=150] (5,6);
		\draw[fill] (5,7) circle(.1);
		\draw[fill] (5,8) circle(.1);
		\draw (4,5) to [out=30,in=150] (5,7);
		\draw (5,7) -- (5,8);
		\draw (5,6) to [out=30,in=30] (5,8);

\node[below] at (6,4) {$w^4UV$};
		
		\draw[fill] (6,4) circle(.1);
		\draw[fill] (6,5) circle(.1);
		\draw (6,4) -- (6,5);
		\draw[fill] (6,6) circle(.1);
		\draw (6,4) to [out=150,in=150] (6,6);
		\draw[fill] (6,7) circle(.1);
		\draw[fill] (7,7) circle(.1);
		\draw (6,6) -- (6,7);
		\draw (6,5) to [out=30,in=150] (7,7);
		\draw[fill] (7,8) circle(.1);
		\draw (7,7) -- (7,8);
		\draw (6,6) to [out=30,in=150] (7,8);
		\draw[fill] (7,9) circle(.1);
		\draw[fill] (7,10) circle(.1);
		\draw (6,7) to [out=30,in=150] (7,9);
		\draw (7,9) -- (7,10);
		\draw (7,8) to [out=30,in=30] (7,10);

		\node[below] at (8,4) {$a^4UV$};
		
		\draw[fill] (8,4) circle(.1);
		\draw[fill] (8,5) circle(.1);
		\draw (8,4) -- (8,5);
		\draw[fill] (8,6) circle(.1);
		\draw (8,4) to [out=150,in=150] (8,6);
		\draw[fill] (8,7) circle(.1);
		\draw[fill] (9,7) circle(.1);
		\draw (8,6) -- (8,7);
		\draw (8,5) to [out=30,in=150] (9,7);
		\draw[fill] (9,8) circle(.1);
		\draw (9,7) -- (9,8);
		\draw (8,6) to [out=30,in=150] (9,8);
		\draw[fill] (9,9) circle(.1);
		\draw[fill] (9,10) circle(.1);
		\draw (8,7) to [out=30,in=150] (9,9);
		\draw (9,9) -- (9,10);
		\draw (9,8) to [out=30,in=30] (9,10);

		\node at (10,3) {$\bigotimes$};
		
		\node[below] at (12,0) {$1$};

\draw[fill] (12,0) circle(.1);

\node[below] at (12,2) {$x_2$};

\draw[fill] (12,2) circle(.1);
\draw[fill] (12,3) circle(.1);
\draw[fill] (12,4) circle(.1);
\draw (12,2) -- (12,3);
\draw (12,2) to [out=150,in=150] (12,4);
\draw[fill] (13,5) circle(.1);
\draw[fill] (13,6) circle(.1);
\draw (13,5) -- (13,6);
\draw (12,3) to [out=30,in=150] (13,5);
\draw (12,4) to [out=30,in=150] (13,6);
\draw[fill] (14,5) circle(.1);

\node[below] at (14,5) {$x_2x_3$};

\draw (14,5) -- (13,6);
\draw[fill] (14,7) circle(.1);
\draw (14,5) to [out=30,in=30] (14,7);
\draw[fill] (14,8) circle(.1);
\draw (14,7) -- (14,8);
\draw[fill] (14,10) circle(.1);
\draw (14,8) to [out=150,in=150] (14,10);

\node at (-2,-10) {$=$};

\node[below] at (0,-15) {$UV$};
		
		\draw[fill] (0,-15) circle(.1);
		\draw[fill] (0,-14) circle(.1);
		\draw (0,-15) -- (0,-14);
		\draw[fill] (0,-13) circle(.1);
		\draw (0,-15) to [out=150,in=150] (0,-13);
		\draw[fill] (0,-12) circle(.1);
		\draw[fill] (1,-12) circle(.1);
		\draw (0,-13) -- (0,-12);
		\draw (0,-14) to [out=30,in=150] (1,-12);
		\draw[fill] (1,-11) circle(.1);
		\draw (1,-12) -- (1,-11);
		\draw (0,-13) to [out=30,in=150] (1,-11);
		\draw[fill] (1,-10) circle(.1);
		\draw[fill] (1,-9) circle(.1);
		\draw (0,-12) to [out=30,in=150] (1,-10);
		\draw (1,-10) -- (1,-9);
		\draw (1,-11) to [out=30,in=30] (1,-9);

\node[below] at (2,-14) {$aUV$};

\draw[fill] (2,-14) circle(.1);
\draw[fill] (2,-13) circle(.1);
\draw (2,-14) -- (2,-13);
\draw[fill] (2,-12) circle(.1);
\draw[fill] (2,-11) circle(.1);
\draw (2,-12) -- (2,-11);
\draw (2,-13) to [out=150,in=150] (2,-11);
\draw[fill] (2,-10) circle(.1);
\draw (2,-12) to [out=30,in=30] (2,-10);
\draw[fill] (2,-9) circle(.1);
\draw (2,-10) -- (2,-9);

\node[below] at (4,-13) {$w^2UV$};
		
		\draw[fill] (4,-13) circle(.1);
		\draw[fill] (4,-12) circle(.1);
		\draw (4,-13) -- (4,-12);
		\draw[fill] (4,-11) circle(.1);
		\draw (4,-13) to [out=150,in=150] (4,-11);
		\draw[fill] (4,-10) circle(.1);
		\draw[fill] (5,-10) circle(.1);
		\draw (4,-11) -- (4,-10);
		\draw (4,-12) to [out=30,in=150] (5,-10);
		\draw[fill] (5,-9) circle(.1);
		\draw (5,-10) -- (5,-9);
		\draw (4,-11) to [out=30,in=150] (5,-9);
		\draw[fill] (5,-8) circle(.1);
		\draw[fill] (5,-7) circle(.1);
		\draw (4,-10) to [out=30,in=150] (5,-8);
		\draw (5,-8) -- (5,-7);
		\draw (5,-9) to [out=30,in=30] (5,-7);

\node[below] at (6,-11) {$w^4UV$};
		
		\draw[fill] (6,-11) circle(.1);
		\draw[fill] (6,-10) circle(.1);
		\draw (6,-11) -- (6,-10);
		\draw[fill] (6,-9) circle(.1);
		\draw (6,-11) to [out=150,in=150] (6,-9);
		\draw[fill] (6,-8) circle(.1);
		\draw[fill] (7,-8) circle(.1);
		\draw (6,-9) -- (6,-8);
		\draw (6,-10) to [out=30,in=150] (7,-8);
		\draw[fill] (7,-7) circle(.1);
		\draw (7,-8) -- (7,-7);
		\draw (6,-9) to [out=30,in=150] (7,-7);
		\draw[fill] (7,-6) circle(.1);
		\draw[fill] (7,-5) circle(.1);
		\draw (6,-8) to [out=30,in=150] (7,-6);
		\draw (7,-6) -- (7,-5);
		\draw (7,-7) to [out=30,in=30] (7,-5);

		\node[below] at (8,-11) {$a^4UV$};
		
		\draw[fill] (8,-11) circle(.1);
		\draw[fill] (8,-10) circle(.1);
		\draw (8,-11) -- (8,-10);
		\draw[fill] (8,-9) circle(.1);
		\draw (8,-11) to [out=150,in=150] (8,-9);
		\draw[fill] (8,-8) circle(.1);
		\draw[fill] (9,-8) circle(.1);
		\draw (8,-9) -- (8,-8);
		\draw (8,-10) to [out=30,in=150] (9,-8);
		\draw[fill] (9,-7) circle(.1);
		\draw (9,-8) -- (9,-7);
		\draw (8,-9) to [out=30,in=150] (9,-7);
		\draw[fill] (9,-6) circle(.1);
		\draw[fill] (9,-5) circle(.1);
		\draw (8,-8) to [out=30,in=150] (9,-6);
		\draw (9,-6) -- (9,-5);
		\draw (9,-7) to [out=30,in=30] (9,-5);

		\node[below] at (10,-12) {$ax_2UV$};

\draw[fill] (10,-12) circle(.1);
\draw[fill] (10,-11) circle(.1);
\draw[fill] (10,-10) circle(.1);
\draw (10,-12) -- (10,-11);
\draw (10,-12) to [out=150,in=150] (10,-10);
\draw[fill] (11,-9) circle(.1);
\draw[fill] (11,-8) circle(.1);
\draw (11,-9) -- (11,-8);
\draw (10,-11) to [out=30,in=150] (11,-9);
\draw (10,-10) to [out=30,in=150] (11,-8);
\draw[fill] (10,-9) circle(.1);
\draw (10,-10) -- (10,-9);
\draw[fill] (10,-8) circle(.1);
\draw[fill] (10,-7) circle(.1);
\draw (10,-9) to [out=30,in=30] (10,-7);
\draw (10,-8) -- (10,-7);
\draw[fill] (10,-6) circle(.1);
\draw (10,-8) to [out=150,in=150] (10,-6);

\node[below] at (12,-11) {$ax_3UV$};
 
\draw[fill] (12,-11) circle(.1);
\draw[fill] (12,-10) circle(.1);
\draw (12,-11) -- (12,-10);
\draw[fill] (12,-9) circle(.1);
\draw (12,-11) to [out=150,in=150] (12,-9);
\draw[fill] (13,-8) circle(.1);
\draw (12,-10) to [out=30,in=150] (13,-8);
\draw[fill] (12,-8) circle(.1);
\draw (12,-9) -- (12,-8);
\draw[fill] (13,-7) circle(.1);
\draw (13,-8) -- (13,-7);
\draw (12,-9) to [out=30,in=150] (13,-7);
\draw[fill] (13,-6) circle(.1);
\draw (12,-8) to [out=30,in=150] (13,-6);
\draw[fill] (13,-5) circle(.1);
\draw (13,-6) -- (13,-5);
\draw (13,-7) to [out=30,in=30] (13,-5); 

\node[below] at (14,-10) {$a^3x_2UV$};
 
\draw[fill] (14,-10) circle(.1);
\draw[fill] (14,-9) circle(.1);
\draw (14,-10) -- (14,-9);
\draw[fill] (14,-8) circle(.1);
\draw (14,-10) to [out=150,in=150] (14,-8);
\draw[fill] (15,-7) circle(.1);
\draw (14,-9) to [out=30,in=150] (15,-7);
\draw[fill] (14,-7) circle(.1);
\draw (14,-8) -- (14,-7);
\draw[fill] (15,-6) circle(.1);
\draw (15,-7) -- (15,-6);
\draw (14,-8) to [out=30,in=150] (15,-6);
\draw[fill] (15,-5) circle(.1);
\draw (14,-7) to [out=30,in=150] (15,-5);
\draw[fill] (15,-4) circle(.1);
\draw (15,-5) -- (15,-4);
\draw (15,-6) to [out=30,in=30] (15,-4);

\node[below] at (0,-26) {$x_2UV$};

\draw[fill] (0,-26) circle(.1);
		\draw[fill] (0,-25) circle(.1);
		\draw (0,-26) -- (0,-25);
		\draw[fill] (0,-24) circle(.1);
		\draw (0,-26) to [out=150,in=150] (0,-24);
		\draw[fill] (0,-23) circle(.1);
		\draw[fill] (1,-23) circle(.1);
		\draw (0,-24) -- (0,-23);
		\draw (0,-25) to [out=30,in=150] (1,-23);
		\draw[fill] (1,-22) circle(.1);
		\draw (1,-23) -- (1,-22);
		\draw (0,-24) to [out=30,in=150] (1,-22);
		\draw[fill] (1,-21) circle(.1);
		\draw[fill] (1,-20) circle(.1);
		\draw (0,-23) to [out=30,in=150] (1,-21);
		\draw (1,-21) -- (1,-20);
		\draw (1,-22) to [out=30,in=30] (1,-20);
		
		\node[below] at (2,-25) {$x_3UV$};

\draw[fill] (2,-25) circle(.1);
		\draw[fill] (2,-24) circle(.1);
		\draw (2,-25) -- (2,-24);
		\draw[fill] (2,-23) circle(.1);
		\draw (2,-25) to [out=150,in=150] (2,-23);
		\draw[fill] (2,-22) circle(.1);
		\draw[fill] (3,-22) circle(.1);
		\draw (2,-23) -- (2,-22);
		\draw (2,-24) to [out=30,in=150] (3,-22);
		\draw[fill] (3,-21) circle(.1);
		\draw (3,-22) -- (3,-21);
		\draw (2,-23) to [out=30,in=150] (3,-21);
		\draw[fill] (3,-20) circle(.1);
		\draw[fill] (3,-19) circle(.1);
		\draw (2,-22) to [out=30,in=150] (3,-20);
		\draw (3,-20) -- (3,-19);
		\draw (3,-21) to [out=30,in=30] (3,-19);

		\node[below] at (4,-24) {$x_2^2UV$};

\draw[fill] (4,-24) circle(.1);
		\draw[fill] (4,-23) circle(.1);
		\draw (4,-24) -- (4,-23);
		\draw[fill] (4,-22) circle(.1);
		\draw (4,-24) to [out=150,in=150] (4,-22);
		\draw[fill] (4,-21) circle(.1);
		\draw[fill] (5,-21) circle(.1);
		\draw (4,-22) -- (4,-21);
		\draw (4,-23) to [out=30,in=150] (5,-21);
		\draw[fill] (5,-20) circle(.1);
		\draw (5,-21) -- (5,-20);
		\draw (4,-22) to [out=30,in=150] (5,-20);
		\draw[fill] (5,-19) circle(.1);
		\draw[fill] (5,-18) circle(.1);
		\draw (4,-21) to [out=30,in=150] (5,-19);
		\draw (5,-19) -- (5,-18);
		\draw (5,-20) to [out=30,in=30] (5,-18);
		
		\node[below] at (6,-23) {$x_2x_3UV$};

\draw[fill] (6,-23) circle(.1);
		\draw[fill] (6,-22) circle(.1);
		\draw (6,-23) -- (6,-22);
		\draw[fill] (6,-21) circle(.1);
		\draw (6,-23) to [out=150,in=150] (6,-21);
		\draw[fill] (6,-20) circle(.1);
		\draw[fill] (7,-20) circle(.1);
		\draw (6,-21) -- (6,-20);
		\draw (6,-22) to [out=30,in=150] (7,-20);
		\draw[fill] (7,-19) circle(.1);
		\draw (7,-20) -- (7,-19);
		\draw (6,-21) to [out=30,in=150] (7,-19);
		\draw[fill] (7,-18) circle(.1);
		\draw[fill] (7,-17) circle(.1);
		\draw (6,-20) to [out=30,in=150] (7,-18);
		\draw (7,-18) -- (7,-17);
		\draw (7,-19) to [out=30,in=30] (7,-17);

\node[below] at (8+.5,-23) {$x_5UV$};

\draw[fill] (8,-23) circle(.1);
		\draw[fill] (8,-22) circle(.1);
		\draw (8,-23) -- (8,-22);
		\draw[fill] (8,-21) circle(.1);
		\draw (8,-23) to [out=150,in=150] (8,-21);
		\draw[fill] (8,-20) circle(.1);
		\draw[fill] (9,-20) circle(.1);
		\draw (8,-21) -- (8,-20);
		\draw (8,-22) to [out=30,in=150] (9,-20);
		\draw[fill] (9,-19) circle(.1);
		\draw (9,-20) -- (9,-19);
		\draw (8,-21) to [out=30,in=150] (9,-19);
		\draw[fill] (9,-18) circle(.1);
		\draw[fill] (9,-17) circle(.1);
		\draw (8,-20) to [out=30,in=150] (9,-18);
		\draw (9,-18) -- (9,-17);
		\draw (9,-19) to [out=30,in=30] (9,-17);

		\node[below] at (10,-24) {$w^2x_2UV$};

\draw[fill] (10,-24) circle(.1);
		\draw[fill] (10,-23) circle(.1);
		\draw (10,-24) -- (10,-23);
		\draw[fill] (10,-22) circle(.1);
		\draw (10,-24) to [out=150,in=150] (10,-22);
		\draw[fill] (10,-21) circle(.1);
		\draw[fill] (11,-21) circle(.1);
		\draw (10,-22) -- (10,-21);
		\draw (10,-23) to [out=30,in=150] (11,-21);
		\draw[fill] (11,-20) circle(.1);
		\draw (11,-21) -- (11,-20);
		\draw (10,-22) to [out=30,in=150] (11,-20);
		\draw[fill] (11,-19) circle(.1);
		\draw[fill] (11,-18) circle(.1);
		\draw (10,-21) to [out=30,in=150] (11,-19);
		\draw (11,-19) -- (11,-18);
		\draw (11,-20) to [out=30,in=30] (11,-18);
		
		\node[below] at (12,-23) {$w^2x_3UV$};

\draw[fill] (12,-23) circle(.1);
		\draw[fill] (12,-22) circle(.1);
		\draw (12,-23) -- (12,-22);
		\draw[fill] (12,-21) circle(.1);
		\draw (12,-23) to [out=150,in=150] (12,-21);
		\draw[fill] (12,-20) circle(.1);
		\draw[fill] (13,-20) circle(.1);
		\draw (12,-21) -- (12,-20);
		\draw (12,-22) to [out=30,in=150] (13,-20);
		\draw[fill] (13,-19) circle(.1);
		\draw (13,-20) -- (13,-19);
		\draw (12,-21) to [out=30,in=150] (13,-19);
		\draw[fill] (13,-18) circle(.1);
		\draw[fill] (13,-17) circle(.1);
		\draw (12,-20) to [out=30,in=150] (13,-18);
		\draw (13,-18) -- (13,-17);
		\draw (13,-19) to [out=30,in=30] (13,-17);

		\end{tikzpicture}
	\end{center}
	\caption{The $\A_2(1)$ module structure of $\H^{*-1}(MT\O(1),\Z_2)\otimes\H^{*+1}(M\O(1),\Z_2)\otimes\H^*(\B^2\Z_2,\Z_2)$.}
	\label{fig:A_2(1)MTO1MO1B2Z2}
\end{figure}
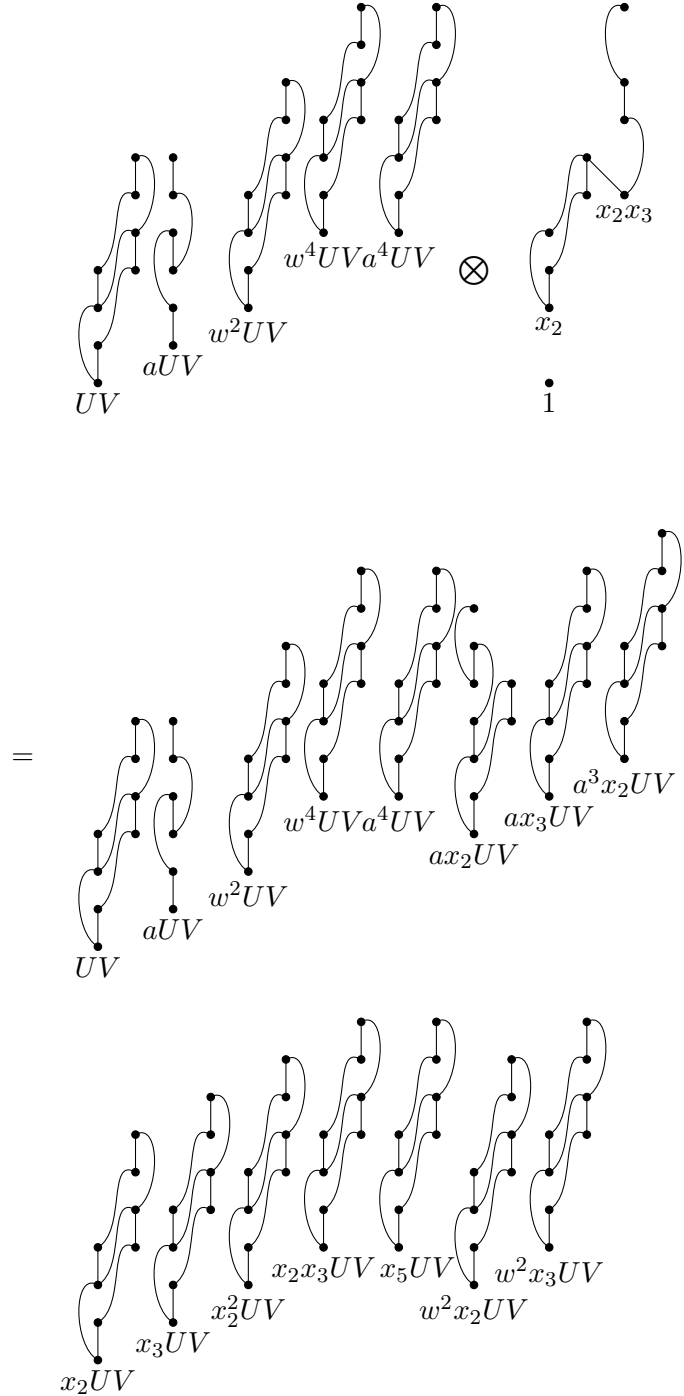

The $\A_2(1)$ module structure of $\H^{*-1}(MT\O(1),\Z_2)\otimes\H^{*+1}(M\O(1),\Z_2)\otimes\H^*(\B^2\Z_2,\Z_2)$ is shown in Figure \ref{fig:A_2(1)MTO1MO1B2Z2}.

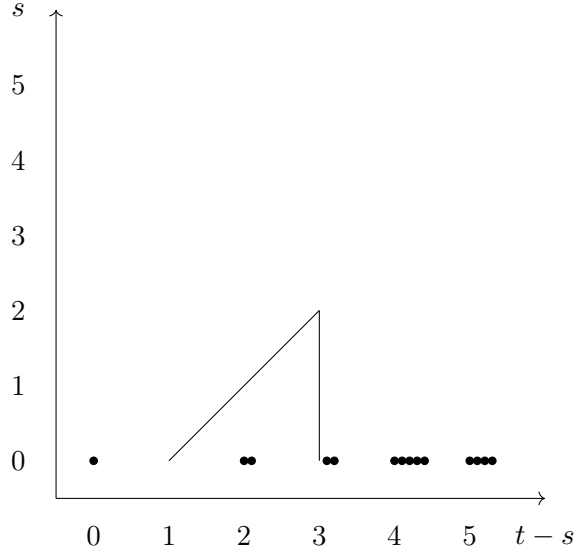
\begin{figure}[H]
	\begin{center}
		\begin{tikzpicture}
		\node at (0,-1) {0};
		\node at (1,-1) {1};
		\node at (2,-1) {2};
		\node at (3,-1) {3};
		\node at (4,-1) {4};
		\node at (5,-1) {5};
		\node at (6,-1) {$t-s$};
		\node at (-1,0) {0};
		\node at (-1,1) {1};
		\node at (-1,2) {2};
		\node at (-1,3) {3};
		\node at (-1,4) {4};
		\node at (-1,5) {5};
		\node at (-1,6) {$s$};
		
		\draw[->] (-0.5,-0.5) -- (-0.5,6);
		\draw[->] (-0.5,-0.5) -- (6,-0.5);
		
		\draw[fill] (0,0) circle(.05);
	\draw (1,0) -- (3,2);
	\draw (3,0) -- (3,2);
	\draw[fill] (2,0) circle(.05);
	\draw[fill] (2.1,0) circle(.05);
	\draw[fill] (3.1,0) circle(.05);
	\draw[fill] (3.2,0) circle(.05);
	\draw[fill] (4,0) circle(.05);	
	\draw[fill] (4.1,0) circle(.05);
	\draw[fill] (4.2,0) circle(.05);	
		\draw[fill] (4.3,0) circle(.05);
		\draw[fill] (4.4,0) circle(.05);
		\draw[fill] (5,0) circle(.05);
		\draw[fill] (5.1,0) circle(.05);
		\draw[fill] (5.2,0) circle(.05);
		\draw[fill] (5.3,0) circle(.05);
		\end{tikzpicture}
	\end{center}
	\caption{$\Omega_*^{\DPin\times\B\Z_2}$}
	\label{fig:E_2DPinBZ2}
\end{figure}

The $E_2$ page of the Adams spectral sequence is shown in Figure \ref{fig:E_2DPinBZ2}.

\begin{table}[H]
	\centering
	\begin{tabular}{ c c c c}
		\hline
		\multicolumn{4}{c}{Bordism and Cobordism group}\\
		\hline
		$d$ & 
		$\Omega^{\DPin\times\B\Z_2}_d$ & $\TP^{\DPin\times\B\Z_2}_d$
		& Cobordism Invariant \\
		\hline
		0& $\Z_2$ & $\Z_2$ &\\
		1& $\Z_2$ & $\Z_2$ & $\tilde\eta$\\
		2& $\Z_2^3$ & $\Z_2^3$ & $x_2,w^2,a\tilde\eta$ \\
		3 & $\Z_2^2\times\Z_8$ & $\Z_2^2\times\Z_8$ & $ax_2,x_3,a\text{ABK}$\\
		4 & $\Z_2^5$ & $\Z_2^5$ & $w^4,a^4,ax_3,x_2^2,w^2x_2$  \\
		5 & $\Z_2^4$ & $\Z_2^4$&  $a^3x_2,x_2x_3,x_5,w^2x_3$ \\
		\hline
	\end{tabular}
	\caption{Bordism group. Here $a$ and $w$ are explained before.
	 $\tilde\eta$ is the mod 2 index of 1d Dirac operator, ABK is the 2d Arf-Brown-Kervaire invariant. $x_2$ is the generator of $\H^2(\B^2\Z_2,\Z_2)$, $x_3=\Sq^1x_2$, $x_5=\Sq^2x_3$. 
	}
	\label{table:DPinBZ2Bordism}
\end{table}

\bibliographystyle{Yang-Mills.bst}
\bibliography{HAHS-II-Gauge.bib}

\end{document}